\documentclass[paper]{JHEP3}

\usepackage[dvips]{graphicx}
\usepackage{epsfig}
\usepackage{amssymb}
\usepackage{amsmath}
\usepackage{cite}

\def\vb{\begingroup\obeyspaces\u}
\def\u#1{\tt#1\endgroup}

\preprint{Cavendish--HEP--09/05 \\ CERN-PH-TH-2009-043}
\title{Phenomenology of Production and Decay of Spinning Extra-Dimensional
Black Holes at Hadron Colliders%
\footnote{Work supported in part by the UK Science and Technology Facilities
Council.}}
\author{James A.\ Frost$^1$, Jonathan R.\ Gaunt$^1$, Marco O.P.\ Sampaio$^1$,
Marc Casals$^2$, Sam R.\ Dolan$^3$, M.\ Andrew Parker$^1$,
and Bryan R.\ Webber$^{1,4}$\\ 
  $^1$Cavendish Laboratory, University of Cambridge,
  J.J.\ Thomson Avenue, Cambridge CB3 0HE, U.K.\\
  $^2$CENTRA - Instituto Superior T\'ecnico, Av. Rovisco Pais 1, 1049-001 Lisbon, Portugal and School of Mathematical Sciences, Dublin City University,
  Glasnevin, Dublin 9, Ireland\\
  $^3$School of Mathematical Sciences, University College Dublin,
  Belfield, Dublin 4, Ireland\\
  $^4$Theory Group, Physics Department, CERN, 1211 Geneva 23, Switzerland \\ \\ E-mail: frost@hep.phy.cam.ac.uk, gaunt@hep.phy.cam.ac.uk, sampaio@hep.phy.cam.ac.uk, marc.casals@dcu.ie, sam.dolan@ucd.ie, parker@hep.phy.cam.ac.uk, webber@hep.phy.cam.ac.uk}
\abstract{We present results of \vb{CHARYBDIS2}, a new Monte Carlo simulation of black
hole production and decay at hadron colliders in theories with large extra dimensions and
TeV-scale gravity.  The main new feature of \vb{CHARYBDIS2} is a full treatment of the
spin-down phase of the decay process using the angular and
energy distributions of the associated Hawking radiation.
Also included are improved modelling of the loss of angular
momentum and energy in the production process as well as a wider range
of options for the Planck-scale termination of the decay.
The new features allow us to study the effects 
of black hole spin and the feasibility of its observation in such theories.
The code and documentation can be found at {\tt http://projects.hepforge.org/charybdis2/}.} 


\begin{document}


\section{Introduction}\label{sec:intro}

The idea that extra spatial dimensions may provide a solution
to the hierarchy problem by lowering the scale of gravity to the
TeV region~\cite{Antoniadis:1990ew,Arkani-Hamed:1998rs,Antoniadis:1998ig,Randall:1999ee,Randall:1999vf}
also raises the exciting possibility of black hole (BH) production in elementary particle
collisions at high energies~\cite{Giddings:2001bu,Dimopoulos:2001hw}.
For a sufficient number of ``large'' extra dimensions (three or more), the current
experimental and astrophysical limits~\cite{Hannestad:2001xi,Satheeshkumar:2008fb}
do not rule out such a process occurring at the energy scales accessible to the Large Hadron Collider (LHC).
Even if not realized at the LHC, the production and decay of
extra-dimensional black holes poses an important problem in theoretical
physics.  Given the number of dimensions $D=4+n$ and the
fundamental Planck scale $M_D$, one should be able to describe the
production of black holes at collision energies well above $M_D$
using classical general relativity extended to $(4+n)$-dimensions~\cite{Banks:1999gd,Solodukhin:2002ui,Hsu:2002bd,Jevicki:2002fq}.
In addition, at least at black hole masses well above $M_D$, the decay of a black
hole so formed should be describable in terms of the Hawking
radiation~\cite{Hawking:1974sw,Hawking:2002qc} expected from such an object.

Although many important questions remain unanswered, there has been
steady progress in the theoretical understanding of these issues in
recent years.  Limits have been placed on the fractions of the
collision energy and angular momentum that can end up in the
black hole, as functions of the impact parameter of the
collision~\cite{Eardley:2002re,Kohlprath:2002yh,Yoshino:2002br,Yoshino:2002tx,Yoshino:2005hi}.
The fact that the black hole will in general have non-zero angular
momentum has led to detailed studies of the Hawking radiation from
rotating higher-dimensional black holes~\cite{Ida:2002ez,Ida:2005,Harris:2005jx,Duffy:2005ns,Casals:2005sa,Ida:2006tf,Casals:2006xp,Creek:2007sy,Creek:2007tw,Creek:2007pw,Kobayashi:2008,Casals:2008}, which have revealed
interesting features such as differences in the angular distributions
of emitted particles with different spins, as well as modifications of the energy distributions.

In parallel with the purely theoretical work, efforts have been made to incorporate the results obtained into Monte Carlo
programs~\cite{Dimopoulos:2001hw,Harris:2003db,Cavaglia:2006uk,Dai:2007ki,Dai:2009by}
that generate simulated black hole events, for use in
studies of the experimental detectability of this process.
In the present paper we report on progress in refining the widely-used
\vb{CHARYBDIS} event generator~\cite{Harris:2003db,Harris:2004xt,Harris:2004mf},
to take into account the recent
theoretical work on the production and decay of rotating black holes.
This enables us to study the effects of rotation on experimentally
observable quantities, assuming that black hole production does
indeed take place.

In the following section we summarize the current theoretical results
on the formation of black holes in particle collisions and the implied
constraints on their masses and angular momenta.  We also describe
how these results are incorporated into the \vb{CHARYBDIS2}\footnote{We
will refer from now on to the new release as \vb{CHARYBDIS2} and will reserve
\vb{CHARYBDIS} for earlier versions.  The particular version
described here is \vb{CHARYBDIS2.0}.} simulation,
with various models for the joint probability distribution of mass and
spin within the allowed region.  Then, in Sect.~\ref{sec:spin}, we
discuss the spin-down and decay of the black hole through Hawking
emission of Standard-Model (SM) particles confined to the physical 3-brane.
Here again the simulation is refined to take full account of the
theoretical results, and options are included for the modelling of
aspects that are not well understood, such as back-reaction effects.

In any simulation of black hole decay in theories with low-scale gravity,
one eventually reaches the stage at which the mass and/or temperature of
the black hole are comparable with the Planck scale, whereupon a complete
theory of quantum gravity is required.  In the absence of such a theory,
various models for the termination of black hole decay have been
suggested~\cite{Dimopoulos:2001qe,Koch:2005ks,Stoecker:2006we,Meade:2007sz,LorenteEspin:2007gz,Scardigli:2008jn}.
In Sect.~\ref{sec:term} we explain the extended range
of options we have included for this phase of the simulation.

Section~\ref{sec:res} presents some results of the new \vb{CHARYBDIS2}
simulation, with particular emphasis on consequences of black hole
rotation.  Our purpose here is not to perform a detailed experimental analysis, but to illustrate new features that will need to be taken into account
in attempts to extract fundamental parameters from experimental data,
should this process actually occur.

Finally in Sect.~\ref{sec:conc}  we draw some conclusions and make suggestions for further work.
We also discuss the comparative features of \vb{CHARYBDIS2} and the black hole event generator
\vb{BlackMax}~\cite{Dai:2007ki,Dai:2009by}, which is currently the only other simulation that takes black hole
rotation into account. 


\section{Black hole production}\label{sec:prod}
Production of a black hole in a hadron-hadron collision occurs when two of the colliding partons pass sufficiently closely that they become trapped by their mutual gravitational attraction. Prior to the collision, we may take these partons to be travelling in anti-parallel directions, with their directions of motion separated by an impact parameter $b$. A complex shaped event horizon forms, which must quickly relax to one of the
axisymmetric stationary black hole solutions of Einstein's equations according to the `no hair' theorem of classical general relativity~\cite{Hawking:2002qc}. For that reason this is sometimes called the balding phase. An emission of gravitational and gauge radiation accompanies the production event.  

It has been argued~\cite{Giddings:2001bu} that the typical time scale for loss of such asymmetries is of the order of the horizon radius $r_H$ (note we are using natural units -- see appendix \ref{app:conventions}). This is physically reasonable because the asymmetries are related to a distortion of the geometry (with respect to the stationary solution); regardless of the details of the interaction responsible for removing them, it will involve a signal propagating over a region of typical size $r_H$. On the other hand the time scale $\Delta t$ for evaporation of a black hole with a mass well above the Planck scale is controlled by the Hawking energy flux. Since we are only interested in an order of magnitude, we can estimate it by using the $D$-dimensional Schwarzschild case combined with dimensional analysis (see for example Sect.~3 of \cite{Giddings:2001bu}) to obtain
\begin{equation}\label{T_evap_estimate}
\dfrac{\Delta t}{r_H}\propto \left(\dfrac{M}{M_D}\right)^{\frac{D-2}{D-3}} \ .
\end{equation}
The prefactor is a constant of order unity containing the dimensionally reduced energy flux and other convention-dependent constants, and $M_D$ is the Planck mass in $D$ dimensions\footnote{See Appendix~\ref{app:conventions} for our convention on the definition of $M_D$.}. For $M\gg M_D$ the time for evaporation will be much longer than for balding.

 For the theoretical calculations involved in the model for formation and for the Hawking fluxes used in the evaporation, we assume that the gauge charges of the incoming partons are completely discharged during the production/balding phase by Schwinger emission~\cite{Schwinger:1951nm,Gibbons:1975kk}. Further, we assume that the black hole solution formed is always of the Myers-Perry type~\cite{Myers:1986qa} (rather than the alternative `black ring' type, which is known to exist in five dimensions~\cite{Emparan:2002qb}, and for which there is strong evidence for $D \ge 6$~\cite{Ida:2002qf}). Given these assumptions, two questions are relevant to the theoretical modelling of the production phase. The first is how small the impact parameter $b$ has to be before two partons will produce a black hole, or equivalently, what is the parton-level cross section for black hole formation. The second is what fractions of the initial state mass and angular momentum are trapped within the Myers-Perry black hole during the production phase.

Theoretical techniques used in recent years have provided more rigorous and complete answers to these questions than were available at the last release of \vb{CHARYBDIS}. Some of these techniques are discussed in Sect.~\ref{sec:prodtheor}. Their incorporation into the program, to produce a more accurate simulation of the production phase, is discussed in Sect.~\ref{sec:prodmeth}. 

\subsection{Theoretical studies of the production phase} \label{sec:prodtheor}

In the production phase, the system may be described reasonably well using classical physics, provided the parton collision energy is sufficiently far above the Planck scale~\cite{Banks:1999gd,Solodukhin:2002ui,Hsu:2002bd,Jevicki:2002fq}.  In principle, one could find reasonably accurate answers to the questions posed above by solving for the spacetime in the future of two colliding partons. However, this has not been achieved, even numerically in simplified scenarios. 

The standard approach to studying the production phase has been the trapped surface method~\cite{Eardley:2002re,Kohlprath:2002yh,Yoshino:2002br,Yoshino:2002tx,Yoshino:2005hi}. This method utilises the fact that the black hole horizon begins forming in the spacetime region outside the future lightcone of the collision event, where we can solve for the geometry. By studying apparent horizons in spacetime slices of this region, and in particular looking at certain areas associated with the apparent horizon as $b$ is varied, one can set bounds on the parton-level cross section and the mass $M$ and angular momentum $J$ trapped for given $b$.

Few trapped surface calculations conducted thus far have attempted to obtain results for nonzero $b$, making them of limited use to our simulation. The most detailed that has, is that of Yoshino and Rychkov~\cite{Yoshino:2005hi} who considered all of the numbers of dimensions used by \vb{CHARYBDIS2}. They produced $(M,J)$ bounds up to the maximum impact parameters which give rise to apparent horizons in their method. The results of this calculation are therefore the primary theoretical input to our simulation of the production phase.  A more detailed discussion of the Yoshino-Rychkov method is given in Appendix \ref{app:TS_method}.

It should be noted that the Yoshino-Rychkov calculation models the colliding partons as boosted Schwarzschild-Tangherlini black holes, and so neglects the effects of the spin, charge, and finite size of the colliding partons. For each effect individually, trapped surface calculations have been carried out~\cite{Gingrich:2006qh,Yoshino:2007ph,Yoshino:2007yk} - but all are only for $b=0$. One important observation from the calculations is that charge effects may be significant. These will be included in \vb{CHARYBDIS2} when the relevant calculations are extended to nonzero impact parameter.

Alongside the trapped surface method, alternative techniques have been developed which use a perturbative approach and/or other approximations to estimate directly the mass lost in the production phase. In one setup~\cite{Berti:2003si}, the collision is modelled as an ultra-relativistic particle falling into a Schwarzschild-Tangherlini black hole, and the gravitational emission is calculated by assuming that the gravitational effects of the in-falling particle may be treated as a perturbation on top of the Schwarzschild-Tangherlini metric. Another calculation~\cite{Cardoso:2005jq} also uses a perturbative approach and assumes that the collision is instantaneous. As a final example, D'Eath and Payne~\cite{D'Eath:1992hb,D'Eath:1992hd,D'Eath:1992qu} have estimated the mass loss in the $D=4$ axisymmetric collision case by finding the first two terms of Bondi's news function, and then extrapolating off axis. Here some assumptions about the angular dependence of the radiation are made. 

\begin{figure}[t]
  \hspace{2cm}
\includegraphics[scale=0.9,clip=true,bb=3cm 15.5cm 17cm 23.1cm]{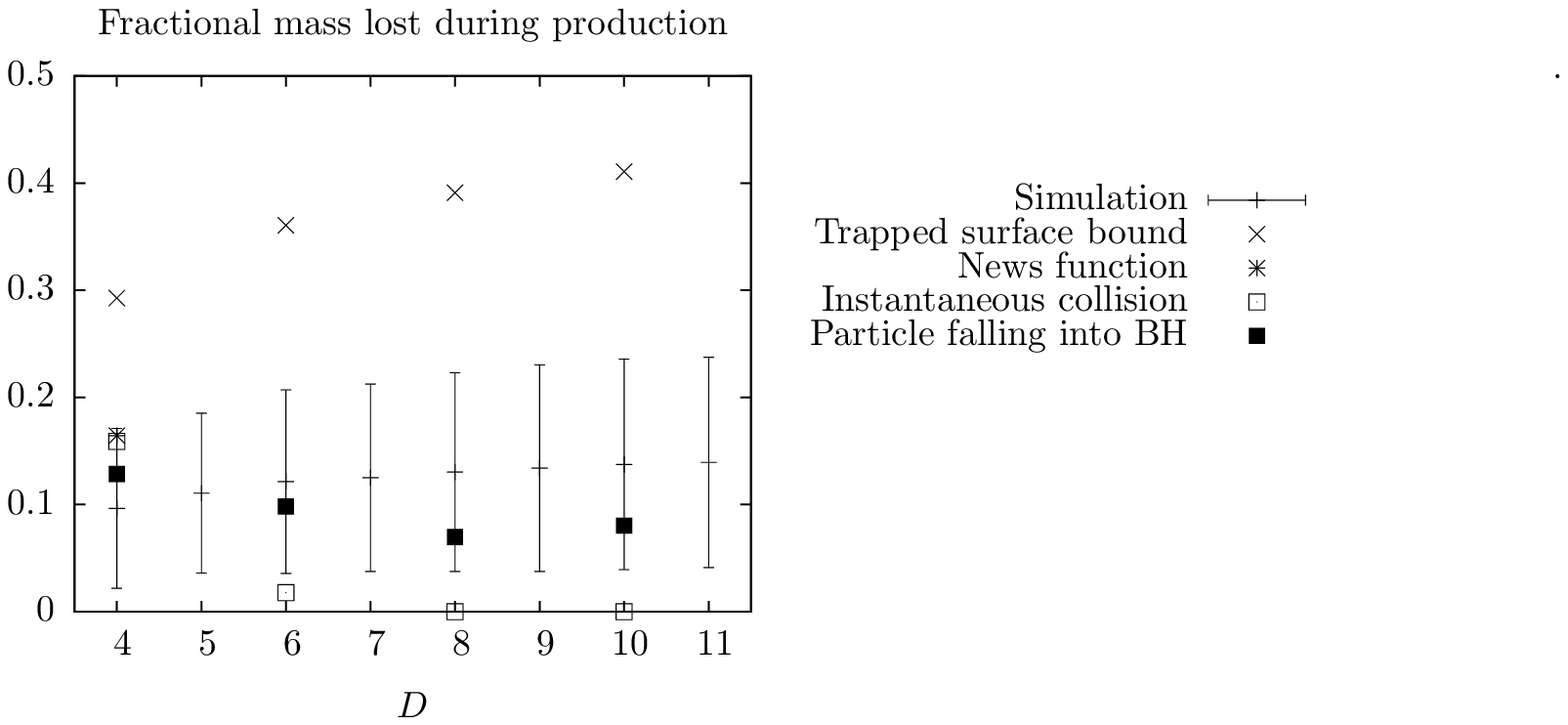} 
   \caption{\label{fig:b0comp} A comparison of various theoretical
     results for the production phase mass loss in the $b=0$ case, with
     the average mass loss produced for $b=0$ by our simulation. Black
     squares: `Particle falling into black hole' results from~\cite{Berti:2003si}. Open squares: `Instantaneous collision'
     results from~\cite{Cardoso:2005jq}. Asterisk: `Bondi's news
     function' result from~\cite{D'Eath:1992hb,D'Eath:1992hd,D'Eath:1992qu}. Crosses: `Trapped
     surface method' upper bound on mass loss. Points with error bars:
     average $b=0$ mass loss from our simulation. The error bars
     represent the standard deviation in the $b=0$ output.}
\end{figure}

The results produced so far by these methods have been limited to $b=0$ and certain values of $D$. The $b=0$ results from different techniques are compared with the trapped surface bound in Fig.~\ref{fig:b0comp}.
 The general indication from these is that much less mass is lost during the production phase than the Yoshino-Rychkov upper limits indicate.

\subsection{Incorporation of the results into \vb{CHARYBDIS2}} \label{sec:prodmeth}

\subsubsection*{Cross section}

In earlier versions of \vb{CHARYBDIS}, parton-level cross sections for different $D$ values were calculated according to the simple formula $\sigma = \pi r_s^2(\sqrt{s})$ which is based on Thorne's hoop conjecture~\cite{Thorne:1972qe}. Here $r_s(\sqrt{s})$ is the radius of the $D$-dimensional Schwarzschild black hole with mass $\sqrt{s}$. Incorporation of the Yoshino-Rychkov cross section results simply requires multiplying these $\sigma$ values by the `formation factors' given in Table~II of~\cite{Yoshino:2005hi}. The increase in $\sigma$ ranges from a factor of $1.5$ at $D=5$ to $3.2$ at $D=11$. The maximum impact parameters for black hole production, $b_{max}$, to be generated in \vb{CHARYBDIS2}, is adjusted accordingly (the two are related through $\sigma = \pi b_{max}^2$).

\subsubsection*{Mass and angular momentum loss}

\begin{figure}[ht]
  \centering 
   \includegraphics[scale=0.68]{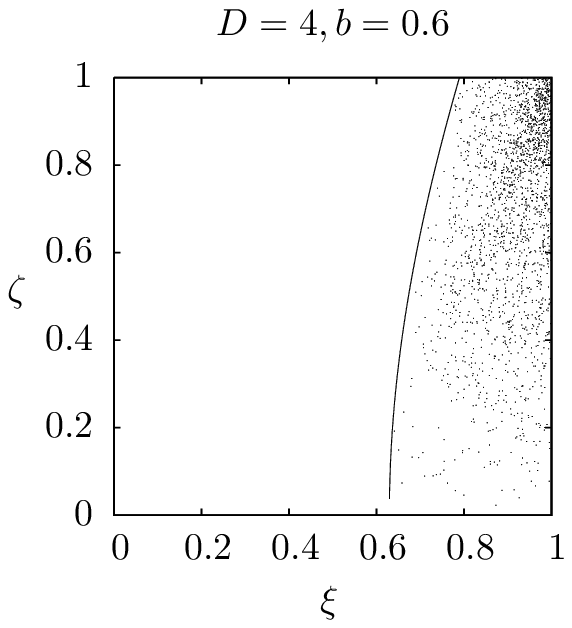}
   \includegraphics[scale=0.68]{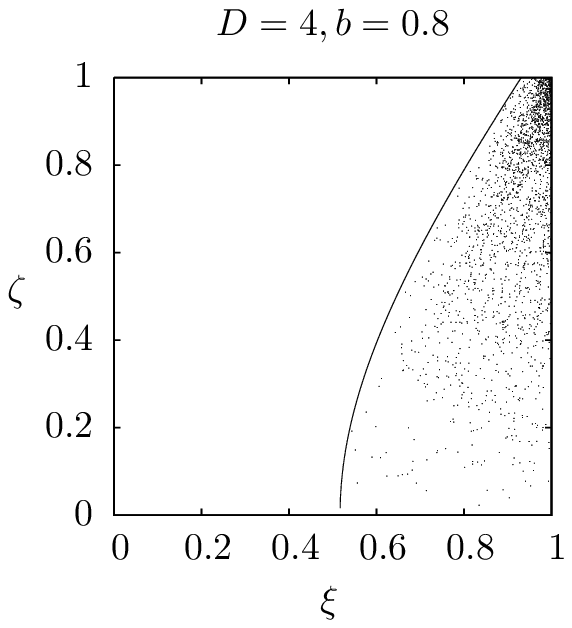}\vspace{3mm} \\
   \includegraphics[scale=0.68]{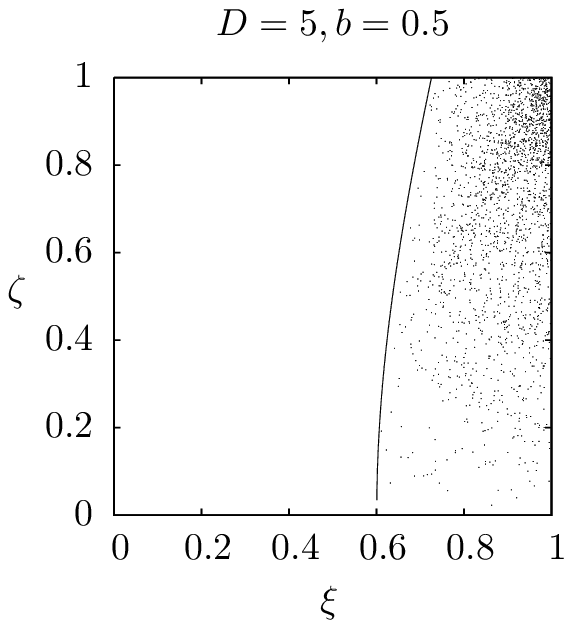} 
   \includegraphics[scale=0.68]{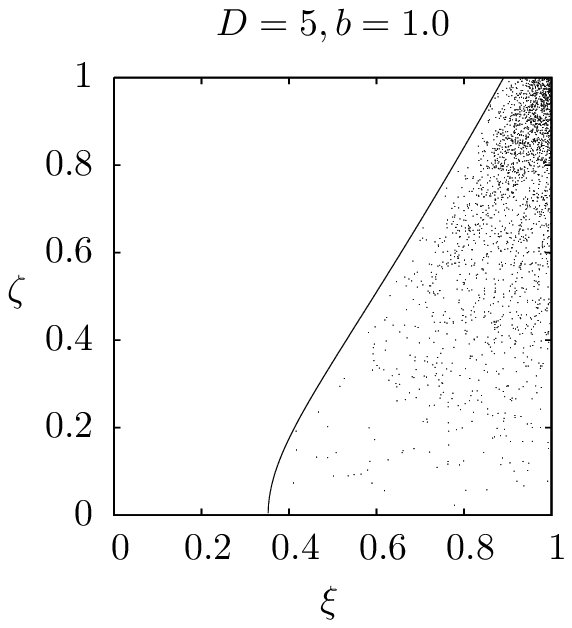}
   \includegraphics[scale=0.68]{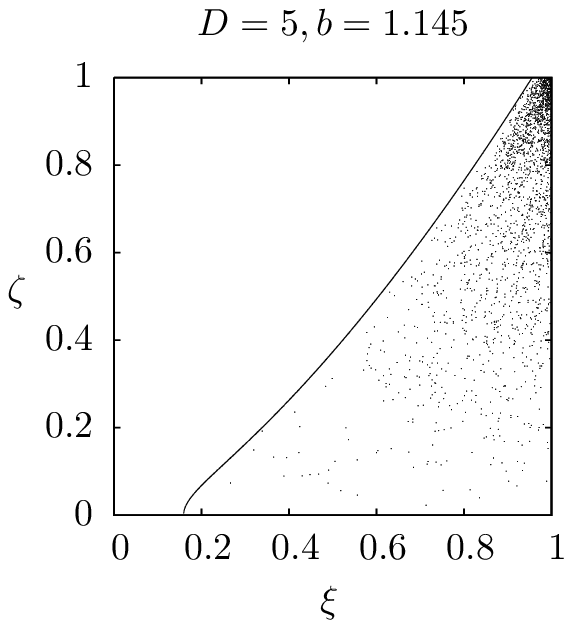}\vspace{3mm} \\
   \includegraphics[scale=0.68]{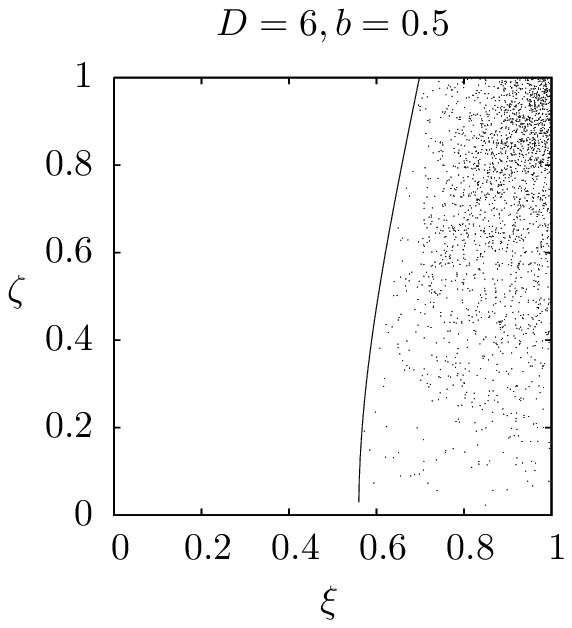} 
   \includegraphics[scale=0.68]{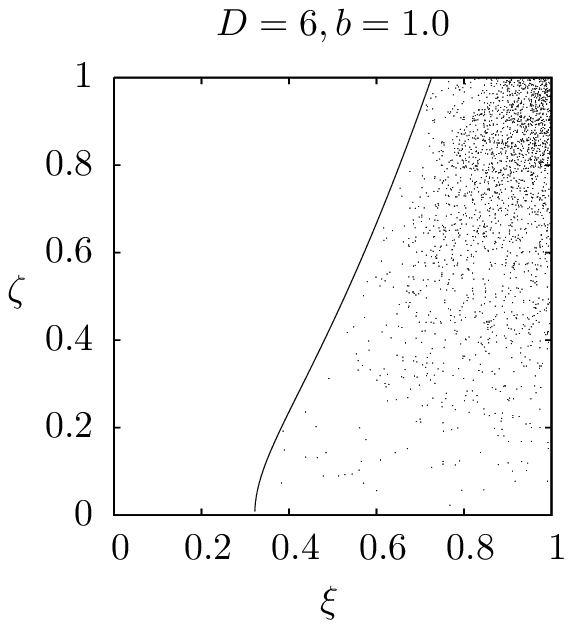}
   \includegraphics[scale=0.68]{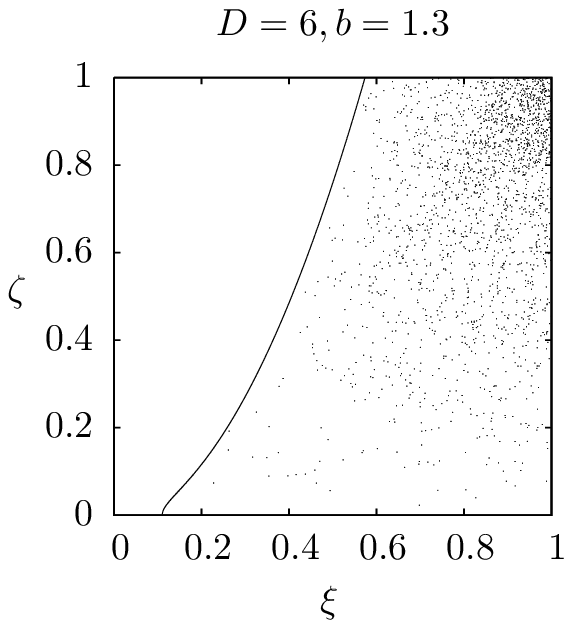}\vspace{3mm} \\
   \includegraphics[scale=0.68]{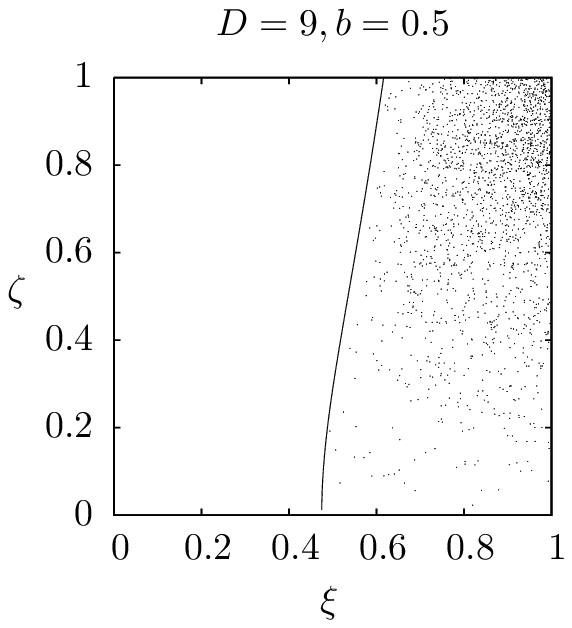} 
   \includegraphics[scale=0.68]{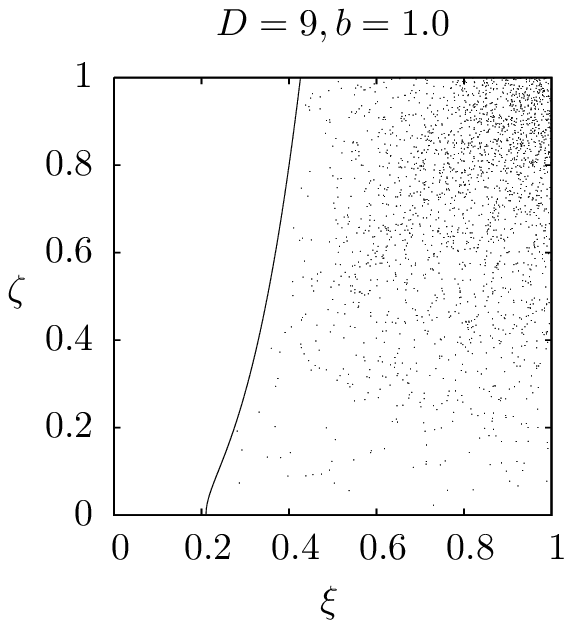}
   \includegraphics[scale=0.68]{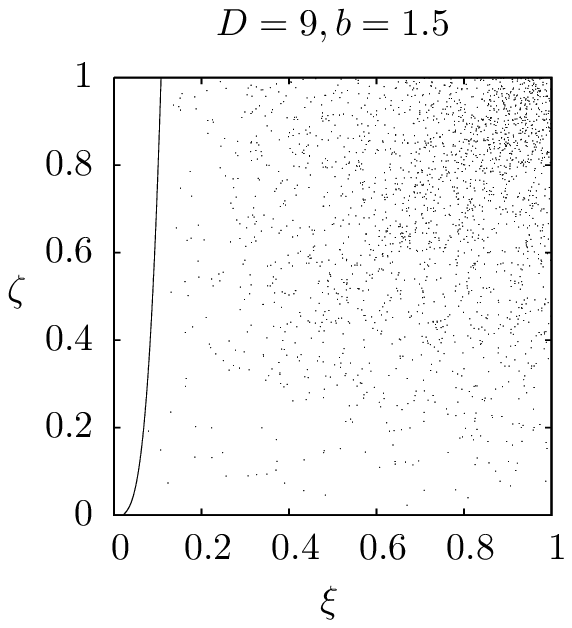}
   \caption{\label{fig:YRplots} Plots displaying the Yoshino-Rychkov bound (solid line) and some output $(\xi, \zeta)$ points from our simulation of the production phase (dots), for selected $D$ and $b$ values (these are given above the plots in each case). Each plot contains 2000 sample output points, which have been generated with \vb{CVBIAS} set to \vb{.TRUE.}.}
\end{figure}

Following Yoshino and Rychkov, we denote the fractions of the initial state mass and angular momentum, trapped after production, by $\xi$ and $\zeta$ respectively. For a given number of dimensions $D$ and impact parameter $b$, the Yoshino-Rychkov bound on these quantities is a curve in the $(\xi,\zeta)$ plane. Examples of such curves for various $D$ and $b$ values are given as the solid lines in Fig.~\ref{fig:YRplots}. 
A boundary curve $\xi_{b}(\zeta)$ always possesses the following two key properties. First, it always passes through $\zeta=0$ with a $\xi$ value between $0$ and $1$, where $\xi_{b}(0)=\xi_{lb}(b,D)$ in the language of~\cite{Yoshino:2005hi}. Second, it increases monotonically after this, passing through $\zeta=1$ with a value satisfying $\xi_{lb} < \xi_{b}(1) \leq 1$. The allowed region is then delimited by this curve and the lines $\zeta=0$, $\zeta=1$, and $\xi=1$.

The new simulation of mass and angular momentum loss based on these curves consists of a point being generated at random in the square $0 \leq \xi \leq 1$, $0 \leq \zeta \leq 1$. The probability distribution for generating this point goes to zero along the Yoshino-Rychkov bound corresponding to the $D$ and $b$ values of the event, such that the generated point is always inside the bound. The $\xi$ and $\zeta$ coordinates of this point are then taken as the fractional mass and angular momentum trapped during the production phase for that event.

The precise rules for the generation of the $(\xi, \zeta)$ point are as follows. First, the $\zeta$ value for the point, $\zeta^*$, is generated, according to a linear ramp distribution. This distribution extends between $\zeta=0$ and $\zeta=1$, with value $0$ at $\zeta=0$ and value $2$ at $\zeta=1$. The $\xi$ value for the point is then also generated. The distribution in this case is similar, except that now it extends between $\xi=\xi_b(\zeta^*)$ and $\xi=1$, ensuring that the point ends up inside the Yoshino-Rychkov bound. The details of how the program calculates $\xi_b(\zeta^*)$ for the $D$ and $b$ appropriate to the event are given in Appendix \ref{app:TS_method}.

The decision to implement a probability distribution favouring smaller mass losses than the Yoshino-Rychkov upper bound was made based on the results from the direct calculations given in Fig.~\ref{fig:b0comp}. In this figure we have plotted the mean mass lost in an event with $b=0$ using the above probability distribution. The error bars represent the standard deviation in the $b=0$ mass loss. We observe a reasonably good agreement between the mean values obtained with our chosen probability distribution and the estimation method results, especially in the important $D=4$ case where the estimation method results agree closely. Given that we favour smaller mass losses, it then seems sensible to ensure that the probability distribution also favours smaller angular momentum losses -- hence the ramp distributions in $\xi$ and $\zeta$.\footnote{Recent results of simulations in four dimensions~\cite{Shibata:2008rq} indicate that $\sim 25\%$ of mass and $\sim 65\%$ of angular momentum are lost in collisions at the maximum impact parameter for black hole formation.  However, the value obtained for the maximum impact parameter
in this case is  $\sim 50\%$ above the Yoshino-Rychkov lower bound, corresponding to an initial-state angular momentum that is more than double the maximum value possible for the black hole that is formed.  Therefore an angular momentum loss greater than 50\% is inevitable in this case.  In $D>5$ dimensions there is no upper limit on the angular momentum of a black hole and such a large loss is not required.}

One possible picture of the production phase is one in which the production phase radiation is `flung out' radially in a frame co-rotating with the forming event horizon. In this scenario, the angular velocity of the event horizon does not change during the production process (where we have to regard the forming black hole as a pseudo-Myers-Perry solution at all points during the production phase for this statement to have any meaning).

Based on this picture, we have included an option to bias the above
probability distribution such that $(\xi,\zeta)$ points corresponding
to smaller changes in the horizon angular velocity are more likely. An
additional condition is added to this bias -- for the points to have
their chance of being picked enhanced, they must also have an
associated value of the oblateness parameter $a_*$, which is sufficiently close to that of the initial state. This is to remedy the problem that, for $D>5$, there are two curves with the same angular velocity as the initial state in the square $0 \leq \xi \leq 1$, $0 \leq \zeta \leq 1$, but only one is connected to the initial state. The bias may be turned on or off using the user switch \vb{CVBIAS}, and the details of its implementation are discussed in Appendix \ref{app:prodprobdist}. 

In each of the plots in Fig.~\ref{fig:YRplots}, there are 2000 $(\xi,\zeta)$ points generated for the appropriate $D$ and $b$ using the new simulation with the bias applied. One can see in each case that there is an increased density of points around the `constant horizon angular velocity' curve.

The mass-energy lost during the production phase is distributed between radiation and the kinetic energy of the formed black hole. The production phase simulation must account for this. On the basis of several calculations~\cite{Cardoso:2003cn,Casadio:2002yj,Casadio:2003iv}, which indicate that gauge radiation is negligible compared to gravitational radiation in the production phase, we assume that all of the radiation is in the form of gravitons. Given that gravitons are missing energy, it is sufficient for the simulation to represent the entire radiation pattern using a `net graviton' with a four-momentum equal to the sum of those of the emitted particles.

The net graviton has an invariant mass $\mu_{g}$, which may potentially lie anywhere between $0$ and $1-\xi$ (in units of the initial state mass). An invariant mass of $1-\xi$ corresponds to a completely symmetric emission of gravitons, whilst lower values correspond to steadily more antisymmetric emissions (which might result if a small number of gravitons is released, and by chance they are emitted in similar directions). In \vb{CHARYBDIS2}, the invariant mass is randomly generated per event from a power distribution, $P(\mu_{g})\propto\mu_{g}^p$. The mean of this distribution is set equal to \vb{FMLOST}$\times (1-\xi)$ by the quantity \vb{FMLOST}$=(p+1)/(p+2)$ (default value $0.99$, corresponding to $p=98$). The simulation of the production phase emission is then a two body decay from initial state object into formed black hole plus net graviton, which is isotropic in the centre of mass frame of the initial state object.

\subsection{Adding the intrinsic spin of the colliding particles}
The results in the previous sections give a model for the angular momentum of the black hole after formation, which is based on using incoming particles with zero spin. Angular momentum conservation requires us to include the intrinsic spin of the incoming particles falling into the black hole. Since the results in the literature taking this effect into account are limited to special cases (see for example~\cite{Yoshino:2007ph}) we assume a simple model where first we combine the spin states of the incoming particles into a state
\begin{equation}
\left|s_1,h_1\right>_z\otimes \left|s_2,h_2\right>_z = \left|s,s_z\right>_z \ .
\end{equation}
 The collision axis is denoted by $z$, $s_i,h_i$ are the spin and helicity of the particles\footnote{Note we are assuming the massless limit where $h_i=\pm s_i$} and $s,s_z$ are the angular momentum quantum numbers of the combined state in the rest frame. Since we have unpolarised beams we give equal weight to each helicity combination. Then this angular momentum state is combined with the orbital contribution obtained from the model for angular momentum loss. We denote it by
\begin{equation}
\left|L,L\right>_{z^\prime} \ ,
\end{equation}
where $L$ is the nearest integer to $J$.
Note that $z^{\prime}$ is an axis in the plane perpendicular to the beam axis ($z$-axis) chosen with uniform probability. Finally, using a Wigner rotation~\cite{Chaichian:1998kd} followed by a tensor product decomposition using Clebsch-Gordan coefficients, we obtain 
\begin{eqnarray}\label{combined_J}
\left|L,L\right>_{z^\prime}\otimes \left|s,s_z\right>_z&=&\left|L,L\right>_{z^\prime}\otimes \sum_{s_z^{\prime}=-s}^{s}d^{(s)}_{s^{\prime}_z,s_z}\left(\cos\frac{\pi}{2}\right)\left|s,s_{z^{\prime}}\right>_{z^{\prime}} \nonumber \\
&=&\sum_{s_z^{\prime}=-s}^{s}d^{(s)}_{s^{\prime}_z,s_z}\left(0\right)\sum_{J=|L-s|}^{L+s}C_{J,L,L,s,s_{z^{\prime}}}\left|J,L+s_{z^{\prime}},L,L,s,s_{z^{\prime}}\right>_{z^\prime} \ ,
\end{eqnarray}
where $d^{(s)}_{s_{z^{\prime}},s_z}$ is a Wigner function and $C_{J,L,L,s,s_{s^{\prime}}}$ is a Clebsch-Gordan coefficient for the tensor product decomposition of $\left|L,L\right>_{z^\prime} \otimes \left|s,s_{z^{\prime}}\right>_{z^{\prime}}$. From \eqref{combined_J} it is straightforward to determine the probabilities for all possible combinations of helicities and incoming partons. 

This model introduces a spread in the orientation of the initial black hole angular momentum axis around the plane perpendicular to the $z$-axis. Note that even though the model for angular momentum loss in the previous sections does not include such an effect, in a realistic situation we would not expect the angular momentum to be exactly perpendicular to the beam axis after the production phase.

\section{Decay of spinning black holes}\label{sec:spin}
After the black hole settles down to the Myers-Perry solution, evaporation will start. The stationary background geometry for an $(4+n)$-dimensional black hole with one angular momentum axis on the brane is~\cite{Myers:1986qa}
\begin{multline}\label{MP_metric}
ds^2=\left(1-\dfrac{\mu}{\Sigma r^{n-1}}\right) dt^2+\dfrac{2a \mu \sin^2{\theta}}{\Sigma r^{n-1}}dt d\phi-\dfrac{\Sigma}{\Delta}dr^2-\\ -\Sigma d\theta^2-\left(r^2+a^2+\dfrac{a^2 \mu \sin^2{\theta}}{\Sigma r^{n-1}}\right)\sin^2{\theta}d\phi^2
-r^2\cos^2\theta d\Omega_n^2 \ , \end{multline}
where
\begin{equation}
\Delta=r^2+a^2-\dfrac{\mu}{r^{n-1}}\,, \hspace{1cm} \Sigma=r^2+a^2\cos^2{\theta} \ ,
\end{equation}
$t$ is a time coordinate, $d\Omega_n^2$ is the metric on an $n$-sphere and $\left\{r,\theta,\phi\right\}$ are spatial spheroidal coordinates. This is clear if we transform to the coordinates $\left\{x,y,z\right\}$ which define a spheroid (through their relation to $r$) according to
\begin{equation}\label{spheroid}
\dfrac{x^2+y^2}{r^2+a^2}+\dfrac{z^2}{r^2}=1
\end{equation}
at any space time point, and are expressed in terms of spheroidal coordinates in the asymptotic $r\rightarrow +\infty$ region through 
\begin{equation}\label{spheroidal_coordinates}
\left\{\begin{array}{rcl} 
x&=&\sqrt{r^2+a^2}\sin\theta\cos\left(\phi+\mathcal{O}(\frac{a}{r})\right) \\ 
y&=&\sqrt{r^2+a^2}\sin\theta\sin\left(\phi+\mathcal{O}(\frac{a}{r})\right) \\
z&=&r\cos\theta
\end{array}\right. \ .
\end{equation}
The only two independent parameters in~\eqref{MP_metric} are $\left\{\mu ,a\right\}$ which are related to the physical parameters of the black hole. Those are respectively the extra dimensional generalisations of the ADM mass ($M$) and angular momentum ($J$)\footnote{The expressions are obtained by looking at the metric when $r\rightarrow +\infty$ and comparing it with the metric for a weak localized massive perturbation in Minkowski space-time (see~\cite{Myers:1986qa} for further details).}
\begin{eqnarray}
\dfrac{M}{M_D}&=&\dfrac{(n+2)}{2}S_{2+n}(2\pi)^{-\frac{n(n+1)}{n+2}}\,M_D^{n+1}\mu \ , \\
J&=&S_{2+n}(2\pi)^{-\frac{n(n+1)}{n+2}}\,M_D^{n+2}\,a\,\mu =\dfrac{2}{n+2}\,M a \ .
\end{eqnarray}
Furthermore, we can switch to a third pair of parameters, closely related to the geometrical properties of the black hole: $\left\{r_H,a_*\right\}$. The first parameter is defined by the location of the horizon of the black hole at the largest positive root of $\Delta(r_H)=0$. $r_H$ is directly related to the surface curvature of the horizon and thus (in some sense) has a frame independent meaning. The second is $a_* = a/r_H$. This is easily interpreted as the oblateness of the spheroid at the horizon as seen from~\eqref{spheroid}.

We are working in an ADD scenario where, the Standard-Model fields are confined to a 4-dimensional brane. This type of setup has been proposed both for the case of flat extra dimensions~\cite{Antoniadis:1990ew,Arkani-Hamed:1998rs,Antoniadis:1998ig} (ADD scenario) and curved extra dimensions (for example the Randall-Sundrum models~\cite{Randall:1999ee,Randall:1999vf}). This constraint is necessary to avoid bounds from electroweak precision observables. 
The upper bound on the extra dimensional width for such a brane is~\cite{Flacke:2005hb} $R\simeq (700 \, \mathrm{GeV})^{-1}$. This is many orders of magnitude below the scale needed in a large extra dimensions scenario (for $n\lesssim 15$) to explain the size of the Planck mass in the spirit of~\cite{Arkani-Hamed:1998rs,Antoniadis:1998ig}. More elaborate setups have been proposed\footnote{Such constructions aim to suppress the effects of dangerous operators which, for example, might allow fast proton decay.} where different SM fields are placed on different branes imbedded in a thicker brane of size $R\lesssim 1\, (\mathrm{TeV})^{-1}$~\cite{ArkaniHamed:1999dc,ArkaniHamed:1999za}. In our model we assume black holes of mass well above $1 \, \mathrm{TeV}$, which means that the typical Schwarzschild radius will be above $1\sim 2 \, (\mathrm{TeV})^{-1}$. Hence the minimum diameter should be $3\sim 4 \, (\mathrm{TeV})^{-1}$, which is already well above the upper bound on the width of the thick brane. So compared to the size of the black hole, all branes should be well inside the black hole and for the purpose of evaporation, all fields effectively behave as being emitted from a single brane. Thus, for brane degrees of freedom, we use the 4-dimensional projected version of the metric~\eqref{MP_metric}, where the coordinates on the $n$-sphere are fixed. Higher order corrections from splitting the branes are neglected.

A related issue is that of brane tension. Here the results in the literature concerning transmission factors are limited to the case of a \mbox{codimension-2} brane in six dimensions~\cite{Kaloper:2006ek,Kiley:2007wb,Dai:2006hf,Kobayashi:2007zi,Chen:2007ay}. The main conclusion (see for example~\cite{Kaloper:2006ek,Dai:2006hf,Kobayashi:2007zi}) is that the brane projected metric (and consequently the Hawking spectra) remains unchanged up to a rescaling of the Planck mass, so only the spectra for bulk fields are qualitatively different. These observations, together with the fact that brane emission should be dominant, justify neglecting such effect for 6-dimensions (note that this case is anyway disfavoured from the bounds in~\cite{Hannestad:2001xi,Satheeshkumar:2008fb}). For the phenomenologically favoured cases of codimension larger than 2, there are virtually no detailed studies on brane tension effects. We may hope that a similar effect of rescaling of the Planck mass will occur for brane fields, in which case our model does not need to be adapted, but further work is required to understand if such an assumption holds.  

For our description of the evaporation to hold, we require the black hole to be sitting on a background that is effectively Minkowski spacetime. This means that the typical size and curvature radius of the extra dimensions must be much larger than the horizon radius of the black hole. This holds for the ADD scenario, where the flat extra dimensions are large enough to solve the hierarchy problem, and for the Randall-Sundrum (RS) model with an infinite extra dimension~\cite{Randall:1999vf}, where the hierarchy problem is solved using a large curvature radius.

We assume that the black hole remains stuck on the brane~\cite{Emparan:2000}. The possibility of ejection would come from graviton emission into the bulk \cite{Cardoso:2005mh,Creek:2006ia}. Since the black hole is formed from SM particles which are themselves confined, we assume that even if gravitons are emitted, the extra-dimensional recoiling momentum is absorbed by the brane, avoiding ejection. One could argue that whichever charges keep the black hole confined to the brane, they are lost at the start of the evaporation through Schwinger emission~\cite{Schwinger:1951nm,Gibbons:1975kk}. However it will be very unlikely that all the different gauge charges are simultaneously neutralized at any stage during the evaporation, if we assume that the black hole decays by emitting one quantum at a time. Furthermore, there are a lot more SM degrees of freedom than gravitational ones so even if the unlikely event of exact neutralisation occurs, it will still be unlikely that a graviton is emitted during the brief period of neutrality. Thus we would expect the number of events in which the black hole is ejected into the bulk to be at most a small fraction of the total. 

Even though exact neutralisation seems unlikely, it is well known, at least in four dimensions, that black holes tend to discharge very rapidly compared to the evaporation time. So we would expect the charge to stay low\footnote{For further references regarding discharge see chapter 10 of~\cite{Frolov:1998wf} and references therein.}, and charge effects on the probabilities of emission to be less important. In the generator we use a simplified model based on this observation, such that whenever a charged field is selected for emission, the electric charge of the state is selected as to reduce the total charge of the black hole (unless the BH is neutral, in which case equal probabilities for particles and anti-particles are used). To avoid complications in hadronization, baryon number conservation is assumed, and colours are assigned to ensure that colour singlet formation is possible. 

\subsection{Energy distribution and greybody factors}
\label{sec:En_GB_factors}
Once the evaporation starts, the black hole loses its mass and angular momentum through the emission of Hawking radiation. The radiation is thought to be predominantly in the form of SM fields on the brane, and gravitational radiation into the higher-dimensional bulk. It is believed~\cite{Emparan:2000} that ``black holes radiate mainly on the brane'', primarily due to the high multiplicity of the brane-confined SM fields. The conjecture is supported by studies of the ratio of power emission of scalar fields in brane and bulk channels~\cite{Kobayashi:2008, Casals:2008}. Determining the energy balance between the
brane and bulk channels for graviton emission remains a key open question. The emission of `tensor' gravitational modes into the bulk was recently considered in~\cite{Doukas:2009cx,Kanti:2009sn}.  A comprehensive analysis, including all gravitational modes (`tensor', `vector' and `scalar' \cite{Kodama:2003jz,Ishibashi:2003ap,Kodama:2003kk}) for a simply rotating black hole, has yet to be conducted. 
In the following, we assume that emission on the brane is indeed the dominant decay channel.

We assume further that the emission of Hawking radiation may be treated semi-classically. That is, that emission of radiation is a smooth, quasi-continuous process in which the black hole has time to reach equilibrium between emissions, and the energy of particles emitted is much less than the mass of the BH.
These assumptions are valid as long as the mass of the black hole is much larger than the (higher-dimensional) Planck mass $M_D$, in which case the typical time between emissions is large compared to $r_H$ (see for example \eqref{T_evap_estimate}) and the Hawking temperature, which gives the typical energy scale of the emissions, is below $M_D$.

The Hawking temperature of a Myers-Perry black hole is
\begin{equation}
T_H = \frac{(n + 1) + (n - 1) a_*^2}{4 \pi (1 + a_*^2) r_H}\;.  \label{eq-T-Hawking}
\end{equation}
The particle flux, mass and angular momentum emitted by the BH per unit time and frequency in a single particle species of helicity $h$ are
\begin{equation}
\frac{d^2 \left\{ N , E, J \right\} }{dt d\omega} = \frac{1}{2\pi} \sum_{j=|h|}^\infty \sum_{m = -j}^{j} \frac{\left\{ 1, \omega, m \right\} }{\exp(\tilde{\omega}/T_H) \pm 1} \mathbb{T}^{(D)}_{k}(\omega, a_*)\;,  \label{eq-flux-spectrum}
\end{equation}
where $N$, $E$, and $J$ denote particle number, energy and angular momentum, respectively. 
Here, $k = \{h, j, m, \omega\}$ is shorthand for the numbers, and $\tilde{\omega} = \omega - m \Omega$, where $\Omega = a_* / [(1+a_*^2) r_H]$ is the horizon angular velocity. $\{h,j,m\}$ are the spin weight, total angular momentum and azimuthal quantum numbers of the emission and $\omega$ its energy.  In the denominator, we select $+1$ for fermionic fields ($|h| = 1/2$), and $-1$ for bosonic fields ($|h|= 0, 1, 2$). The dimensionless quantity $\mathbb{T}^{(D)}_{k}(\omega, a_*)$ is the \emph{transmission factor}, also known as the  \emph{greybody factor}. 

In the context of Hawking radiation, a transmission or greybody factor is the proportion of the flux in a given mode that escapes from the horizon to infinity. 
No closed-form expression for these factors is known, although a number of useful approximations have been derived~\cite{Creek:2007tw, Creek:2007sy, Creek:2007pw}.   
Simulation with \vb{CHARYBDIS2} requires accurate transmission factors across a wide parameter space; consequently a numerical approach was taken.

To determine transmission factors one must solve the higher-dimensional Teukolsky equations~\cite{Duffy:2005ns,Casals:2005sa, Casals:2006xp}. After performing a separation of variables for a field on the brane, the resulting radial equation is
\begin{equation} \label{teuk-eq}
\Delta^{-h}\frac{d}{dr}\left(\Delta^{h+1}\frac{d R_k}{dr}\right)+ 
\left[\frac{K^2-ihK\Delta'(r)}{\Delta}+4ih \omega r+h(\Delta''(r)-2)
\delta_{h,|h|} -  \Lambda_k  \right] R_k =0  
\end{equation}
where
$K = (r^2 + a^2) \omega - a m$,
and 
$
\Delta = r^2 + a^2 - (r_H^2 + a^2) r_H^{n-1} /  r^{n-1} .
$
Here $\Lambda_k = \mathcal{A}_k - 2 m a \omega + a^2 \omega^2$, where $\mathcal{A}_k$ is the angular separation constant defined later on in~\eqref{angular_equation}. The transmission factors $\mathbb{T}^{(D)}_{k}$  are found by considering modes which are purely ingoing at the outer horizon (for details see e.g.~\cite{Duffy:2005ns, Casals:2005sa, Casals:2006xp}).

The separation constant $\Lambda_k$ is found by solving the angular equation (\ref{angular_equation}) discussed in the next section. Unlike the radial equation, the angular equation does not depend on the dimensionality $(D = 4+n)$ of the bulk spacetime. A range of methods for finding angular eigenvalues are detailed in~\cite{Berti:2005gp}. For the purpose of calculation of transmission factors, we employed a spectral decomposition method (for example, see Appendix A in~\cite{Hughes:2000}) and a numerical shooting method~\cite{Casals:2005}. We checked our results against known series expansions in $a \omega$~\cite{Berti:2005gp}.

The numerical methods employed to determine the transmission factors are described in detail in~\cite{Duffy:2005ns, Casals:2005sa, Casals:2006xp}. Transmission factors were computed numerically in the parameter range
$n = 1, 2, \ldots 6$, $\omega r_H = 0.05, 0.10, \ldots 5.0$ and $a_* =
0.0, 0.2, \ldots 5.0$ for the angular modes $j = |h|, |h| + 1, \ldots
|h|+12$ and $m = -j \ldots j$. For each point we have computed the flux spectrum using \eqref{eq-flux-spectrum}. This quantity is used in \vb{CHARYBDIS2} as a probability distribution function for the quantum numbers of a particle with a given spin and to determine the relative probability of different spins (through integration of equation~\eqref{eq-flux-spectrum}). For convenience in the Monte Carlo, we have computed the following cumulative distributions from the transmission factors:
\begin{eqnarray}
C_{h,j,m,a_{*},D}(\omega r_H)&=&\int_0^{\omega r_H}\mathrm{d}x\frac{1}{\exp(\tilde{x}/\tau_H) \pm 1} \mathbb{T}^{(D)}_{k}(x, a_*)  \\
C_{h,a_{*},D}(K)&=&\sum_{Q=1}^{K}C_{h,jm,a_{*},D}(\omega r_H\rightarrow \infty) , \label{cumulative_lm}
\end{eqnarray} 
where $x$ is energy in units of $r_H^{-1}$, $\tau_H=T_Hr_H$, $\tilde{x}=x-m\Omega/r_H$ and $Q$ is an integer that counts modes. The modes are ordered with increasing $j$ and within equal $j$ modes they are ordered with increasing $m$.

Cumulative functions are more convenient since they allow for high efficiency when selecting the quantum numbers. This is done by generating a random number in the range $[0,C(\infty)]$, followed by inversion of the corresponding cumulant. In \vb{CHARYBDIS2}, when values of $a_*$ between those mentioned above are called, linear interpolation is used.  When $a_*$ is larger than 5, we use the cumulative functions for $a_*=5$. We have checked that for most of the evaporation such large values are very unlikely. The exception is the final stage, when the black hole mass approaches the Planck mass. Here one of the remnant models takes over, as described in section \ref{sec:term}.

The effect of black hole rotation on the semi-classical Hawking emission spectrum is described in detail in the studies~\cite{Ida:2002ez,Harris:2005jx,Ida:2005,Duffy:2005ns,Casals:2005sa,Ida:2006tf,Casals:2006xp}. Here we briefly recall some qualitative features. 

In the non-rotating (Schwarzschild) phase, the emission spectrum (\ref{eq-flux-spectrum}) is found from a sum over modes, and the modes are degenerate in $m$. In the 4D case, emission is dominated by the lowest angular mode ($j = |h|$), whereas in the higher-dimensional case, higher modes ($j > |h|$) are also significant. The Hawking temperature (\ref{eq-T-Hawking}) increases monotonically with increasing $n$, and hence the total power (per particle species) increases steeply with increasing $n$. 

\begin{figure}[t]
\begin{center}
\includegraphics[scale=0.65,clip=true,trim=2cm 0cm 0cm 0cm]{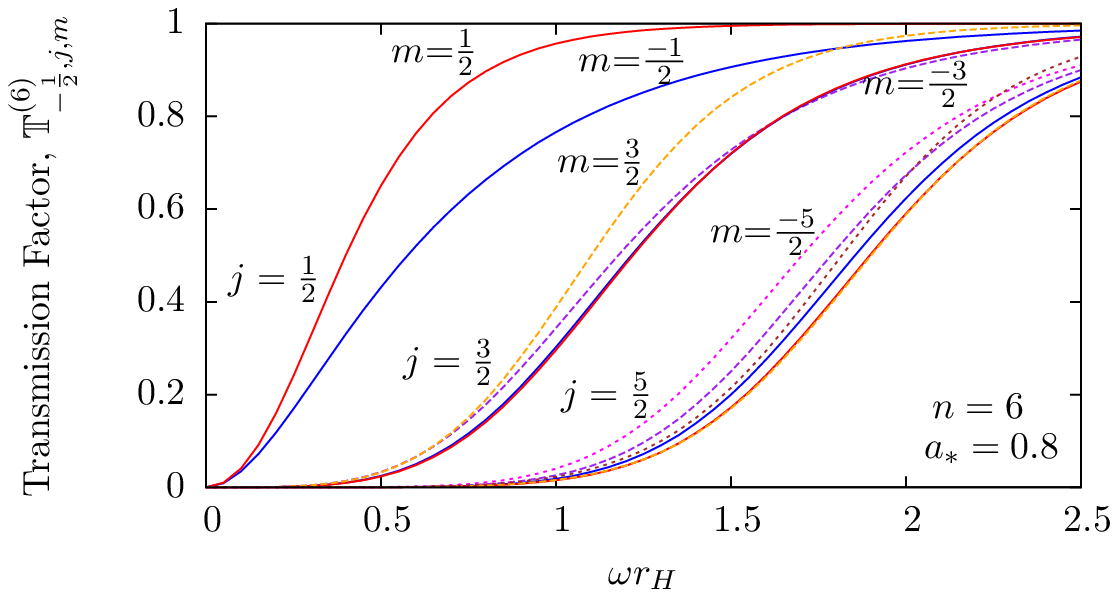}
\includegraphics[scale=0.65,clip=true,trim=2cm 0cm 0cm 0cm]{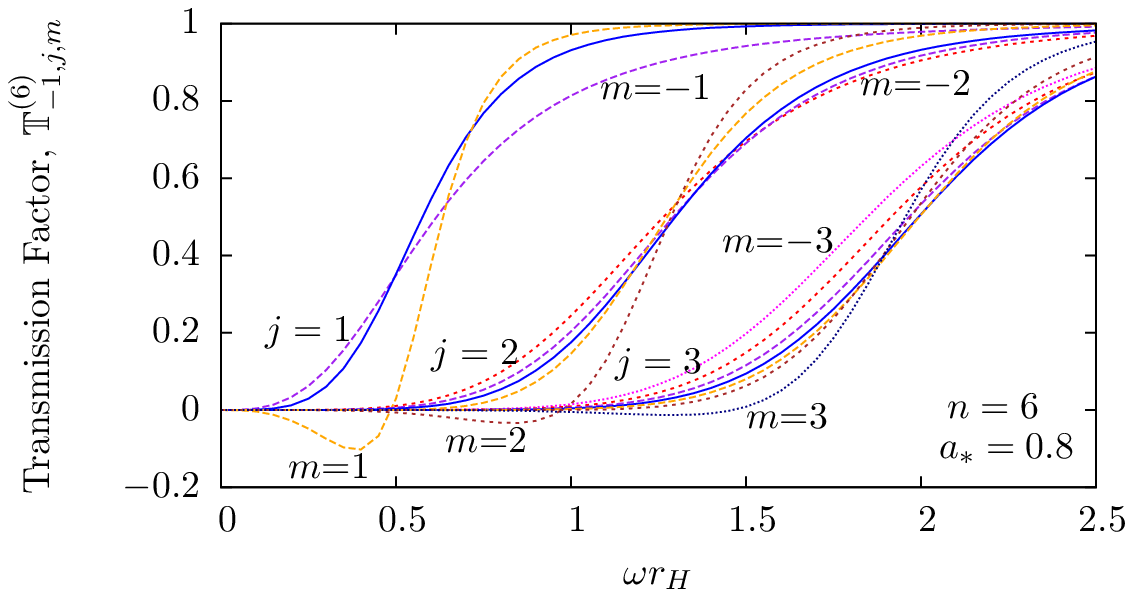}
\includegraphics[scale=0.65,clip=true,trim=2.4cm 0cm -0.3cm 0.2cm]{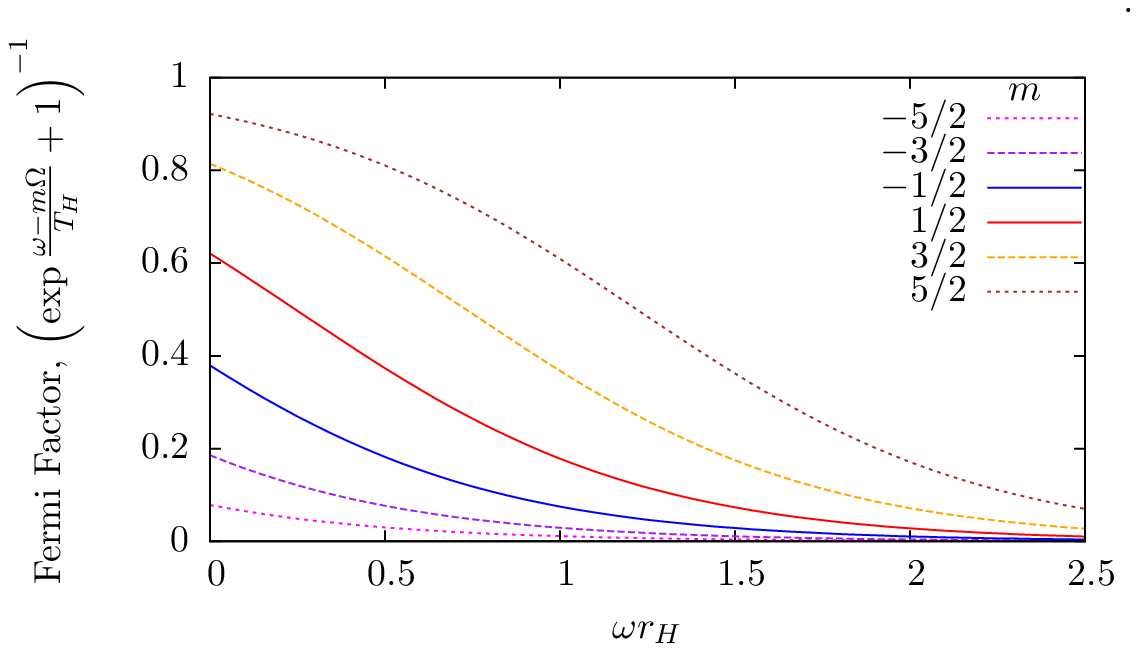} 
\includegraphics[scale=0.65,clip=true,trim=2.3cm 0cm 0.2cm 0.2cm]{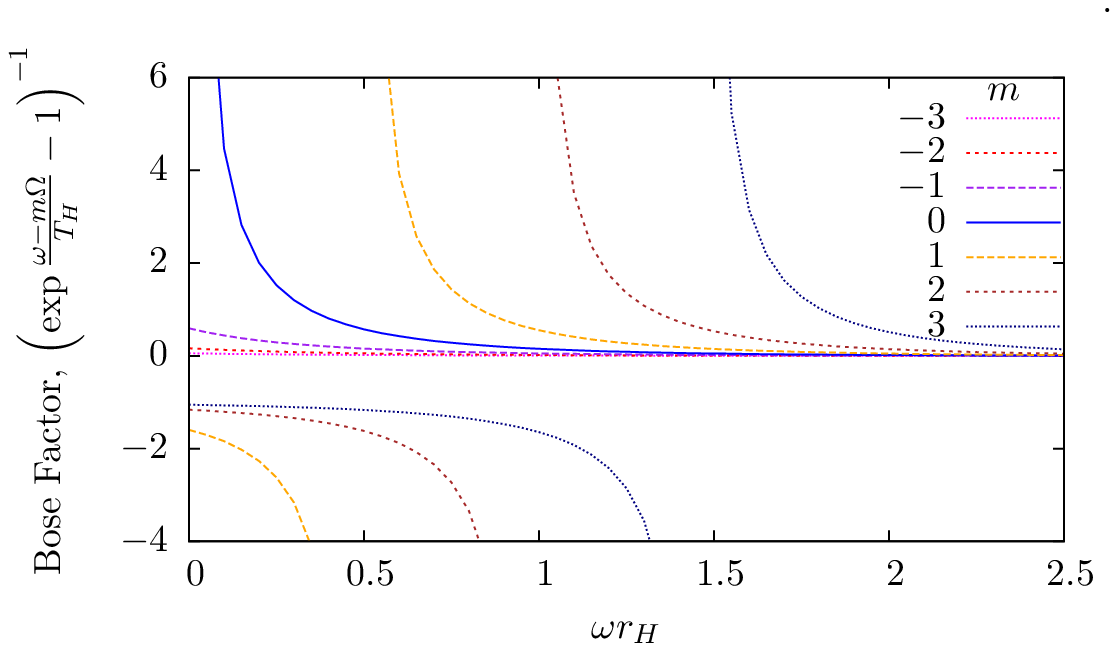}
\end{center}
\caption{\emph{Transmission factors and Planckian factors for $a_* = 0.8$ and $n=6$.} The left plots show spin $1/2$ and the right plots show spin $1$. The top plots show the transmission factors $\mathbb{T}_k$ and the bottom plots show the Planckian factors $\left[ \exp(\tilde{\omega}/T_H) \pm 1 \right]^{-1}$, for a range of $j$ and $m$ modes. Lines with the same $m$ have the same colour and line type.}
\label{fig-trans-factors}
\end{figure}

\begin{figure}[t]
\begin{center}
\includegraphics[scale=0.68,clip=true,trim=0.5cm 0cm 0cm 0cm]{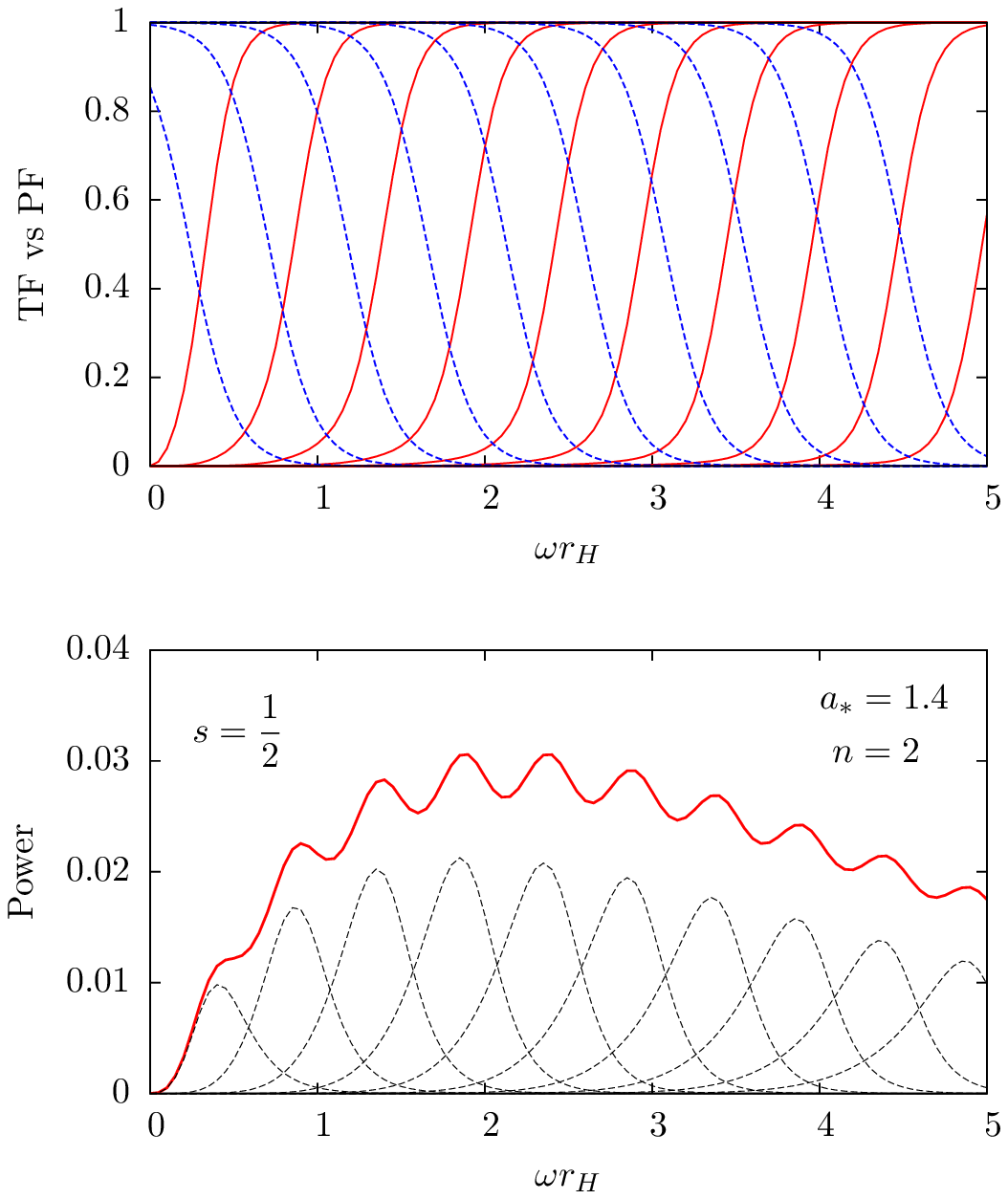} \hspace{3mm}   \includegraphics[scale=0.68,clip=true,trim=0.5cm 0cm 0cm 0cm]{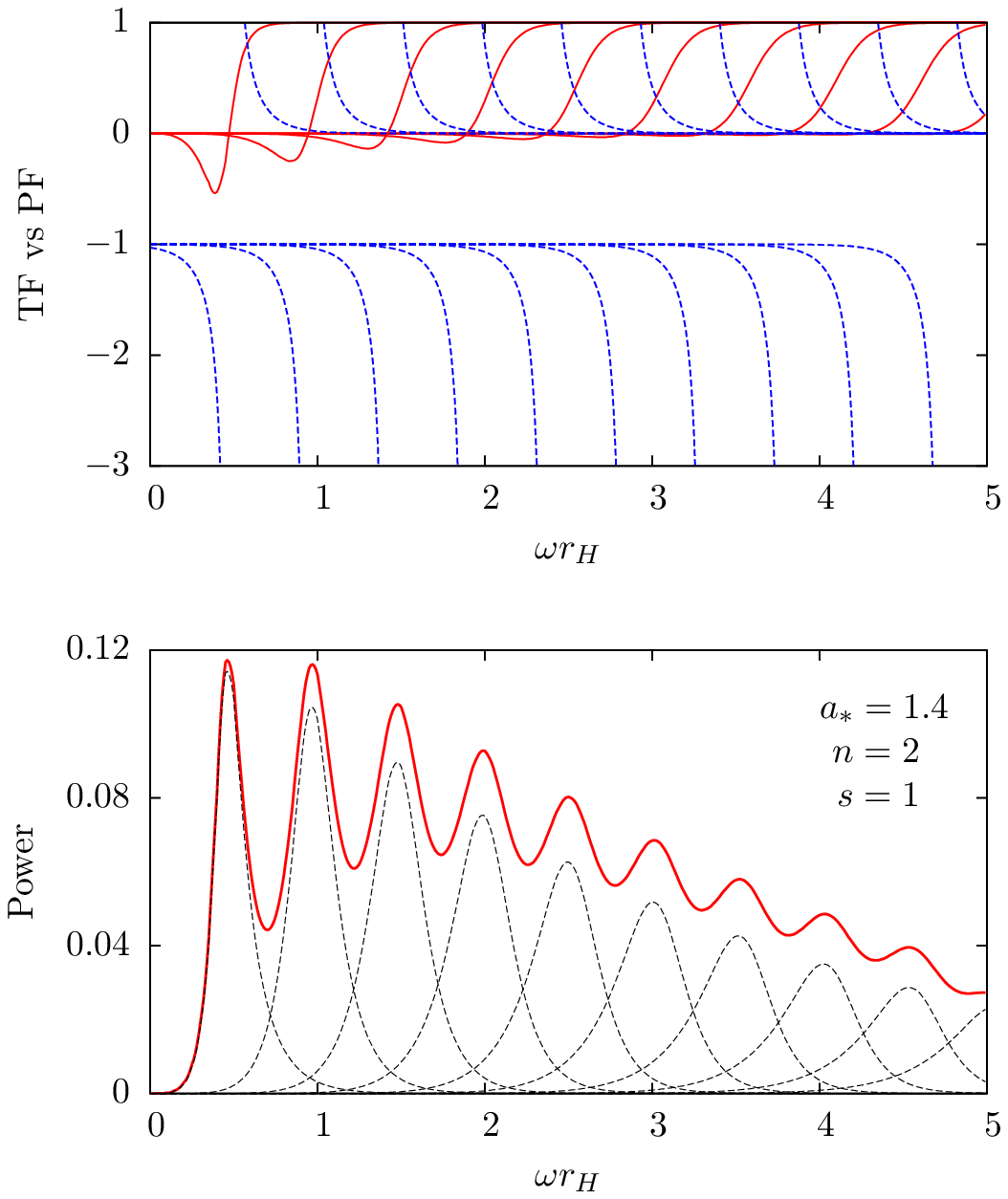}
\end{center}
\caption{\emph{Power spectra for a rotating six-dimensional black hole at $a_*=1.4$.} The top plots show the transmission coefficients (red, solid) and the Planckian factors (blue, dotted) as a function of $\omega r_h$, for the $m = j$ modes up to $j = 19/2$ for fermions (left) and $j=10$ for vector bosons (right). The bottom plots display the power emission spectrum (thick red curve) containing contributions from all modes, together with the curves for the leading $m=j$ modes. The region of overlap between different modes is small, leading to sharply-peaked oscillations in the power emission spectrum.
} 
\label{fig-emission-6d}
\end{figure}

Rotation splits the azimuthal degeneracy; modes of different $m$ are distinguished. Radiation is emitted preferentially into the co-rotating ($m > 0$) modes and causes the BH to lose angular momentum. The importance of co-rotating modes can be understood by examining the interplay between the transmission factors $\mathbb{T}_k$ (TFs) and the so-called Planckian (or thermal) factors ($[\exp(\tilde{\omega}/T_H) \pm 1]^{-1}$) (PFs) in equations (\ref{eq-flux-spectrum}), as illustrated in Figs.~\ref{fig-trans-factors}
and \ref{fig-emission-6d}.
 At fixed $j$, we find the TFs for counter-rotating ($m < 0$) modes generally exceed the TFs for co-rotating modes, implying that (in some loose sense) the counter-rotating modes escape from the vicinity of the hole more easily that the co-rotating modes. However, the PFs are larger for co-rotating modes, implying that more Hawking radiation is generated in co-rotating modes. We find that this latter effect dominates over the former. In general, the effect of rotation is to substantially enhance emission, even though the Hawking temperature depends only weakly on $a_*$ for a fixed BH mass. 

For low dimensionalities ($n = 0, 1, 2$) and fast rotation ($a_* \sim 1$) the emission spectrum is dominated by the maximally-corotating modes ($m = +|j|$). This can lead to an oscillatory or `saw-tooth' emission spectrum, which is clearly visible in Fig.~\ref{fig-emission-6d}. The dominance of $m = +|j|$ modes will obviously lead to a rapid loss of angular momentum. If the number of bulk dimensions is large, $n \gtrsim 6$, then this effect only occurs for very fast rotation $a_* \gg 1$. In general for high $n$, all $m$ modes contribute and the emission profile is much smoother.  

In the case of bosonic fields, black hole rotation induces another effect: the transmission factor $\mathbb{T}_k^{(D)}$ is \textit{negative} for modes such that $\omega\tilde{\omega}<0$ (see right-hand side plots of Figs.~\ref{fig-trans-factors} and \ref{fig-emission-6d}) which can be checked from the Wronskian relations for the radial equation \eqref{teuk-eq}. That is, the transmitted part of `in' wave modes with $\omega\tilde{\omega}<0$ falls into the rotating black hole carrying in negative energy and their reflected part returns to infinity with a gain in energy\footnote{Note that the denominator in Eq.~\mbox{\eqref{eq-flux-spectrum}} is negative for superradiant modes in the bosonic case, thus cancelling out the negativity of the superradiant transmission factor $\mathbb{T}_k^{(D)}$, and so contributing positively to the fluxes.}. This classical phenomenon, which occurs for bosonic but not for fermionic fields, is known as \textit{superradiance}~\cite{Zeldovich:1971,Starobinskii:1973,StarobinskiiChurilov:1973}. Within the context of higher-dimensions, it has been shown~\cite{Casals:2005sa} that spin-1 superradiance on the brane increases with the black hole intrinsic angular momentum $a_*$ (as expected) and with the number $n$ of extra dimensions; see also~\cite{Ida:2005zi,Ida:2005} for studies of scalar superradiance on the brane. It has been suggested~\cite{Frolov:2002xf} that the presence of superradiance might lead to spin-2 radiation into the bulk being the dominant emission channel,
thus disproving the claim that ``black holes radiate mainly on the brane" made in \cite{Emparan:2000}. However, it was shown in~\cite{Casals:2008} (see also~\cite{Nomura:2005mw,Jung:2005pk}) that superradiance for the scalar field emission in the bulk diminishes with the number $n$ of extra dimensions, and that the percentage of the total (i.e., brane plus bulk) scalar power which is emitted into the bulk is always below 35\% for $n=1,2,\dots,6$. In the absence of a solution for the higher-dimensional spin-2 equations in the rotating case, these results for the scalar field seem to indicate that the main emission channel will be into the brane, rather than the bulk, in spite of superradiance. 

\subsection{Angular distribution of Hawking radiation}
The angular distribution of brane fields emitted from a black hole rotating on the physical (3+1)-brane is controlled by the spheroidal wave functions $S_k(c,x=\cos\theta)$, which satisfy the differential equation
\begin{equation}\label{angular_equation}
\left[\dfrac{d}{dx}\left((1-x^2)\dfrac{d}{dx}\right)+c^2x^2-2hcx-\dfrac{(m+hx)^2}{1-x^2}+\mathcal{A}_{k}(c)+h\right]S_{k}(c,x)=0 \ ,
\end{equation}
where $c= a\omega$ and  $\mathcal{A}_k(c)$ is the angular eigenvalue. In the generator, our method of solution for this equation is based on that of Leaver~\cite{Leaver:1985ax}: details may be found in Appendix~\ref{app:spher}.  Given the values of $k$ and $\mathcal{A}_k(c)$, the probability distribution function for an emission with momentum in the direction $\cos\theta$ is then given by the normalized square modulus\footnote{From now on, we will drop the explicit dependence on $x$ and $c$ whenever convenient.} of~$S_k$. This follows from the decomposition of spheroidal one-particle states into plane wave one-particle states, which is analogous to the decomposition of spherical waves into plane waves, for the usual case of scattering off a spherical potential. We work out the case of a massless Dirac field in Appendix~\ref{app:sphroidal_plane}, in order to clarify the assignment of a spheroidal function of a certain spin weight to the correct helicity state in our convention. It turns out that in the convention of \eqref{angular_equation} the physical helicity of the particle is actually $-h$.

\begin{figure}[ht]
  \centering
   \includegraphics[scale=0.78]{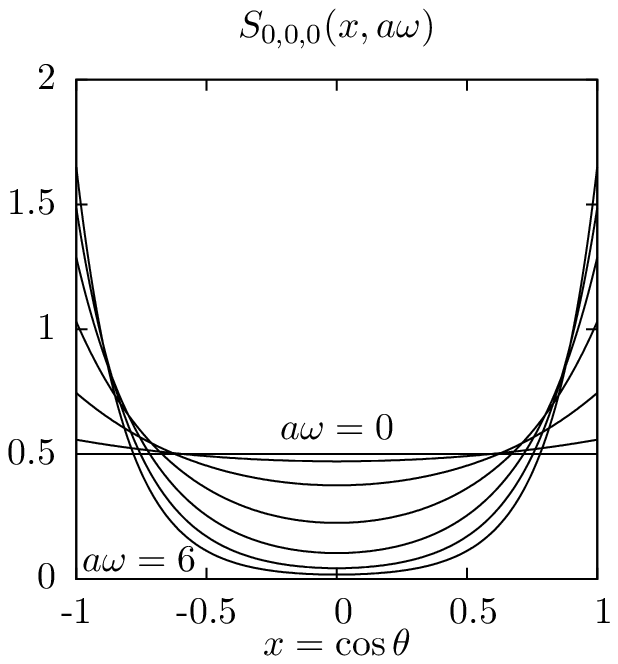} 
   \includegraphics[scale=0.78]{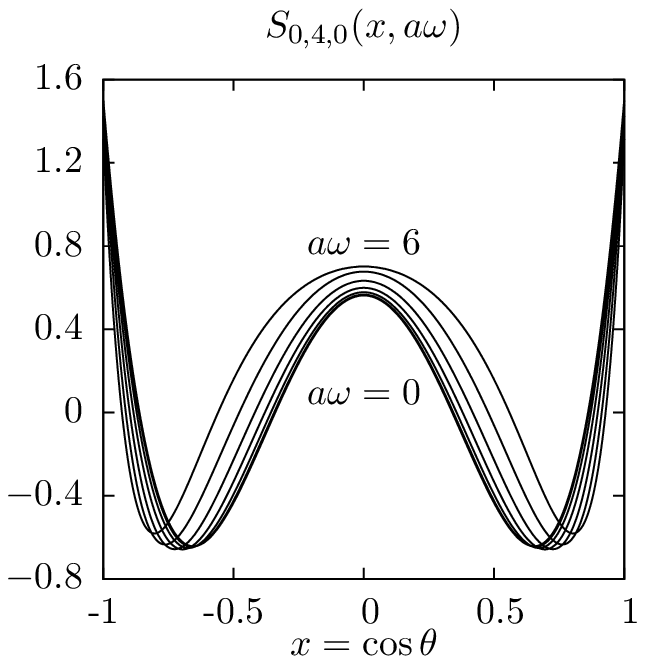}
   \includegraphics[scale=0.78]{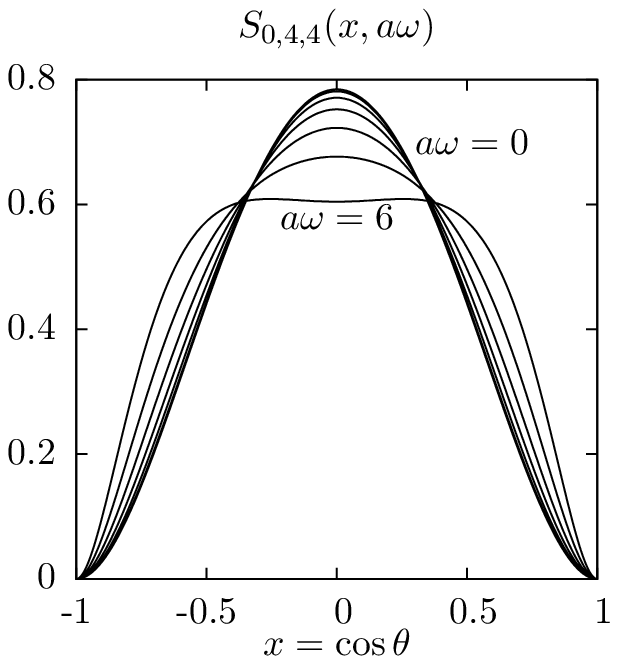}\vspace{3mm} \\
   \includegraphics[scale=0.78]{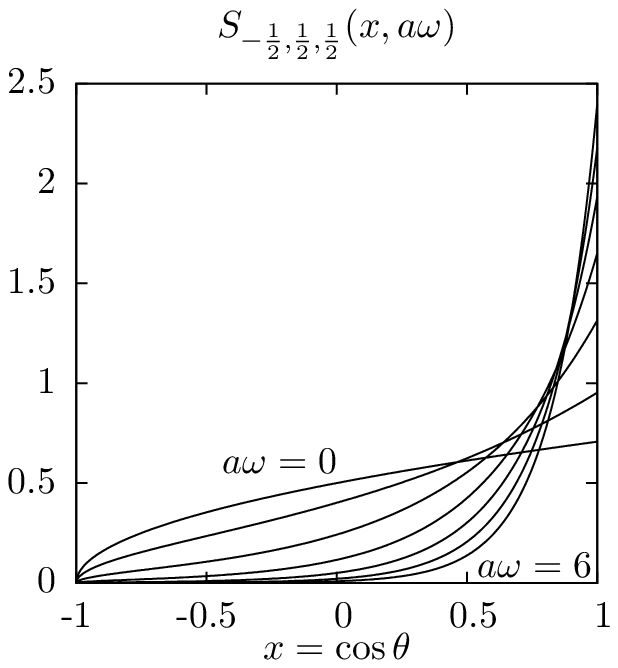} 
   \includegraphics[scale=0.78]{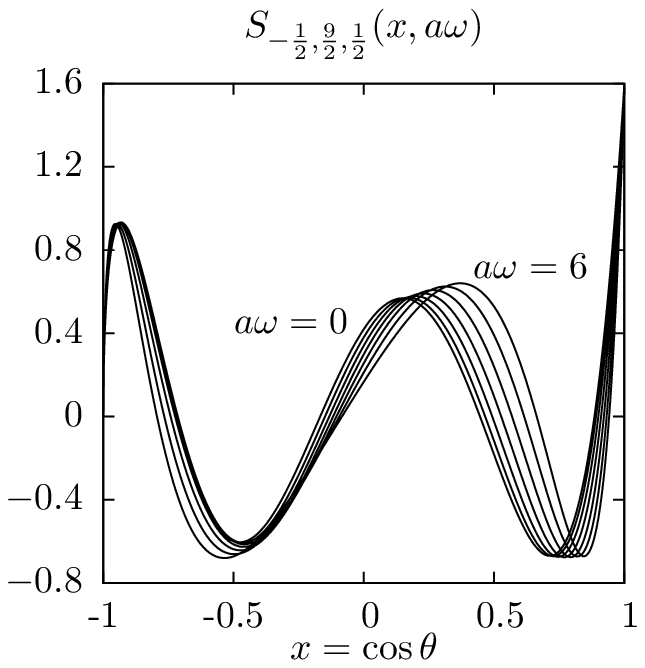}
   \includegraphics[scale=0.78]{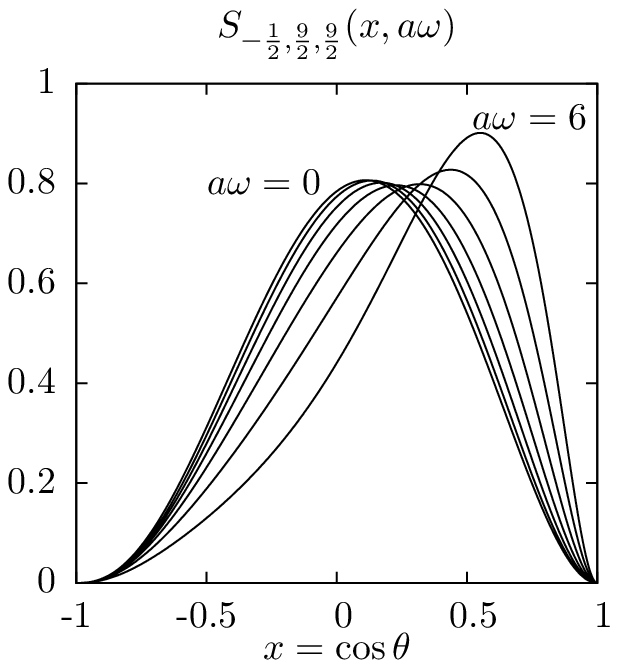}\vspace{3mm} \\
   \includegraphics[scale=0.78]{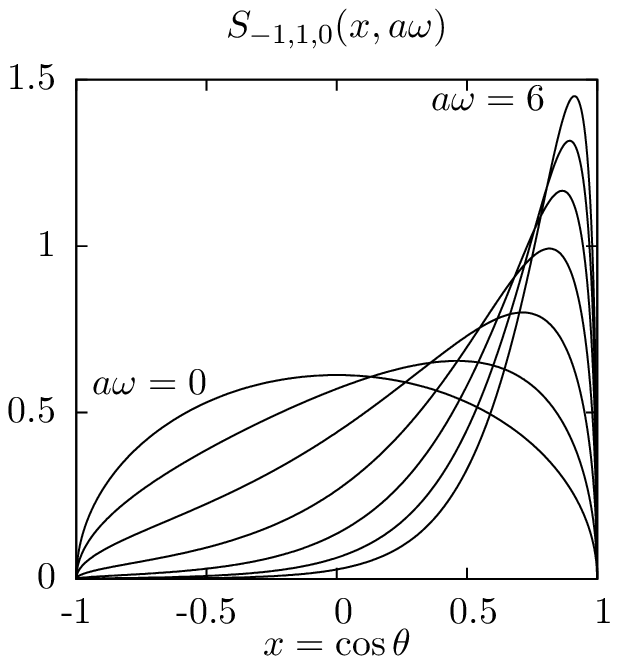} 
   \includegraphics[scale=0.78]{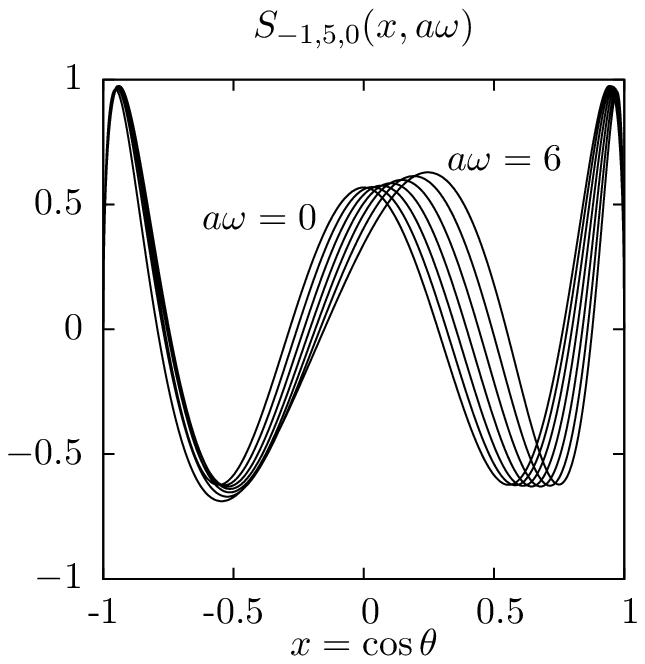}
   \includegraphics[scale=0.78]{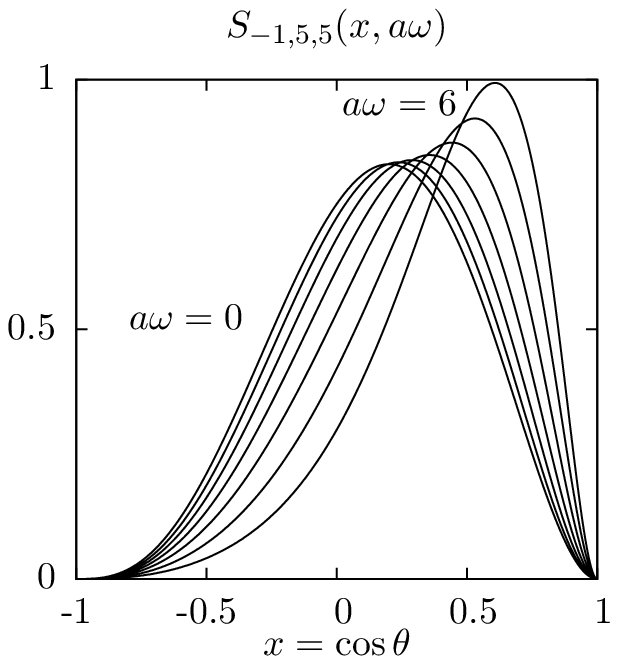}
   \caption{\label{spheroidal_plots}The figure shows various spheroidal wave functions $S_{h,j,m}(x,a\omega)$ as a function of $\cos\theta$. Plots for $a\omega=0,1,2,3,4,5,6$ are shown (only the first and last are indicated, since the curves are regularly ordered). The title of each plot indicates $\{h,j,m\}$.} 
\end{figure}

Some important properties of the angular distributions for our analysis follow from the observation of Fig.~\ref{spheroidal_plots}.
We have verified that in general, for any mode, higher rotation tends to make the spheroidal functions more axial. This means that at low energies (where the modes are departing from being degenerate), the angular distribution of Hawking radiation will tend to become more axial. However, for higher energies, the effect of rotation on the emission spectrum is to favour emission of modes with $m=j$ in order to spin down the black hole. This will produce a more equatorial angular distribution, as we can see from Fig.~\ref{spheroidal_plots} where the $m=j$ mode is always more central (in $x= \cos\theta$) than the $m=0$ mode (for $j\neq 0$). So as we increase the rotation parameter we have a competition between the increase in the angular function's axial character and the increase in probability of emission of more equatorial modes (which are those with larger $j$). At low (high) energies the former (latter) wins.  
This observation is consistent with the energy dependence of the angular profiles shown in, e.g., Fig.~16 of~\cite{Casals:2006xp}. 

Low-energy vector bosons are more likely to be emitted close to the rotation axis, whereas high energy vector bosons are more likely to be emitted in the equatorial plane. A similar but far less pronounced effect exists for spin-half particles. 

Particles with a single helicity, such as the neutrino, will be emitted asymmetrically by a rotating black hole~\cite{Leahy:1979xi}. For example, if the black hole's angular momentum vector is pointing north, then (anti-)neutrinos will be preferentially emitted in the (northern) southern hemisphere. W and Z boson decays will exhibit similar asymmetries in the two hemispheres. 


\subsection{Back-reaction and spin-down}
The difficult problem of studying the back-reaction is interesting, both from the theoretical and the phenomenological point of view, since it should start to influence the evaporation as the mass of the black hole is lowered. 

On the theory side, there is no well established framework to study the evolution of a Hawking-evaporating black hole over the full range of possible initial conditions. The usual approach~\cite{Hawking:1974sw,Page:1976df,Page:1976ki,Page:1977um} is to write down mean value differential equations for the variation of the parameters ($M$ and $J$) such as~\eqref{eq-flux-spectrum} and integrate them with appropriate initial conditions. However, this is only valid for a continuous process of emission where the variation of the parameters is very slow and the symmetry of the background spacetime is kept. Thus, it ignores the momentum recoil of the black hole and the change in orientation of the angular momentum axis between emissions, which are certainly negligible for an ultra-massive black hole, but will start to become important as we approach the Planck mass. Furthermore, since the Hawking spectra at fixed background parameters are used, it also neglects the effect of the backreaction on the metric by the emitted particle. This point has been explored in simplified cases of $j=0$ waves for fields of several spins using the method in~\cite{Parikh:1999mf} with some results regarding the modification of the thermal factors, but a full treatment is still lacking.

In the program, we have included two possible models for the momentum recoil of the black hole set by the switch \vb{RECOIL}, which takes the values 1 or 2.

$\vb{RECOIL}=1$ interprets the selected energy as the energy of the particle in the rest frame of the initial black hole. The momentum orientation is computed in this frame with probability distribution given by the square modulus of the spheroidal function \eqref{angular_equation} and the momentum of the final black hole is worked out from conservation. The argument for this model comes from the observation that particles in the decay are highly relativistic. They propagate close to the speed of light, so the background they see is that of the initial black hole, since no signal of the back-reaction on the metric can propagate outwards faster than light. Thus the momentum of the emission is determined by the background metric in this picture.

$\vb{RECOIL}=2$ takes the energy of the emission as being the loss in mass of the black hole. This corresponds to the usual prescription for computing the rate of mass loss. The orientation of the momentum in the rest frame of the initial black hole is computed as before with a probability distribution given by the square modulus of the spheroidal function \eqref{angular_equation} and the 4-momentum of the emission as well as that of the black hole are worked out. 

Note that for any of the previous options, full polarization information of the emission is kept, as it is generated with the correct angular distribution. This will potentially produce some observable angular asymmetries and correlations, which would not be present if angular distributions averaged over polarizations had been used.

The other quantity we need to evolve is the angular momentum of the black hole. We have two options, controlled by the switch \vb{BHJVAR}. The default $\vb{BHJVAR=.TRUE.}$ uses Clebsch-Gordan coefficients to combine the state of angular momentum $M_z=J$ of the initial black hole (i.e. taking as quantisation axis the rotation axis of the black hole), with the emitted $j,m$ state for the particle. The probability of a certain polar angle and magnitude for the angular momentum of the final black hole is given by the square modulus of the corresponding Clebsch-Gordan coefficient, and the azimuthal angle  is chosen with uniform probability. If \vb{BHJVAR=.FALSE.}, the orientation of the axis remains fixed even though the magnitude will change by subtraction of the $m$ value of the emission.

From previous versions of \vb{CHARYBDIS}, we have kept the switch \vb{TIMVAR} which allows one to fix the parameters of the black hole used in the spectrum (such as the Hawking temperature) throughout the evaporation. This option corresponds to a model where the evaporation is no longer slow enough for the black hole to re-equilibrate between emission, so in effect it represents a simultaneous emission of all the final state particles from the initial black hole without any intermediate states.

\begin{figure}[t]
  \centering
   \includegraphics[scale=0.7]{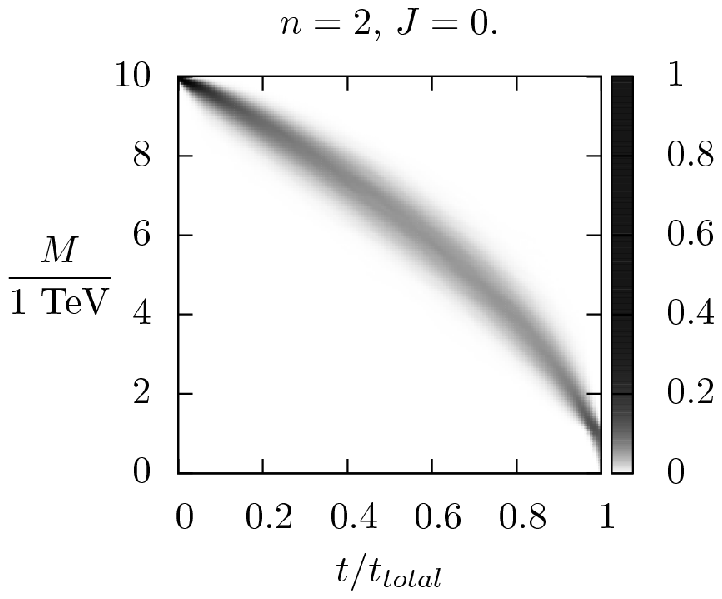} 
   \includegraphics[scale=0.7]{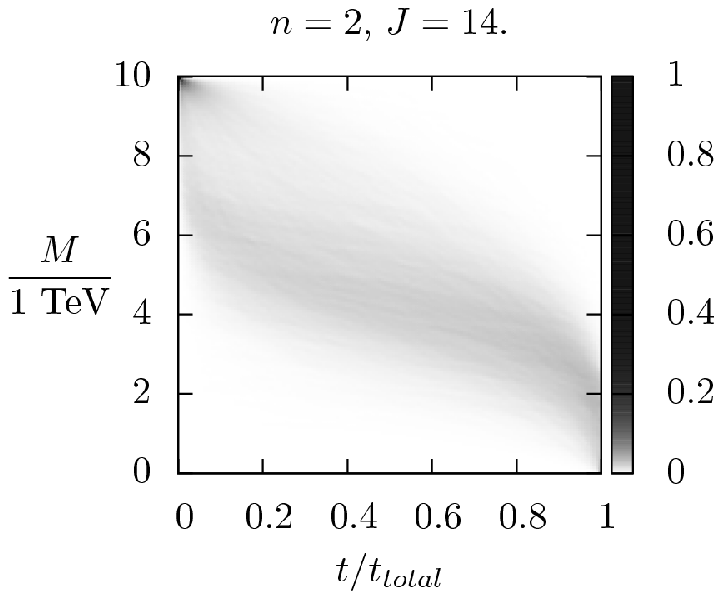} 
   \includegraphics[scale=0.7]{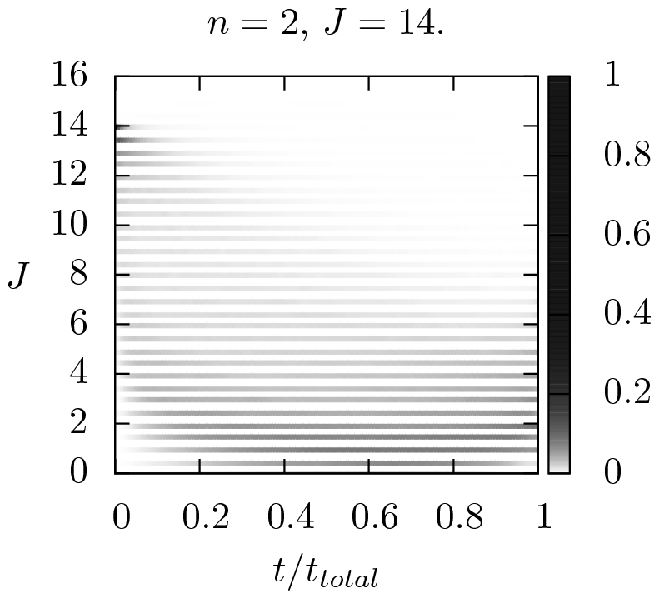}\vspace{3mm} \\
   \includegraphics[scale=0.7]{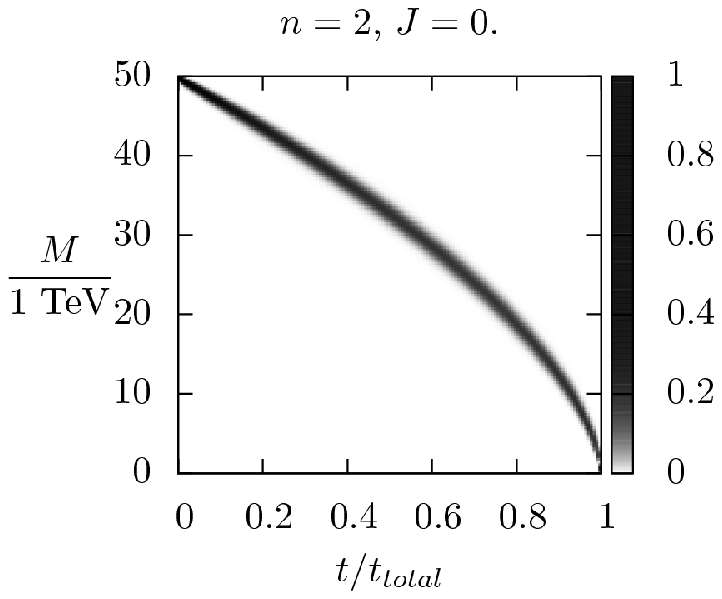} 
   \includegraphics[scale=0.7]{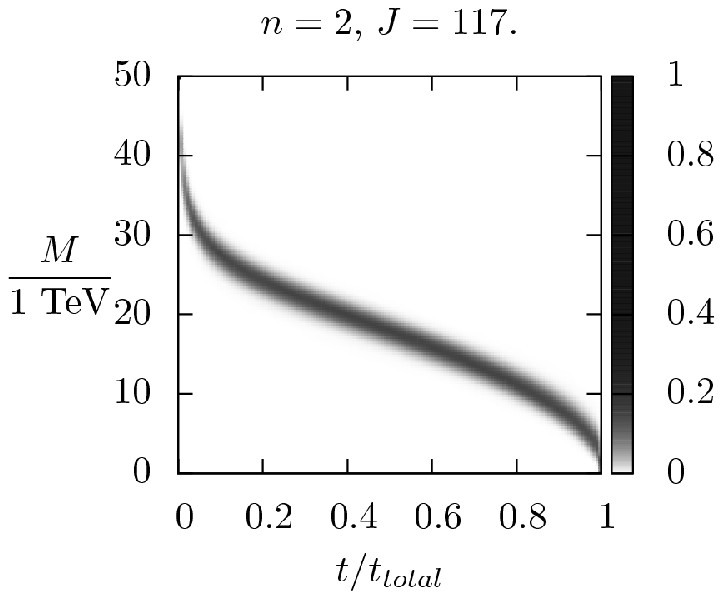} 
   \includegraphics[scale=0.7]{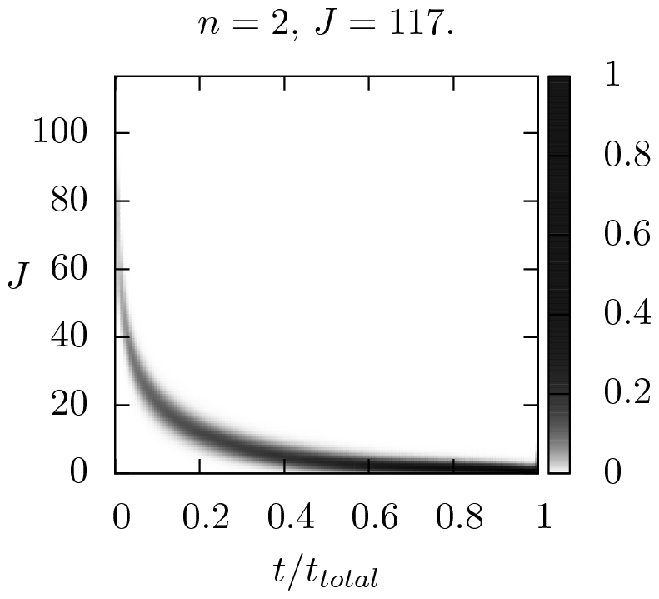}\vspace{3mm} \\
   \includegraphics[scale=0.7]{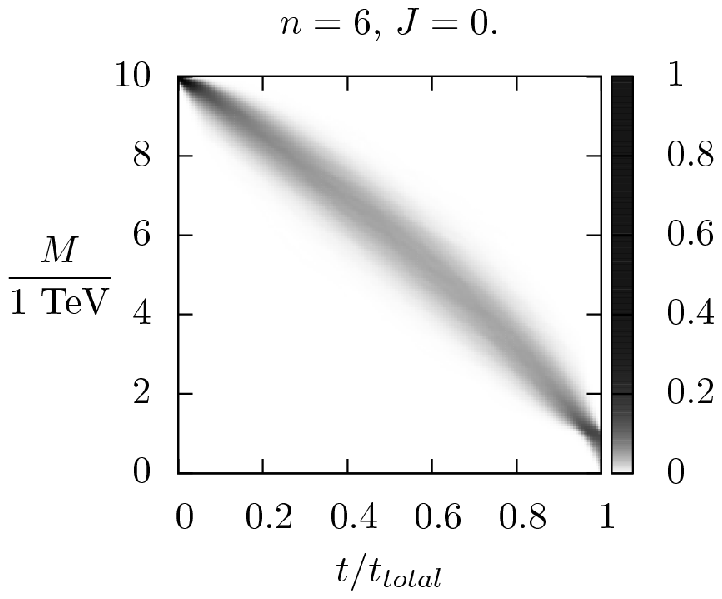} 
   \includegraphics[scale=0.7]{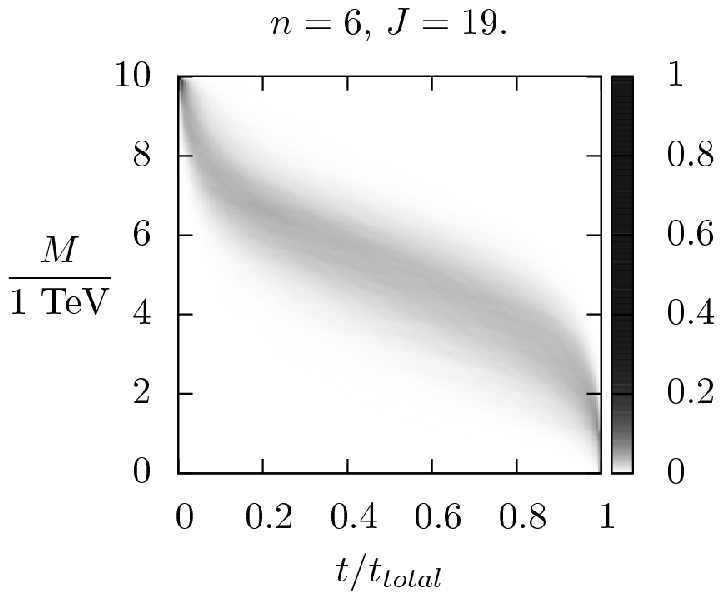} 
   \includegraphics[scale=0.7]{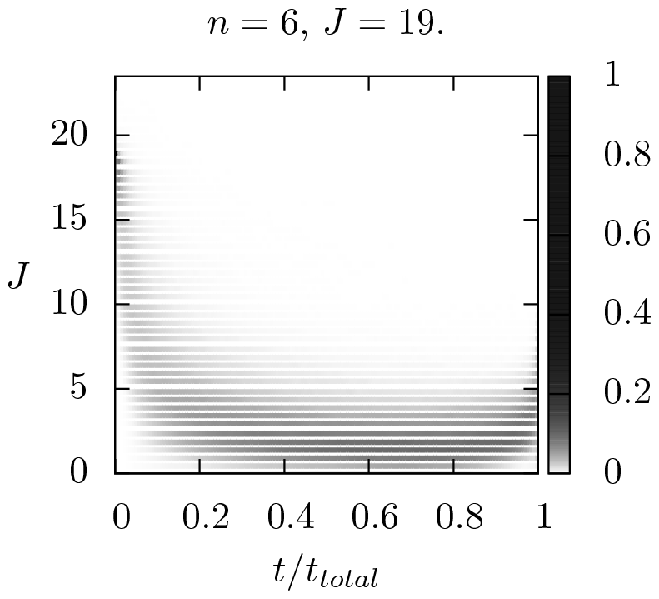}\vspace{3mm} \\
   \includegraphics[scale=0.7]{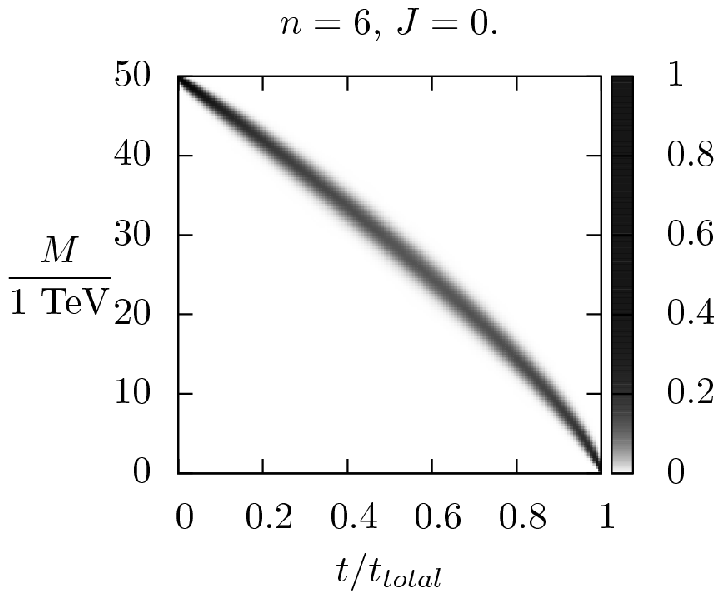} 
   \includegraphics[scale=0.7]{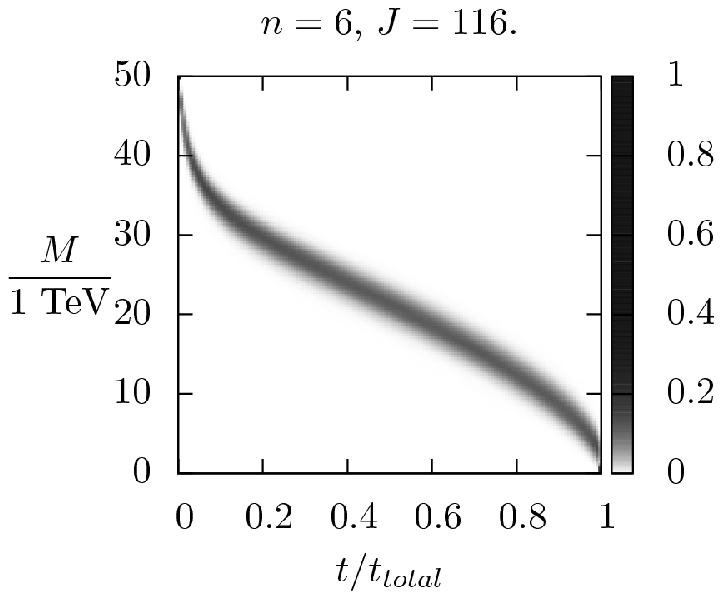} 
   \includegraphics[scale=0.7]{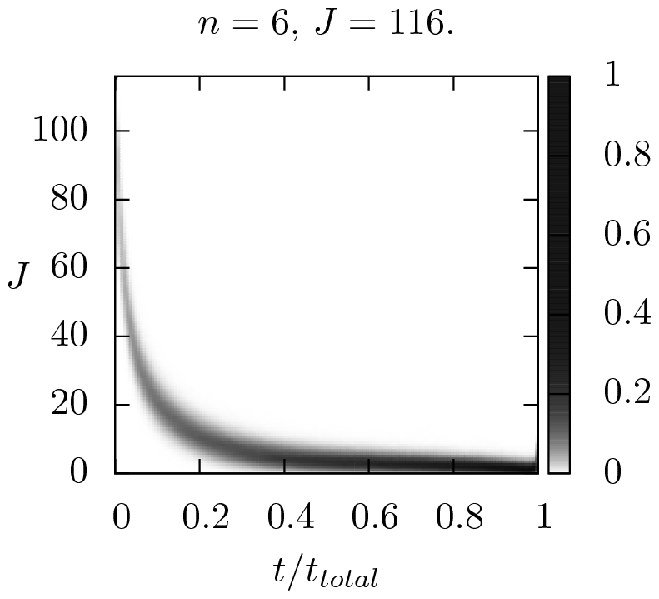}
   \caption{\label{maps_backreaction} Probability maps for physical parameters, constructed from $10^4$ trajectories for different BH events with fixed initial conditions $M,J$ (for each horizontal line). Each trajectory contributes with weight 1 to the bins it crosses on the $\left\{P,t/t_{total}\right\}$ plane where $P$ is the relevant parameter. Note that the time is normalised to the total time for evaporation $t_{total}$. The horizontal lines for the plots on the right are due to the discretisation of J in semi-integers.} 
\end{figure}

\begin{figure}[t]
  \centering
   \includegraphics[scale=0.7]{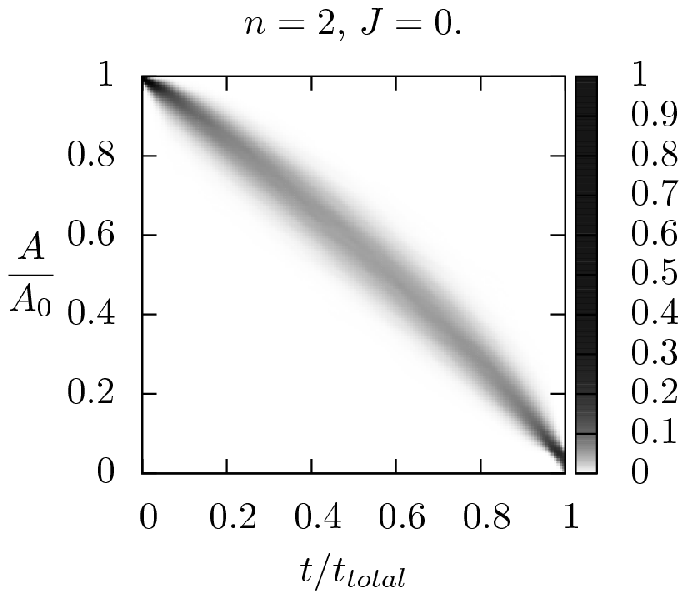} \hspace{1mm}
   \includegraphics[scale=0.7]{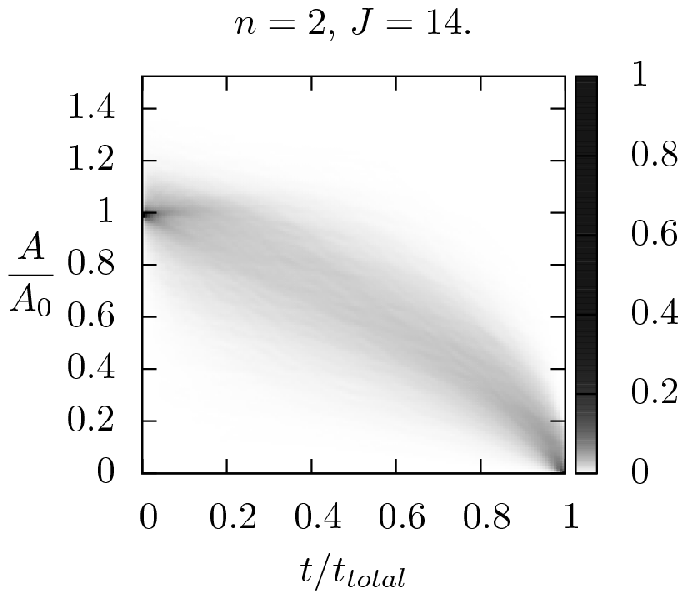} 
   \includegraphics[scale=0.7]{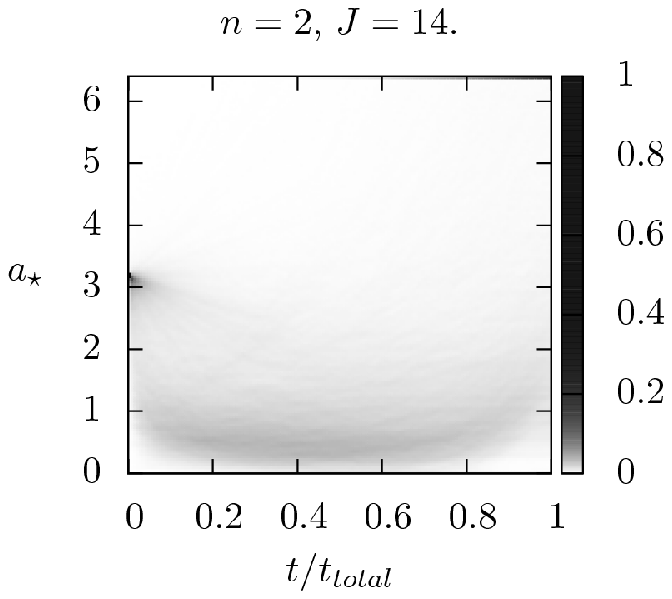}\vspace{3mm} \\
   \includegraphics[scale=0.7]{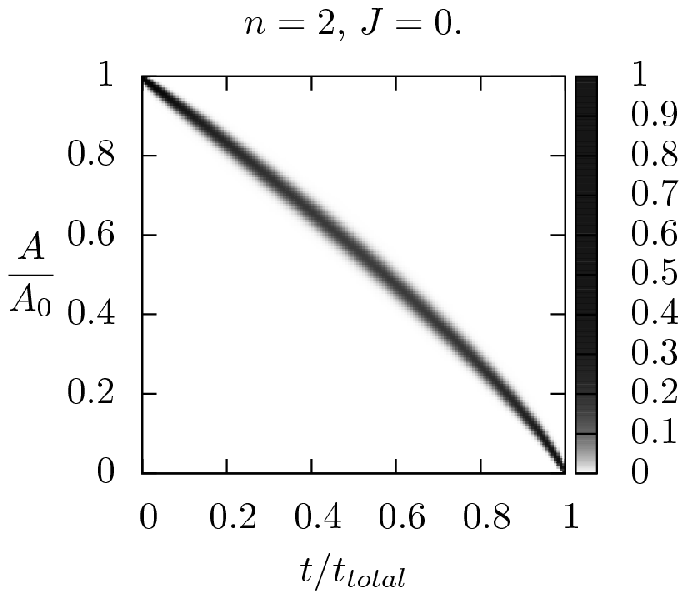}\hspace{2mm} 
   \includegraphics[scale=0.7]{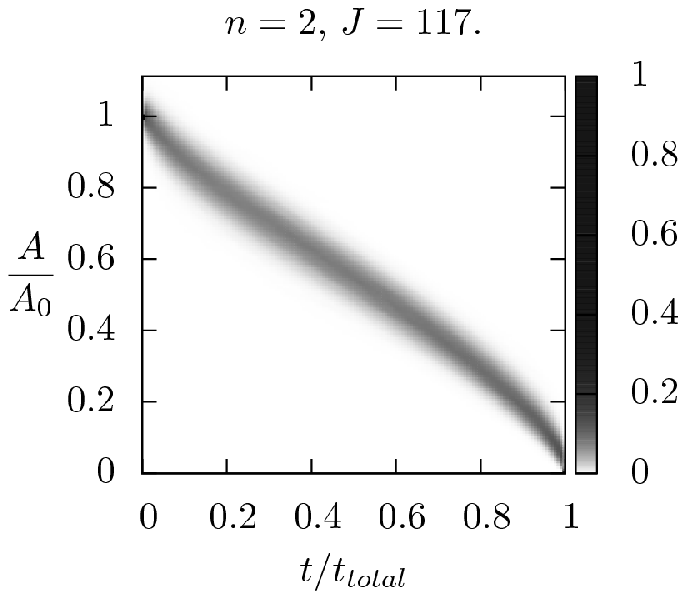} 
   \includegraphics[scale=0.7]{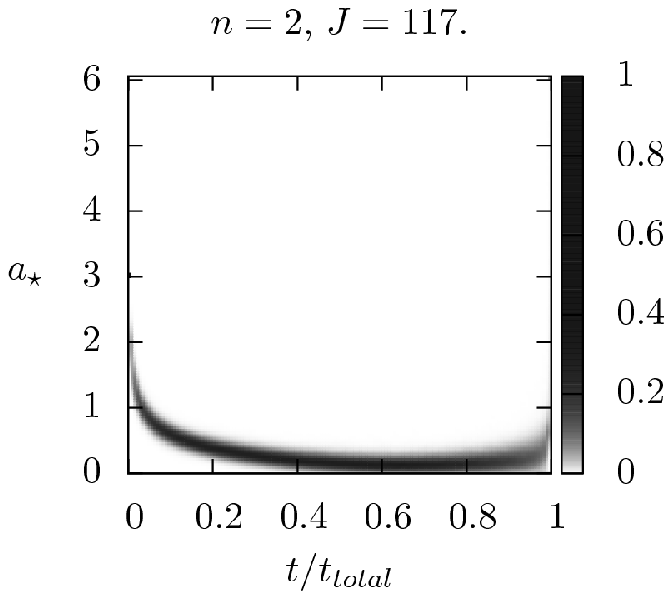}\vspace{3mm} \\
   \includegraphics[scale=0.7]{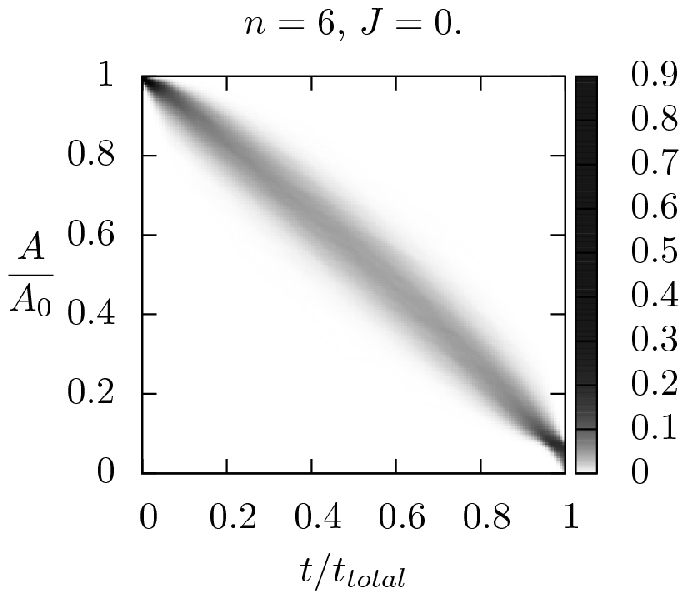} \hspace{1mm}
   \includegraphics[scale=0.7]{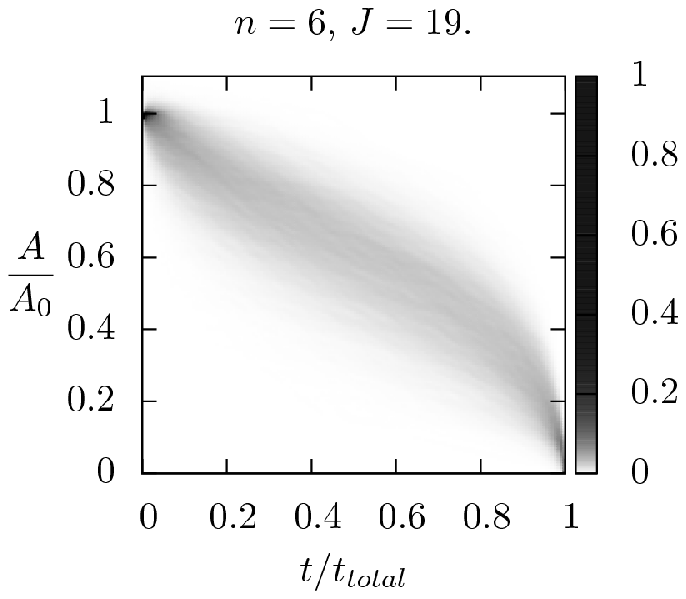} 
   \includegraphics[scale=0.7]{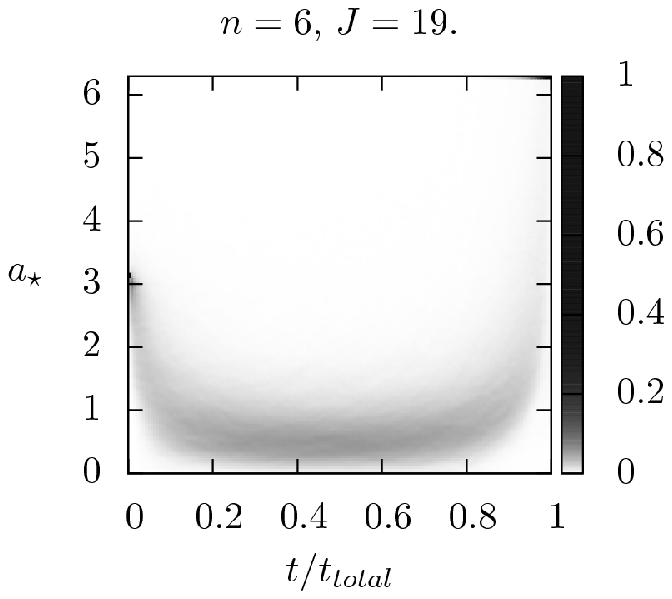}\vspace{3mm} \\
   \includegraphics[scale=0.7]{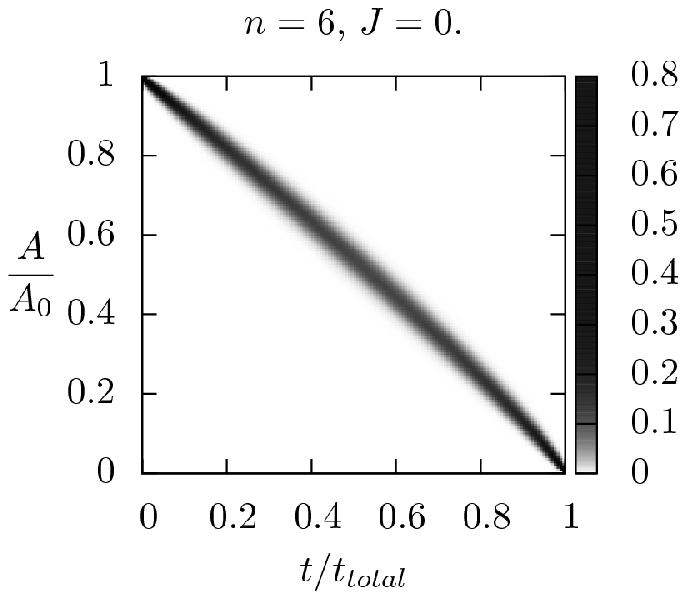} \hspace{1mm}
   \includegraphics[scale=0.7]{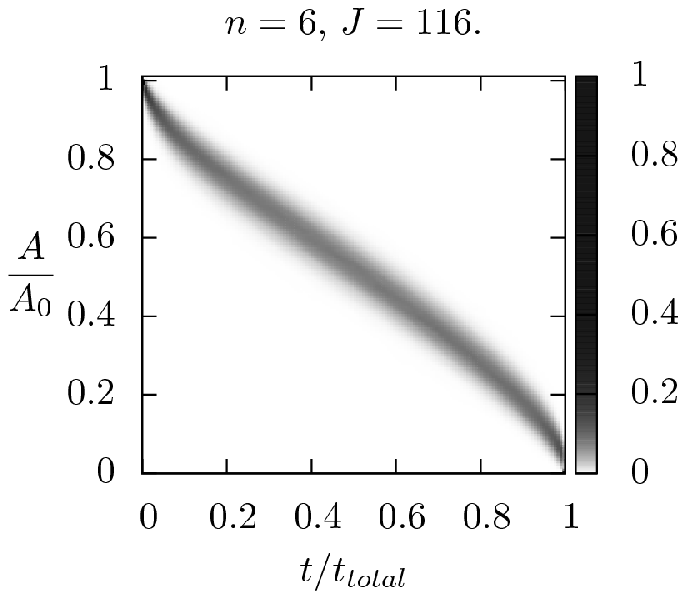} 
   \includegraphics[scale=0.7]{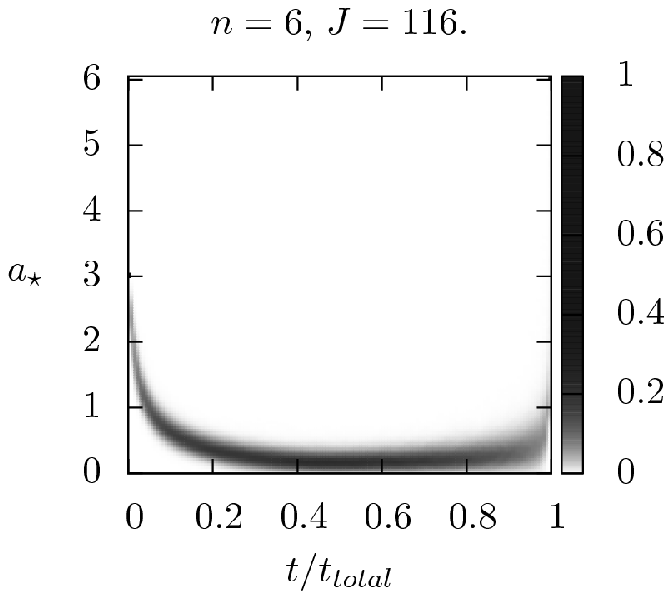}
   \caption{\label{maps_backreaction_geo} Probability maps for the geometrical parameters $A$ and $a_*$ which characterize respectively the size and oblateness of the BH. Note that for the $a_*$ plot in the last $t/t_{total}=1$ line there is very often a jump to very large $a_*$. In these plots we have put all such points in the bin on the upper right corner to avoid squashing the interesting region. Each horizontal line has the same initial $M,J$ as the corresponding one in Fig.~\ref{maps_backreaction}.}  
\end{figure}

In Fig.~\ref{maps_backreaction}
 we plot the evolution of the physical parameters $M$ and $J$ for BH events with fixed initial $M$, in the non-rotating case and the highly rotating case. In Fig.~\ref{maps_backreaction_geo}  we plot the horizon area and oblateness for the same cases as in Fig.~\ref{maps_backreaction}. 

Events were generated for two possible initial masses, 10~TeV (reachable at the LHC) and 50 TeV. The latter serves as a check of the semi-classical limit. We focus on $n=2$ and $n=6$. An important quantity necessary to produce these plots, is the time between emissions. Since our model for the evolution relies on the mean value equation \eqref{eq-flux-spectrum}, before each emission, an average time can be computed (i.e. $\delta t$ for $\delta N=1$):
\begin{equation}
 \nonumber \delta N = \dfrac{dN}{dt}\delta t \ \ \Rightarrow \ \ \delta t =\left[\dfrac{dN}{dt}\right]^{-1} , 
\end{equation} where a sum over all species is assumed (see Sect.~\ref{sec:term} for further details).

Each plot contains $10^4$ trajectories (one per event generated), each contributing with weight 1 to the density plot. The darker areas correspond to higher probability and in all the plots we can discern a tendency line which is sharper for the $50 \ \mathrm{TeV}$ case and more diffuse for $10 \ \mathrm{TeV}$, reflecting the magnitude of the statistical fluctuations. 

The left columns of Figs.~\ref{maps_backreaction} and \ref{maps_backreaction_geo} show respectively the evolution of the mass parameter and horizon area for non-rotating black holes. The centre and right columns show the evolution of the mass and angular momentum, or the horizon area and oblateness, for the highly rotating case ($a_* \simeq 3$). The main features are as follows:
\begin{itemize}
\item{{\it Non-rotating case:} Both $M$ and $A$ decrease approximately linearly with time except for the last $\sim 10-20 \ \%$ when they drop faster. This is directed related to the behaviour of the temperature which increases slowly (approximately linearly) for most of the evaporation and rises sharply near the end. The rates tend to be faster for higher $n$ which is in agreement with the increase in Hawking temperature with $n$.}
\item{{\it Highly-rotating case:} Here the statistical fluctuations tend to smear out the plots for the case of lowest mass. However, the same tendency can be seen as for the $M=50 \ \mathrm{TeV}$ black holes; the latter display better a true semi-classical behaviour. There is in general an initial period of roughly $10-15 \ \%$ of the total time when $M$ drops faster to about $60-70 \%$. At the same time the angular momentum also drops sharply to $20 \ \%$. This corresponds to the usual spin-down phase~\cite{Page:1976ki}. Note that the fluctuations are quite large for the low-mass $n=2$ plots. As for the geometrical parameters, they follow a similar tendency if we make the correspondences $M \leftrightarrow A$ and $J \leftrightarrow a_*$. Again for the low-mass plots the statistical fluctuations smear out the sharper initial drop in area, in particular for $n=2$ in which it can occasionally increase substantially. The $a_*$ plots show how the black hole tends to become more spherical in this spin-down phase. The remainder of the evolution resembles the non-rotating case and can be identified with a Schwarzschild phase. Note however that by the end of the evaporation (when $M$ approaches the Planck mass) this description breaks down and $a_*$ rises again, since even only one unit of angular momentum has a very large effect on this quantity at the Planck scale. This means that $a_*$ ceases to have a well defined geometrical meaning, as we reach the Planck phase. Similarly to the non-rotating case, we have checked that the temperature increases slowly and approximately linearly for most of the evaporation except for a sharp rise as the Planck mass is approached.}
\end{itemize}  

These observations agree with the usual results in four dimensions\footnote{see for example~\cite{Page:1976ki} or chapter 10.5.3 of~\cite{Frolov:1998wf}} and the results of~\cite{Ida:2006tf} in $D$ dimensions. 

\section{Termination of black hole decay}\label{sec:term}
Our model for black hole decay relies heavily on the assumptions that we are in the semi-classical regime and the evaporation is slow (i.e. there is enough time for re-equilibration between emissions)~\cite{Giddings:2001bu,Preskill:1991tb}. However, as the evaporation evolves, we will reach a point where neither of these assumptions will be true. 

In the generator we introduce some options for the final remnant decay based on different physical assumptions. 
First of all, we need a criterion to decide whether or not the remnant stage has been reached. The various options in the program are connected to a departure from semi-classicality. This occurs when the expectation value $\left<N\right>$ for the number of emissions becomes small, which is a sign of the low number of degrees of freedom associated with the black hole. Together with the drop in $\left<N\right>$, the Hawking temperature will rise sharply. This is all related to the approach of the black hole mass to the Planck mass. The options are:
\begin{itemize}
\item \vb{NBODYAVERAGE}=\vb{.TRUE.}: An estimate for the multiplicity of the final state is computed at each step during the evaporation, according to the Hawking spectrum:
\begin{equation}\label{Naverage_estimate}
\left<N\right> \simeq \dfrac{dN}{dt}\delta t \simeq \dfrac{dN}{dt}M\left(\dfrac{dE}{dt}\right)^{-1} =M r_H\dfrac{\sum_{i}g_i\left(\dfrac{1}{r_H}\dfrac{dN}{dt}\right)_{i}}{\sum_{j}g_j\left(\dfrac{dE}{dt}\right)_{j}} \, .
\end{equation} The sums are over all particle species with appropriate degeneracies $g_i$. The integrated flux and power are computed using \eqref{eq-flux-spectrum}. 
 A natural criterion for stopping the evaporation is when this estimate drops below some number close to 1. In the generator we use $\left<N\right>\leq \vb{NBODY}-1$ where \vb{NBODY} gives the average multiplicity of the remnant decay final state (see Sect.~\ref{var_mult} for further comments).  Varying the parameter $\vb{NBODY}$ will give a measure of uncertainties in the remnant model. In addition, if we choose a remnant model that decays, \eqref{Naverage_estimate} gives an estimate of the final state multiplicity for such a decay.  When $\vb{NBODYAVERAGE}=\vb{.FALSE.}$, one of the options below, inherited from earlier versions of \vb{CHARYBDIS}, is used.

\item \vb{KINCUT}=\vb{.TRUE.}: Terminate evaporation if an emission is selected which is not kinematically allowed. This is closely related to the rapid increase in temperature as we approach the Planck mass and consequently the generation of kinematically disallowed energies for the emission. Otherwise if $\vb{KINCUT}=\vb{.FALSE.}$ the kinematically disallowed emissions are rejected and the evaporation terminates when the mass of the black hole drops below the Planck mass\footnote{The Planck mass used in \vb{CHARYBDIS2} to decide on the termination is always the internal one, \vb{INTMPL}, which is obtained by converting the Planck mass input by the user (in a given convention), to the Giddings-Thomas convention -- see Appendix~\ref{app:conventions}.}.
\end{itemize} 
\subsection{Fixed-multiplicity model}
The default remnant decay option is a fixed-multiplicity model similar to that in earlier versions
of \vb{CHARYBDIS}.  At the end of the BH evaporation, the remaining object
is decayed isotropically in its rest frame, into a fixed number
\vb{NBODY} of primary particles, where the parameter \vb{NBODY} is an
integer between 2 and 5.  The decay products are chosen with relative
probabilities appropriate to the final characteristics of the black hole (i.e. weighted according to the integrated Hawking fluxes for each spin). 

The selection of the outgoing momenta of the decay products may be chosen either using pure phase space (\vb{NBODYPHASE=.TRUE.}) or by using the following probability density function in the rest frame of the black hole (\vb{NBODYPHASE=.FALSE.}):
\begin{equation}
\mathrm{d}P\propto \delta^{(4)}\left(\sum_ip_i-P_{BH}\right) \prod_{i}\rho_i\left(E_i,\Omega_i\right)\mathrm{d}^3\mathbf{p}_i \ ,
\end{equation}
which amounts to the usual phase space momentum conservation with an extra weight function for each particle
\begin{equation}
\rho_i\left(E_i,\Omega_i\right)=\frac{\mathbb{T}^{(D)}_{k}(E r_H, a_*)}{\exp(\tilde{E}/T_H) \pm 1} \left|S_{k}(\cos\theta_i)\right|^2 \ ,
\end{equation}
where $k=\left\{j,m\right\}$ are chosen according to the cumulants \eqref{cumulative_lm} combined with angular momentum conservation. Here $E_i,\Omega_i$ are the energy and momentum orientation of the emission in the rest frame of the remnant. This choice treats the final state particles on an equal footing, keeping a gravitational character for the decay (since it uses Hawking spectra), as well as some correlations with the the axis of rotation through the spheroidal function factor. Furthermore, at this stage, slow evaporation should no longer be valid, so it makes sense to perform a simultaneous decay at fixed black hole parameters. This remnant option can be used with any of the criteria for termination.   

\subsection{Variable-multiplicity model}\label{var_mult} 
In addition to a fixed multiplicity final state, an option has been introduced to select the multiplicity of the final state on an event-by-event basis. We follow an idea in~\cite{Bekenstein:1995ju}, which has been used for example in the case of $2\rightarrow 2$ sub-processes in \cite{Meade:2007sz}. Here we implement a more general model for arbitrary multiplicity,
which is invoked by setting the parameter \vb{NBODYVAR}=\vb{.TRUE.}.

As argued previously, when the remnant stage is reached, the black hole should no longer have time to re-equilibrate between emissions. Under this assumption, the probability distributions should become time independent. It is relatively straightforward to prove that under these conditions for a time interval $\delta t$, the multiplicity follows a Poisson distribution~\cite{Bekenstein:1995ju}:
\begin{equation}
P_{\delta t}(n)=e^{-\alpha \delta t}\dfrac{(\alpha \delta t)^n}{n!} \ \ ,
\end{equation} 
with $\alpha$ some constant. From the Hawking flux, we have computed an estimate for the average number of particles emitted during $\delta t$ (i.e. the time interval until all mass disappears), so $\alpha$ is determined from this condition. The final result is
\begin{equation}
P_{\delta t}(n)=e^{-\left<N\right>}\dfrac{\left<N\right>^n}{n!} \ \ ,
\end{equation} where $\left<N\right>$ is the estimate in \eqref{Naverage_estimate}. This expression gives us an estimate for the probability of emission of $n$ particles from the remnant, so we choose to interpret $n+1$ as the multiplicity of the final system. In the generator we have removed the $n=0$ case (i.e. multiplicity 1 final state) since the probability of the remnant to have all the correct quantum numbers and mass of a standard model particle will be vanishingly small.

After the multiplicity is chosen, either the pure phase space decay or the model described in the previous section is used,
according to the value of \vb{NBODYPHASE}.

\subsection{Boiling model}
The boiling remnant model, activated by setting \vb{RMBOIL}=\vb{.TRUE.},
is loosely motivated by the expectation that at the Planck scale
the system becomes like a string ball~\cite{Dimopoulos:2001qe,Gingrich:2008di}, which has a limiting
temperature due to the exponential degeneracy of the string spectrum~\cite{LorenteEspin:2007gz}.
In this model, evaporation of the BH proceeds until the
Hawking temperature for the next emission would exceed a maximum value
set by the parameter \vb{THWMAX}.  From that point on, the temperature is
reset to \vb{THWMAX} and the oblateness is frozen at the current
value.
The remaining object evaporates like a BH with those characteristics,
until its mass falls below a value set by the parameter \vb{RMMINM}.
It then decays into a fixed number \vb{NBODY} of primary particles,
as in the fixed-multiplicity model, or a variable number if the variable-multiplicity model is on.

\subsection{Stable remnant model}
A number of authors have proposed that the endpoint of black
hole evaporation could be a stable remnant~\cite{Koch:2005ks,Stoecker:2006we,Scardigli:2008jn}.
This option is activated by setting \vb{RMSTAB}=\vb{.TRUE.}.
In order for the cluster hadronisation model of \vb{HERWIG} to hadronise
the rest of the final state successfully, the stable remnant must be a colourless object essentially equivalent to a quark-antiquark
bound state.  Therefore it is required to have baryon number $B_R=0$
and charge $Q_R=0$ or $\pm1$.

The stable remnant appears in the event record as
\vb{Remnant0}, \vb{Remnant+} or \vb{Remnant-}, with PDG identity
code 50, 51 or --51, respectively, according to its charge. This object will behave as a heavy fundamental particle with conventional
interactions in the detector. 

If a remnant with $B_R\neq 0$ or $|Q_R|>1$ is generated, the whole BH
evaporation is repeated until $B_R=0$ and $|Q_R|\leq 1$.  This can
make the stable remnant option much slower than the other options,
depending on the length of the black hole decay chain.

\subsection{Straight-to-remnant option}
Recently, there has been discussion of the possibility that the formation of a semi-classical black hole may lie beyond current experimental reach, with low-multiplicity gravitational scattering more likely at the TeV scale~\cite{Meade:2007sz}.  To simulate this scenario, \vb{CHARYBDIS2} provides the option of bypassing the evaporation phase by setting the switch \vb{SKIP2REMNANT}=\vb{.TRUE.} and skipping directly to one of the remnant models presented in the previous sections. This permits the study of a wide range of qualitatively different possibilities, from simple $2\rightarrow 2$ isotropic scattering (fixed multiplicity) to more complicated variable-multiplicity $2\rightarrow N$ sub-processes.  

The $2\rightarrow N$ model is particularly flexible, allowing either a phase-space distribution or one using the Hawking energy and angular spectra (see Sect.~\ref{var_mult}). Apart from this, all particle species are treated on an equal footing consistent with conservation laws. Alternatively the quantum-gravity motivated boiling model can be used. Further work will be presented in future publications exploring the phenomenological consequences of these scenarios.


\section{Results}\label{sec:res}

In this section we present results from \vb{CHARYBDIS2} simulations of black hole production at the LHC. A range of \vb{CHARYBDIS2} samples were produced using \vb{HERWIG 6.510}~\cite{Corcella:2000bw,Corcella:2002jc} to do the parton showering, hadronisation and standard model particle decays. The results of which were then passed through a generic LHC detector simulation, \vb{AcerDET 1.0}~\cite{RichterWas:2002ch}. \vb{CHARYBDIS2} parameter defaults are shown in Table~\ref{tab:charyb}. In all following discussion, the number of extra spatial dimensions is $n=\vb{TOTDIM}-4$. Samples were generated with a 1~TeV Planck mass (in the PDG convention, i.e. $\vb{MSSDEF}=3$) so as to investigate the phenomenologically preferred region accessible at the LHC.
Black holes were generated with a lower mass threshold of 5~TeV such that the semi-classical approximations for production are valid. 

Our settings for \vb{AcerDET 1.0} are as follows: we select electrons and muons with $P_T>15$ GeV and $|\eta|<2.5$. They are considered isolated if they lie at a distance $\Delta R = \sqrt{(\Delta \eta)^2 + (\Delta \phi)^2}>0.4$ from other leptons or jets and if less than 10 GeV of energy was deposited in a cone of $\Delta R = 0.2$ around the central cluster.
The same prescription is followed for photons. 
Jets are reconstructed from clusters using a cone algorithm of $\Delta R = 0.4$, with a lower $P_T$ cut of 20 GeV. Lepton momentum resolutions were parameterised from ATLAS full simulation results published in~\cite{:2008zzm}.\footnote{Electrons are smeared according to a pseudorapidity dependent parameterisation; for muons, we take the resolutions from $|\eta|<1.1$.}
 Where reference is made to reconstructed multiplicities or spectra, the reconstructed objects are either electrons, muons, photons or jets from \vb{AcerDET}. 

\begin{table}[tb]
\caption{\label{tab:charyb}Default \vb{CHARYBDIS2} generator parameters (new parameters in the second set).}  
\centering
\begin{tabular}{c|l|c}
\hline\hline
Name   & Description                                & Default  \\
\hline
\vb{MINMSS} & Minimum parton-parton invariant mass                &  5 TeV  \\
\vb{MAXMSS} & Maximum parton-parton invariant mass                & 14 TeV  \\
\vb{MPLNCK} & Planck scale                               &  1 TeV  \\
\vb{GTSCA} & Use Giddings-Thomas scale for PDFs          &  .FALSE. \\
\vb{MSSDEF} & Convention for Planck scale                & 3       \\
\vb{TOTDIM} & Total number of dimensions                 & 6       \\
\vb{NBODY } & Number of particles in remnant decay       & 2       \\ 
\vb{TIMVAR} & Allow $T_H$ to evolve with BH parameters            & \vb{.TRUE.}    \\
\vb{MSSDEC} & Allowed decay products (3=all SM)     & 3    \\
\vb{GRYBDY} & Include grey-body factors                  & \vb{.TRUE.}    \\
\vb{KINCUT} & Use a kinematic cut-off on the decay       & \vb{.FALSE.}    \\
\vb{THWMAX} & Maximum Hawking temperature & 1 TeV \\
\hline
\vb{BHSPIN} & Simulate rotating black holes & \vb{.TRUE.} \\
\vb{BHJVAR} & Allow black hole spin axis to vary & \vb{.TRUE.} \\
\vb{BHANIS} & Non-uniform angular functions for the evaporation & \vb{.TRUE.} \\
\vb{RECOIL} & Recoil model for evaporation                 & 2 \\
\vb{MJLOST} & Simulation of $M$, $J$ lost in production/balding & \vb{.TRUE.} \\
\vb{CVBIAS} & `Constant angular velocity' bias & \vb{.FALSE.} \\
\vb{FMLOST} & Isotropy of gravitational radiation lost & 0.99 \\
\vb{YRCSC} & Use Yoschino-Rychov cross-section enhancement  & \vb{.TRUE.} \\
\vb{RMSTAB} & Stable remnant model & \vb{.FALSE.} \\
\vb{NBODYAVERAGE} & Use flux criterion for remnant -- see Eq.~\eqref{Naverage_estimate} & \vb{.TRUE.} \\
\vb{NBODYVAR} & Variable-multiplicity remnant model & \vb{.FALSE.} \\
\vb{NBODYPHASE} & Use phase space for remnants & \vb{.FALSE.} \\
\vb{SKIP2REMNANT} & Bypass evaporation phase & \vb{.FALSE.} \\ 
\vb{RMBOIL} & Use boiling remnant model & \vb{.FALSE.} \\
\vb{RMMINM} & Minimum mass for boiling model & 100 GeV \\
\hline\hline
\end{tabular} 
\end{table}

\subsection{Black hole mass and angular momentum}
\begin{figure}[t]
  \centering 
   \includegraphics[scale=0.5]{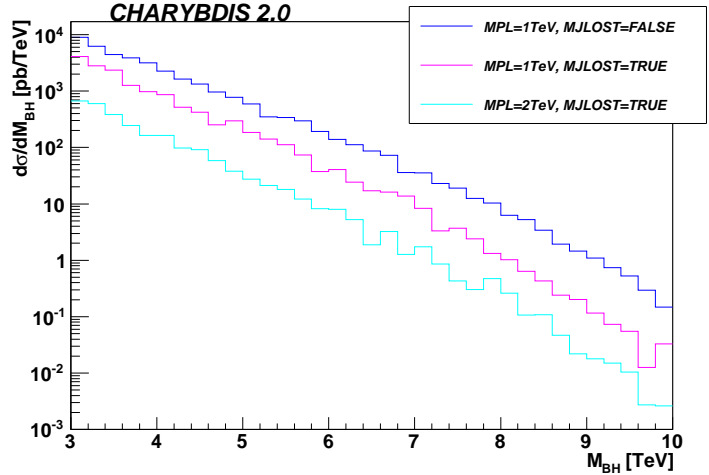}
\caption{Differential cross-sections for black hole production with $n=4$ extra dimensions. They are shown for different settings of the Planck mass and \vb{MJLOST} (simulation of mass and angular momentum lost in production/balding).}
\label{fig:cross-sections}
 \end{figure}

The cross-section for black hole production is a strong function of the Planck mass. Though not affecting the total black hole cross-section, simulating the mass and spin lost during black hole formation does have a large effect on the cross-section for a particular mass range. The differential cross-section will be reduced, for the same input state will produce a black hole of lesser mass, as is illustrated in Fig.~\ref{fig:cross-sections}.

Most collisions with sufficient energy to create a black hole are between two (valence) quarks, however a minority occur in collisions between a quark and a gluon. 
\vb{CHARYBDIS2} adds the spins of the colliding partons when forming a black hole; the initial angular momentum is either integer or half-integer accordingly. An integer loss of orbital angular momentum in the formation process is simulated by the Yoshino-Rychkov model described in Sect.~\ref{sec:prod}.
\begin{figure}[t]
\includegraphics[scale=0.39]{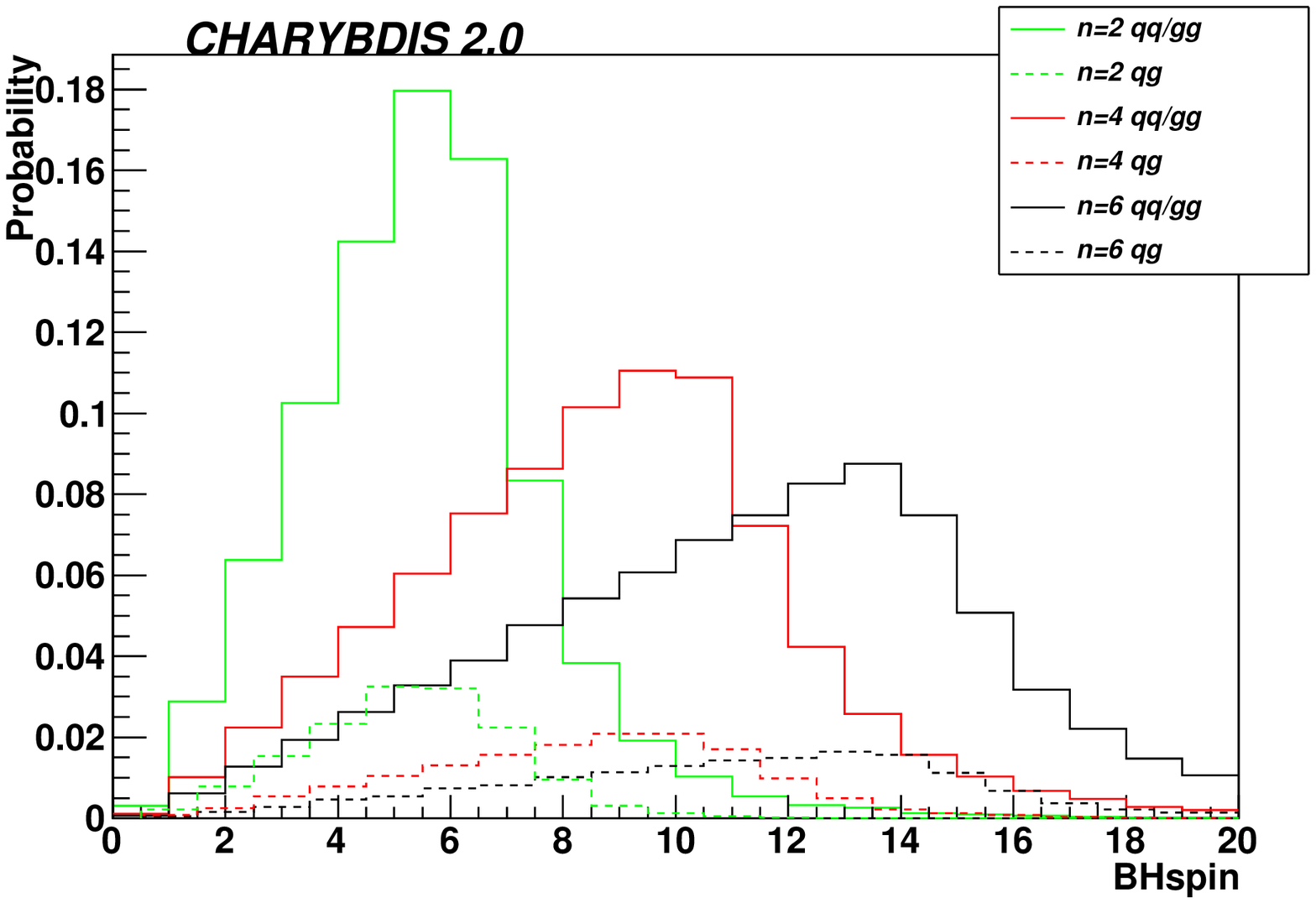} \includegraphics[scale=0.39]{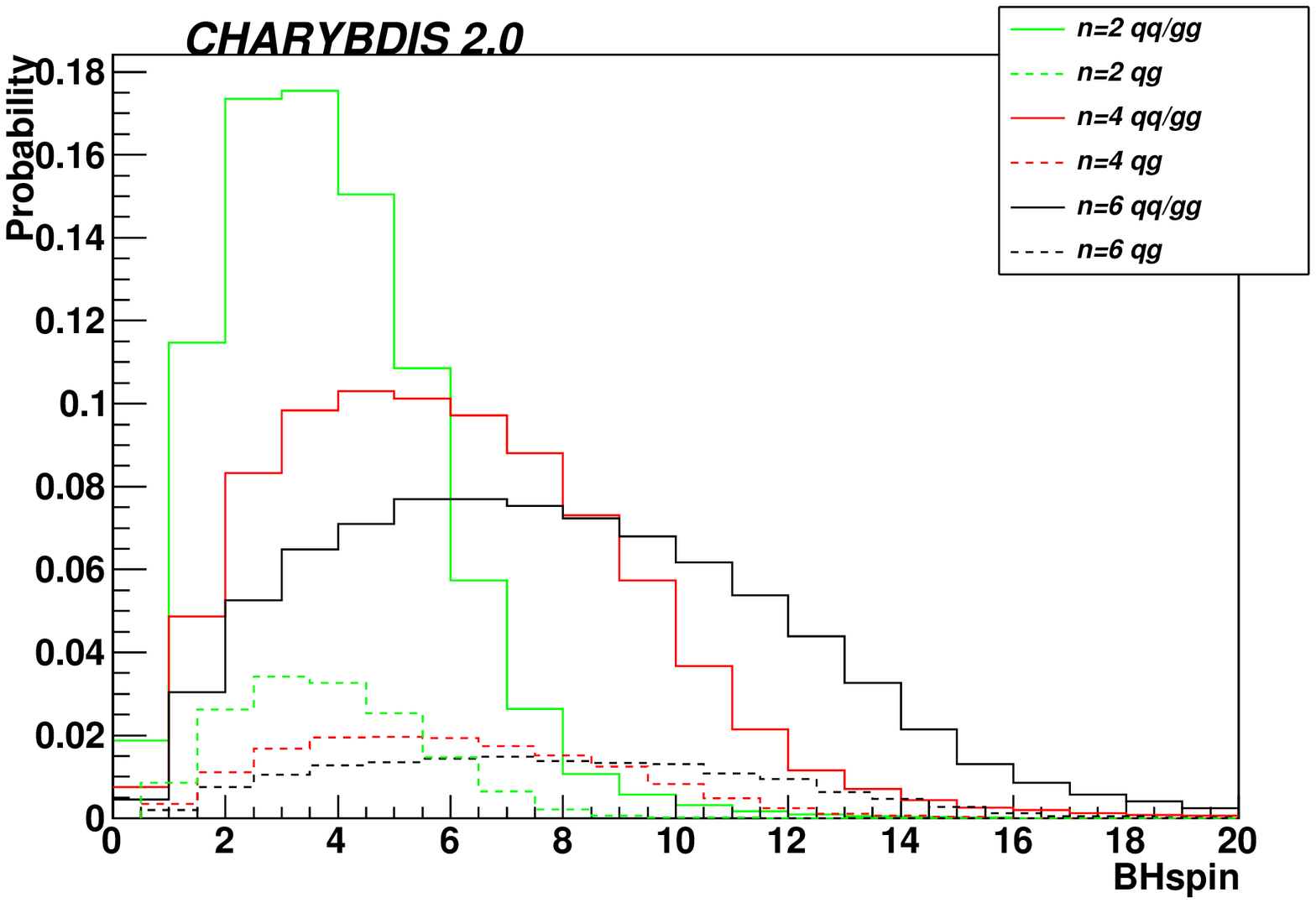} \\ 
\includegraphics[scale=0.39]{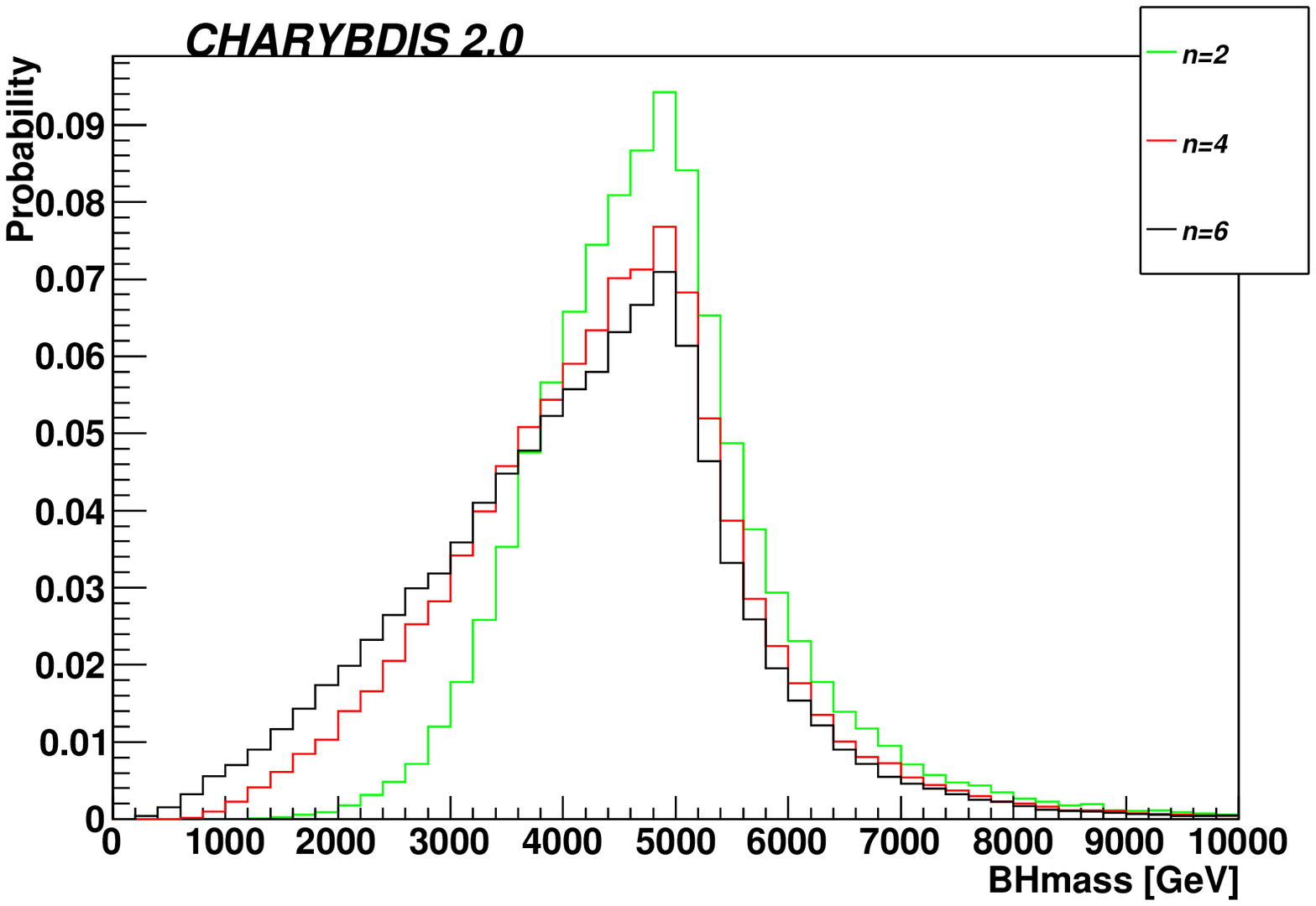} \includegraphics[scale=0.39]{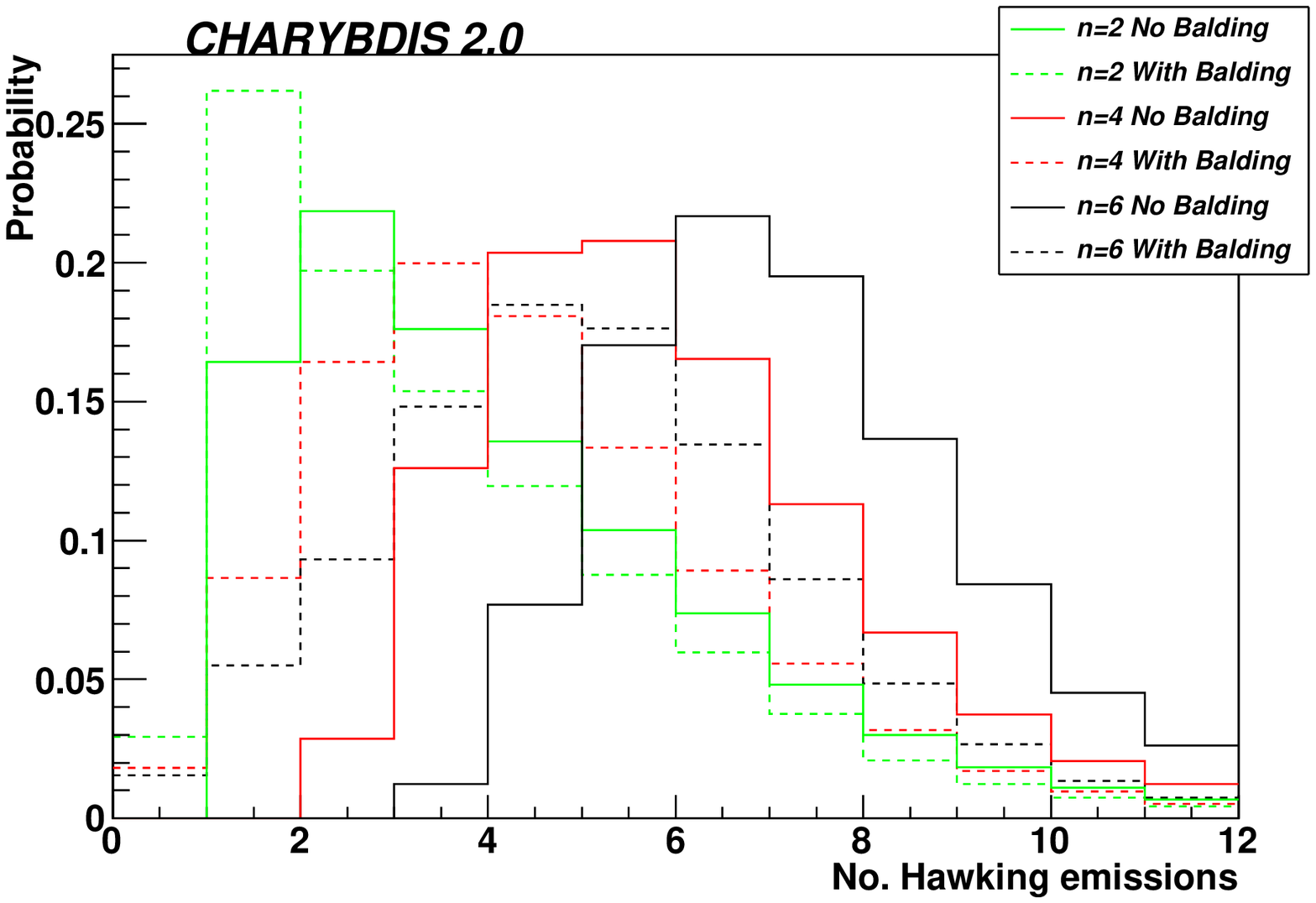}
\caption{Effects of mass and angular momentum loss in the formation/balding phase. The left (right) upper plot shows the angular momentum distribution before (after) this phase, whilst the lower row details the resulting black hole masses and the number of Hawking emissions.}
\label{fig:prod_mjsteps}
\end{figure}

At high $n$ there is a large increase in the 
production of high spin states, as seen in Fig.~\ref{fig:prod_mjsteps}. The average spin of the produced black hole rises from 5.0 units for $n=2$, to 8.1 for $n=4$ and 10.6 for $n=6$. 
The generator switch {\tt MJLOST} toggles a model of the loss of black hole mass and angular momentum in graviton emission during black hole production and balding, as described in Sect.~\ref{sec:prodmeth}. Setting this to \vb{.TRUE.} decreases the spin slightly by an average of 30\% for $n=2$, 4, 6, whilst the mass drops by 18\% ($n=2$) to 30\% ($n=6$), as shown in Fig.~\ref{fig:prod_mjsteps}.

The variation in the number of Hawking emissions is caused primarily by the black hole mass, though the black hole spin and temperature play a role -- a more highly rotating, or higher temperature black hole will emit more energetically. Consequently, the decrease in the number of Hawking emissions follows the drop in mass and is greatest for higher numbers of extra dimensions, with an average of two fewer emissions (or 30\%) manifest for $n=6$.

\begin{figure}[h]
\includegraphics[scale=0.39]{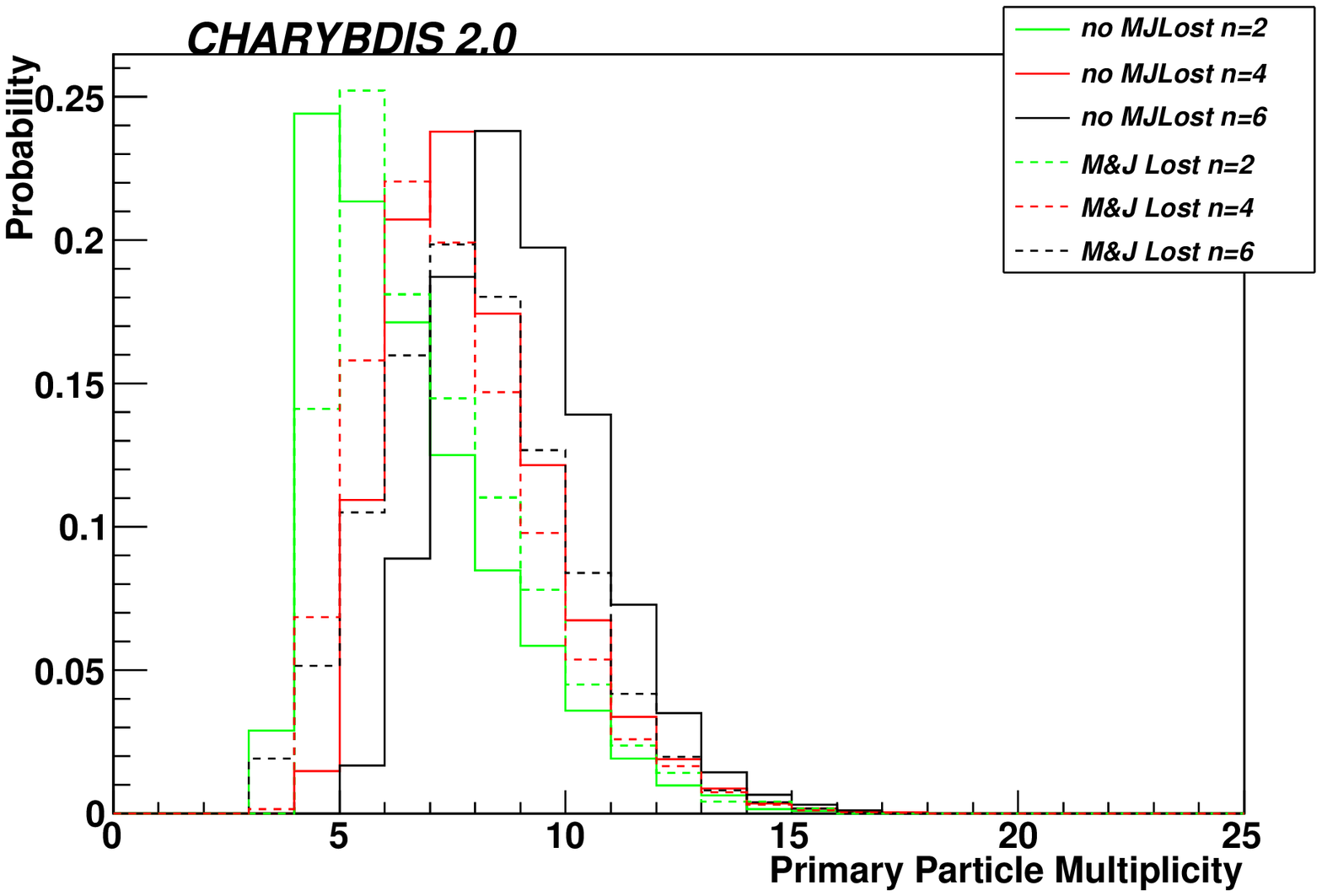} \includegraphics[scale=0.39]{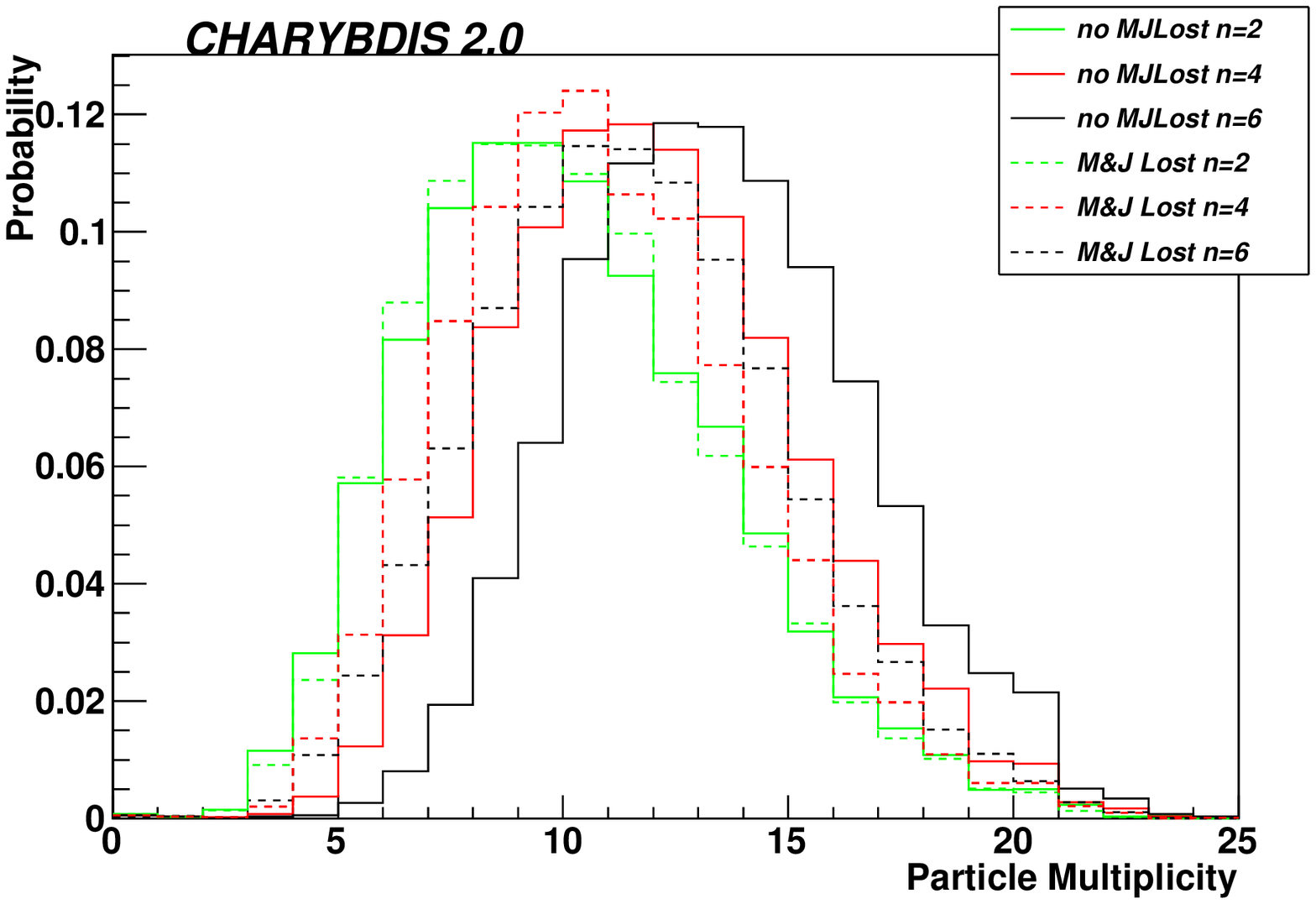} \\ 
\includegraphics[scale=0.39]{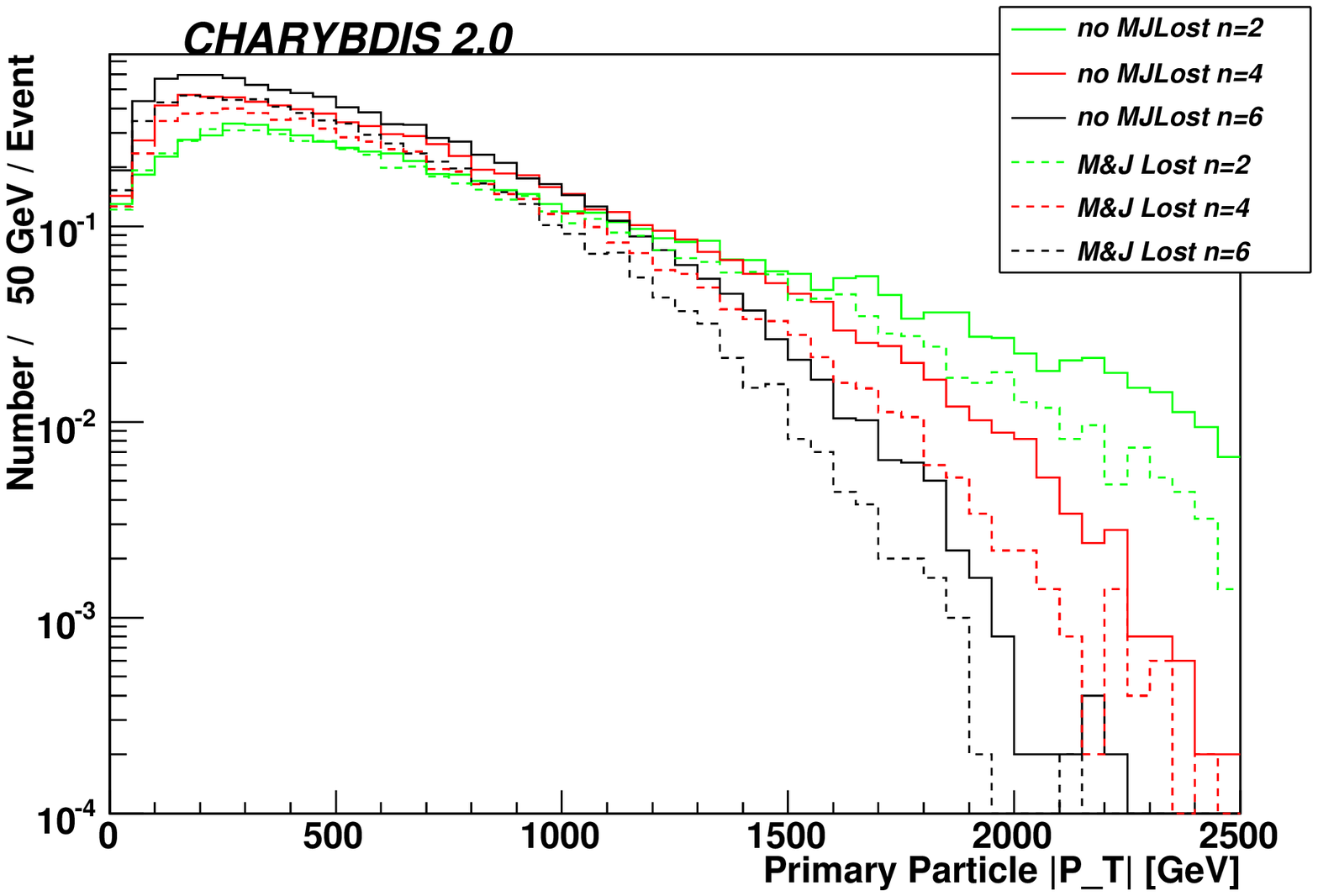} \includegraphics[scale=0.39]{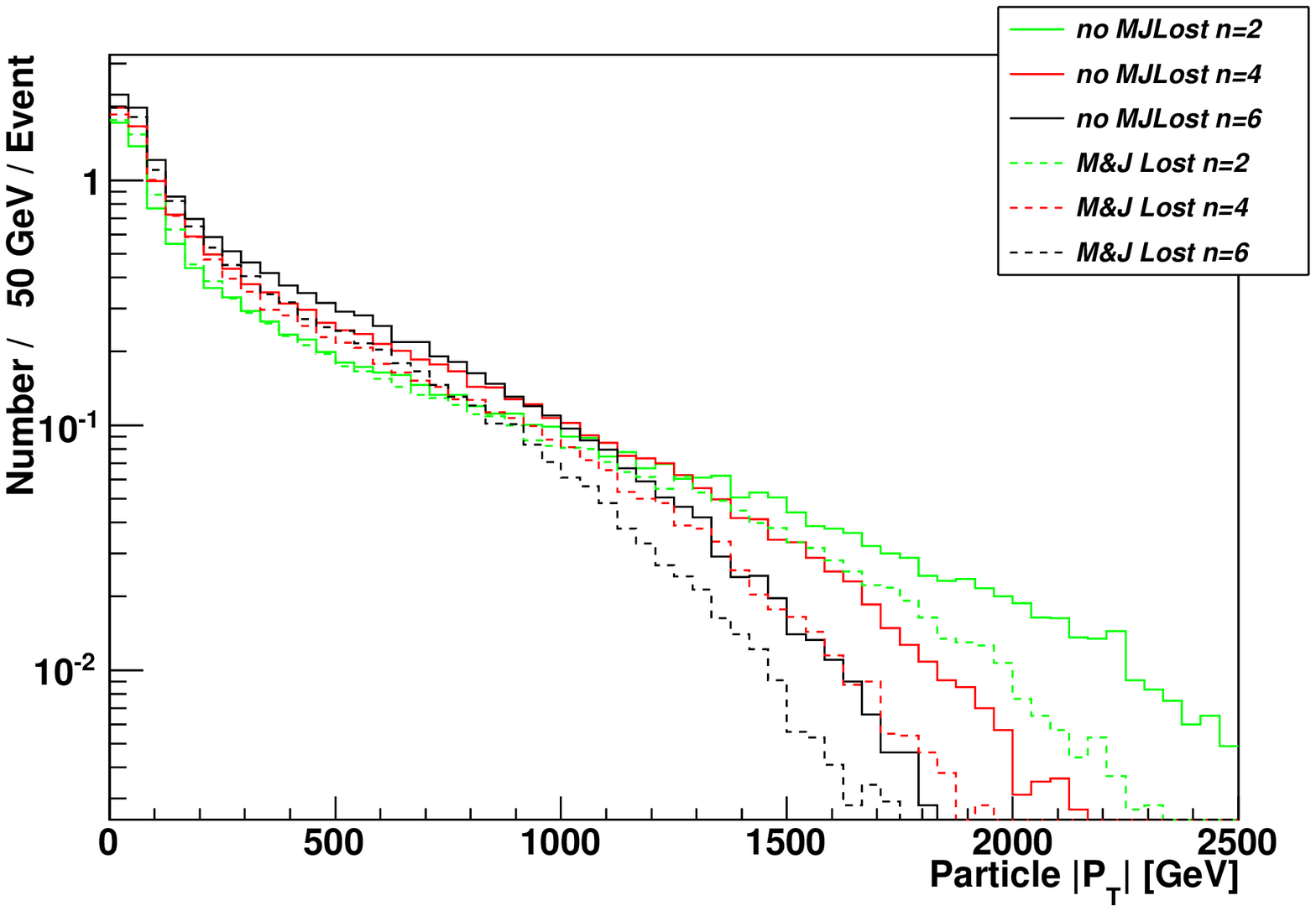}
\caption{Effect of simulating the mass and angular momentum lost in black hole production on particle multiplicity distributions and $P_T$ spectra at generator level (left) and after \vb{AcerDET} detector simulation (right), for a fixed 2-body remnant decay using the option in the second row of Table~\ref{tab:remn}.}
\label{fig:prod_rembasics}
\end{figure}

The simulation of losses from production and balding has several major effects upon the produced particle spectrum. The reduction in the black hole mass and number of Hawking emissions leads to a decrease in the number of particles observed experimentally and to a reduced differentiation between samples with different numbers of dimensions. The decrease in initial black hole mass also leads to a softening of the emitted particle spectrum, with the high energy and transverse momentum tail of the distribution being reduced.  Fig.~\ref{fig:prod_rembasics} shows this for a fixed 2-body remnant decay using the option in the second row of Table~\ref{tab:remn}.

\begin{table}[htb]
\caption{\label{tab:remn}Parameters used for remnant comparison. All samples have $n=2$ and \vb{MJLOST=.FALSE.}}
\centering
\begin{tabular}{c|c|c|c}
\hline\hline
Legend   & Remnant Criterion    & Fixed/Variable   & Remnant No./Mean   \\
\hline
Kincut on & $M<\vb{INTMPL}$ (\vb{KINCUT=.TRUE.})   & Fixed   & 2  \\
Kincut off & $M<\vb{INTMPL}$ (\vb{KINCUT=.FALSE.})   & Fixed    & 2   \\
Nbody2   & Flux (\vb{NBODYAVERAGE=.TRUE.})         & Fixed      &  2   \\
Nbody3   & Flux (\vb{NBODYAVERAGE=.TRUE.})        & Fixed       &  3   \\
Nbody4   & Flux (\vb{NBODYAVERAGE=.TRUE.})        & Fixed       &  4   \\
Nvar2   & Flux  (\vb{NBODYAVERAGE=.TRUE.})       & Variable     &  2   \\
Nvar3   & Flux  (\vb{NBODYAVERAGE=.TRUE.})       & Variable     &  3   \\
Nvar4   & Flux  (\vb{NBODYAVERAGE=.TRUE.})       & Variable     &  4   \\
Boiling   &   $\vb{RMMINM}<M<\vb{INTMPL}$       & Variable   &  2   \\
\hline\hline
\end{tabular} 
\end{table}

\subsection{Rotation effects}

\begin{figure}[ht]
\includegraphics[scale=0.39]{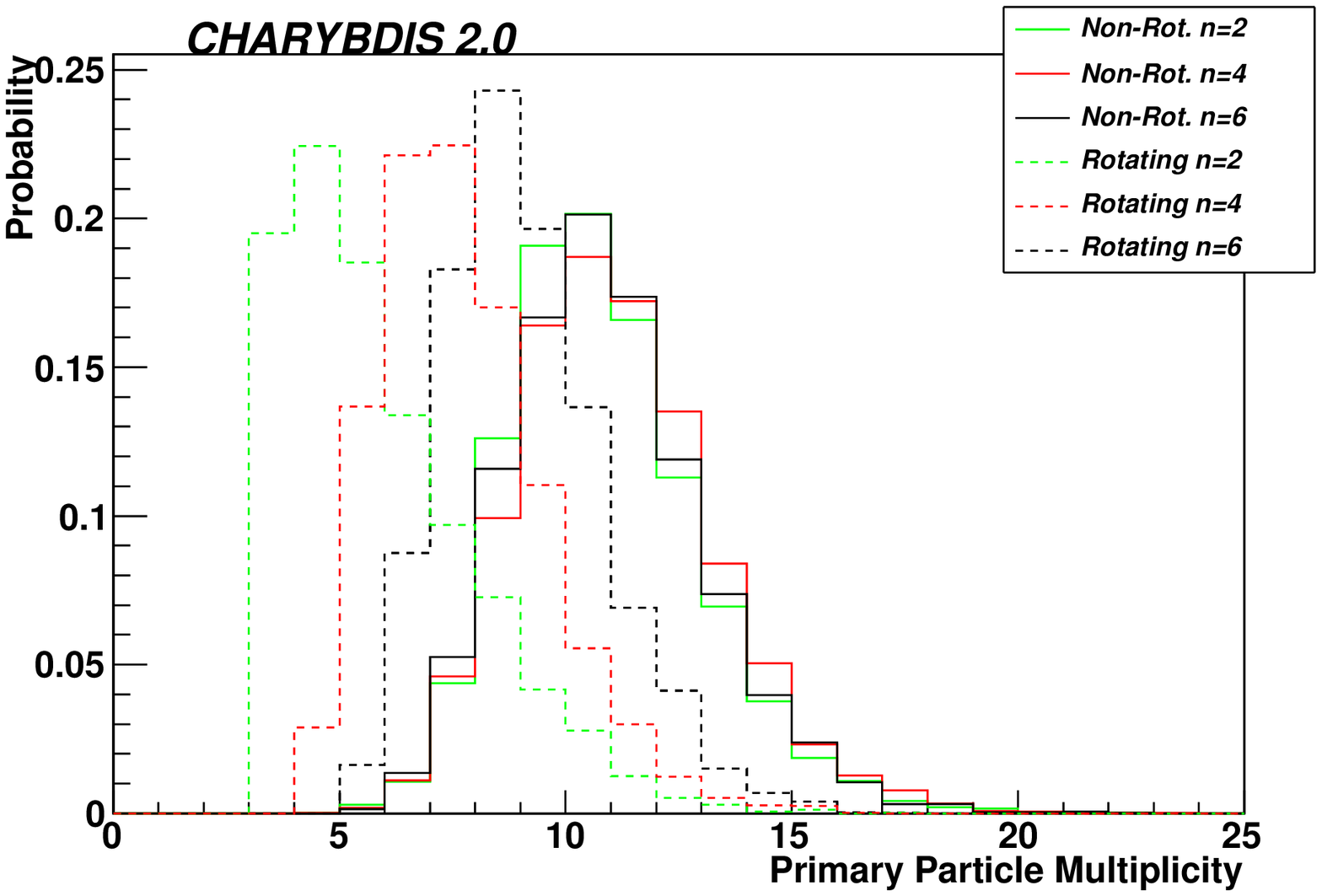} \includegraphics[scale=0.39]{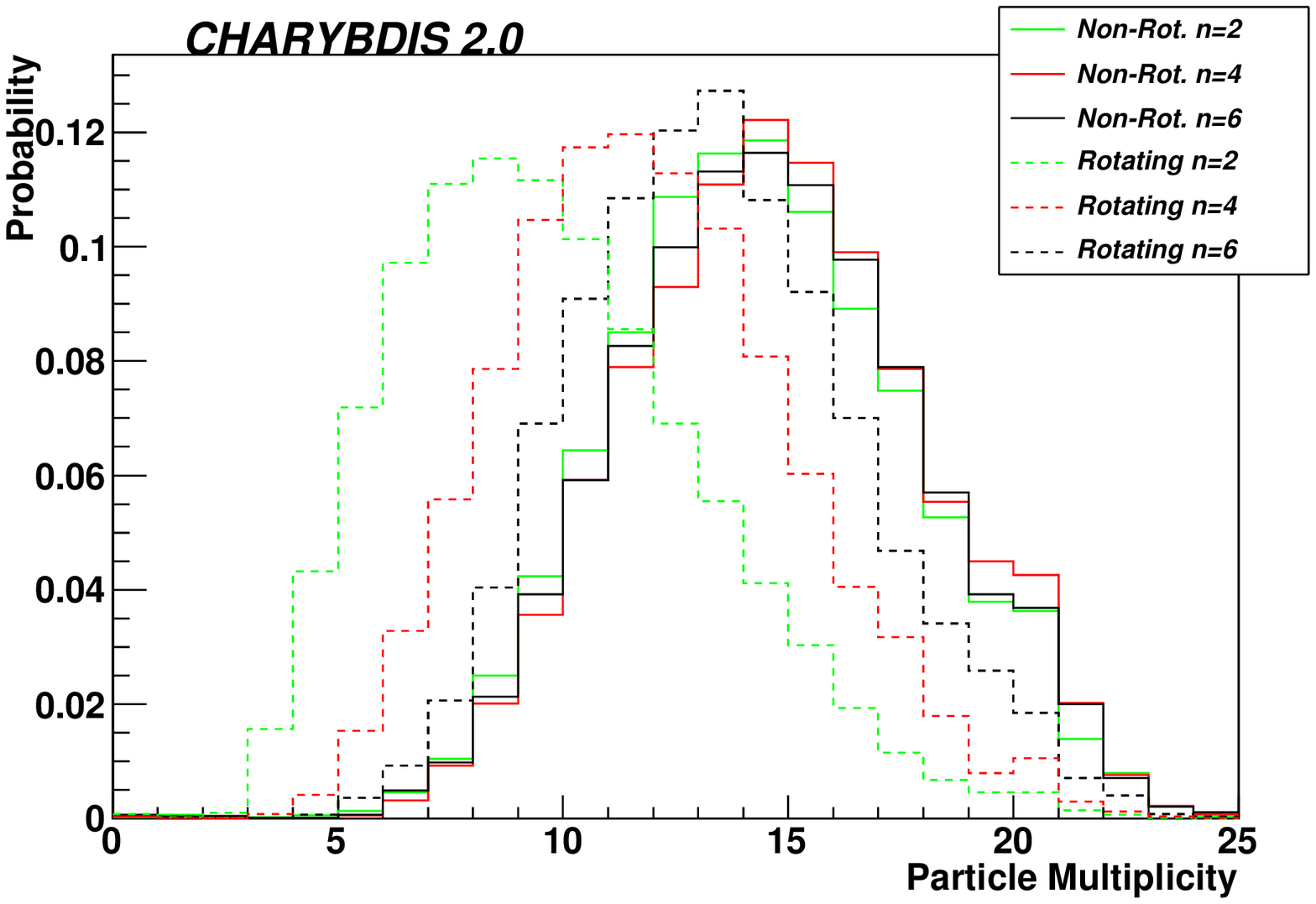} \\ 
\includegraphics[scale=0.39]{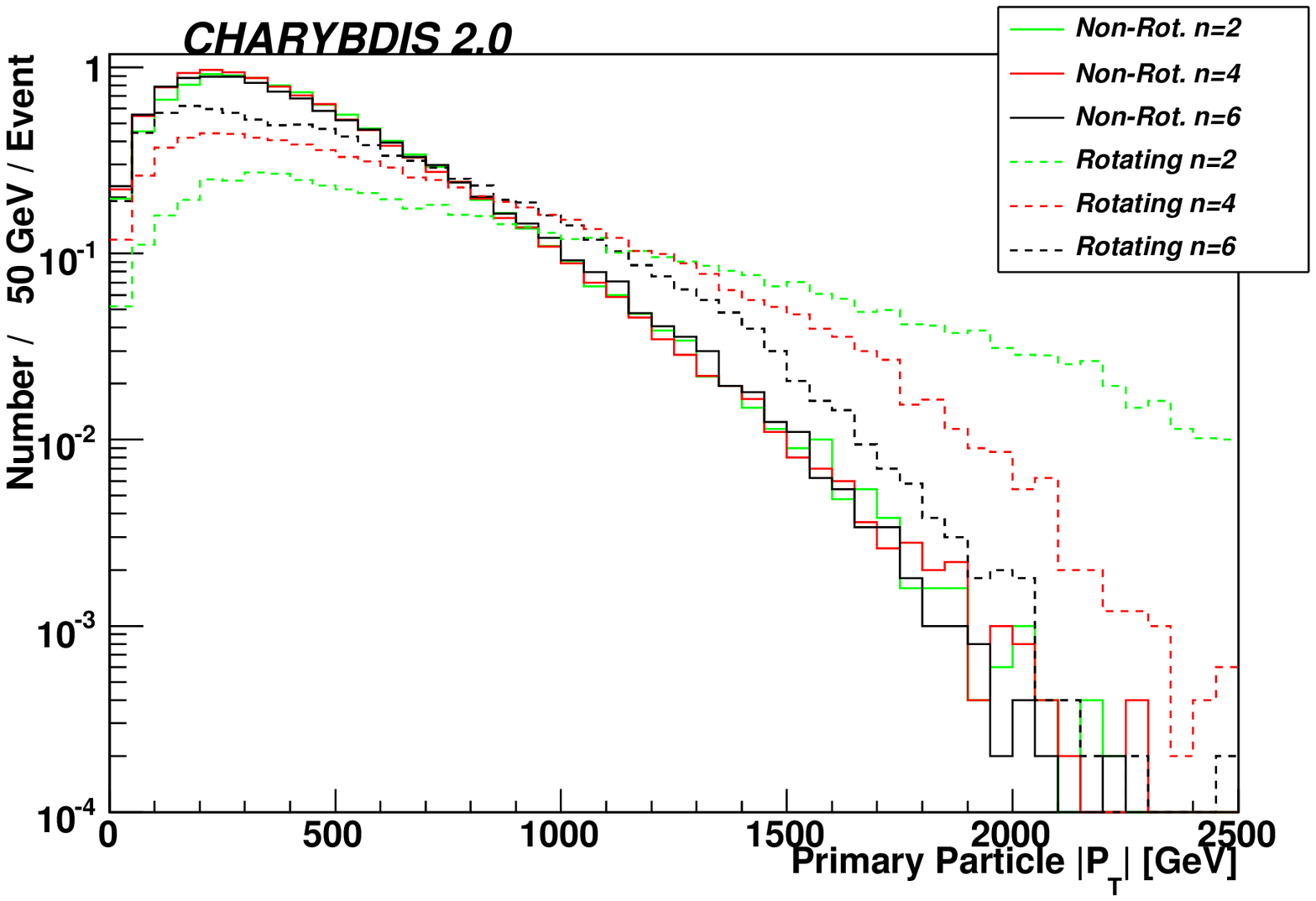} \includegraphics[scale=0.39]{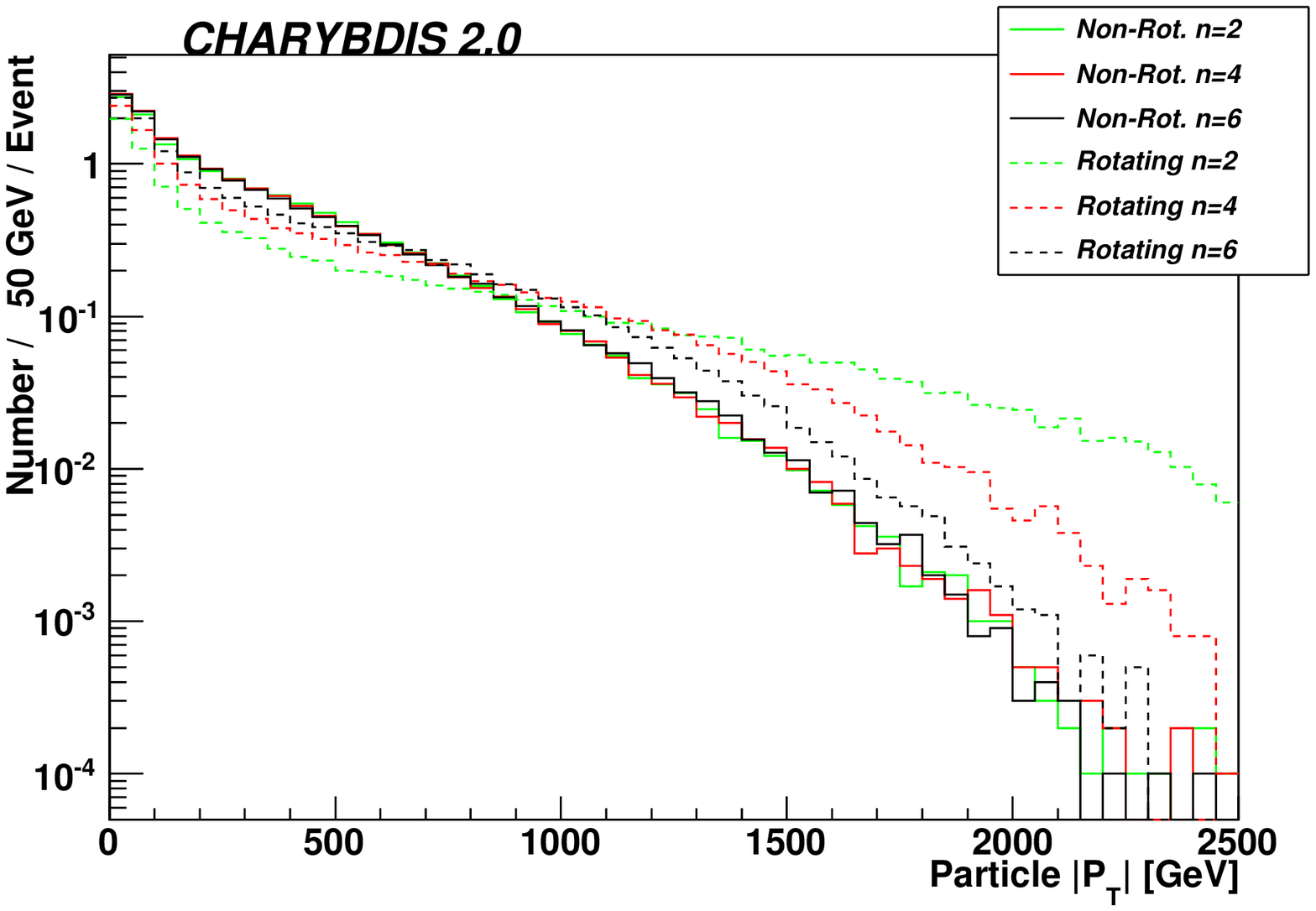}
\caption{Particle multiplicity distributions and $P_T$ spectra at generator level (left) and after \vb{AcerDET} detector simulation (right) for non-rotating and spinning black hole samples, with $n=2$, 4 and 6 extra dimensions and $\vb{MJLOST}=\vb{.FALSE.}$. }
\label{fig:spin_basics}
\end{figure}

The inclusion of black hole angular momentum has several large effects upon the emitted particles and their spectrum. 
 The probability of a highly energetic emission is enhanced for partial waves with high values of the azimuthal quantum number $m$. As discussed at the end of Sect.~\ref{sec:En_GB_factors}, this results from the interplay between Planckian factors and transmission coefficients. The former, which turn out to be the dominant effect, are enhanced strongly for large positive values of $m$ due to the $-m\Omega$ term in the exponential of equation~\eqref{eq-flux-spectrum}, reducing the Planckian suppression. Consequently, the particle energy and transverse momentum ($P_T$) distributions for emissions from a rotating black hole are harder.
The number of primary emissions is correspondingly reduced. Fig.~\ref{fig:spin_basics} shows the emitted particle multiplicity and $P_T$ spectra for different numbers of extra dimensions. 
The effects of black hole rotation are largest for fewest number of extra dimensions, for which the spin term (in the Planckian factors) has greater magnitude. This more than compensates for their slightly lower Hawking temperature. 

The effect of black hole rotation on the pseudorapidity ($\eta$) distribution is more subtle. Assuming no strong spin recoil during the balding phase, the initial black hole formed will have a spin axis perpendicular to the beam direction.

\begin{figure}[t]
\includegraphics[scale=0.39]{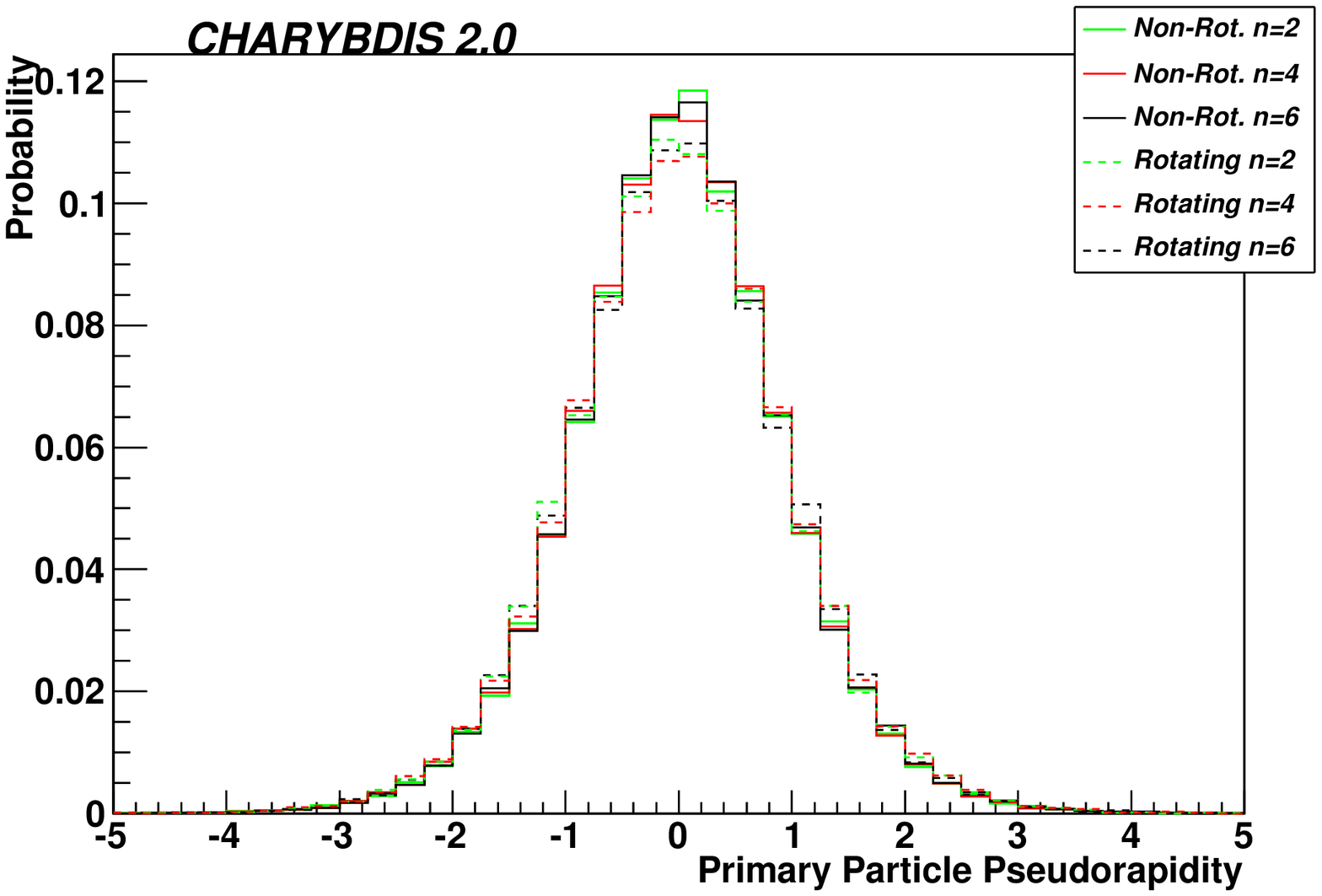} \includegraphics[scale=0.39]{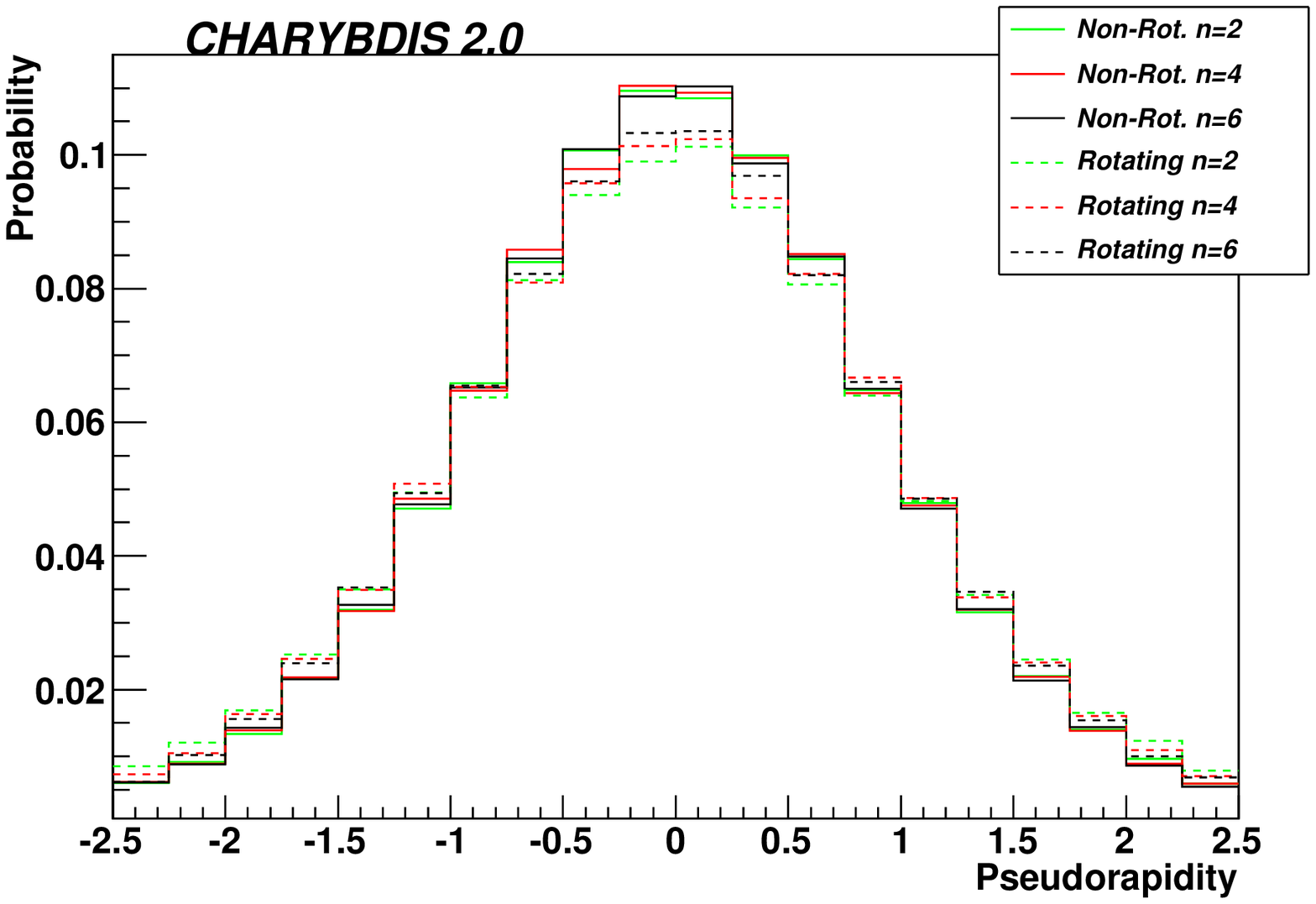}
\caption{Normalised particle $\eta$ distributions at generator level (left) and after \vb{AcerDET} detector simulation (right) from black hole samples with $n$ extra dimensions and $\vb{MJLOST}=\vb{.FALSE.}$.}
\label{fig:spin_eta}
\end{figure}

Since emission in the equatorial plane is favoured, particularly for scalars and fermions, one would expect the component along the beam direction, and hence at high $\eta$, to be enhanced, at least for initial emissions. This effect is seen experimentally in (Fig.~\ref{fig:spin_eta}), but is slight.

\begin{figure}[t]
\includegraphics[scale=0.39]{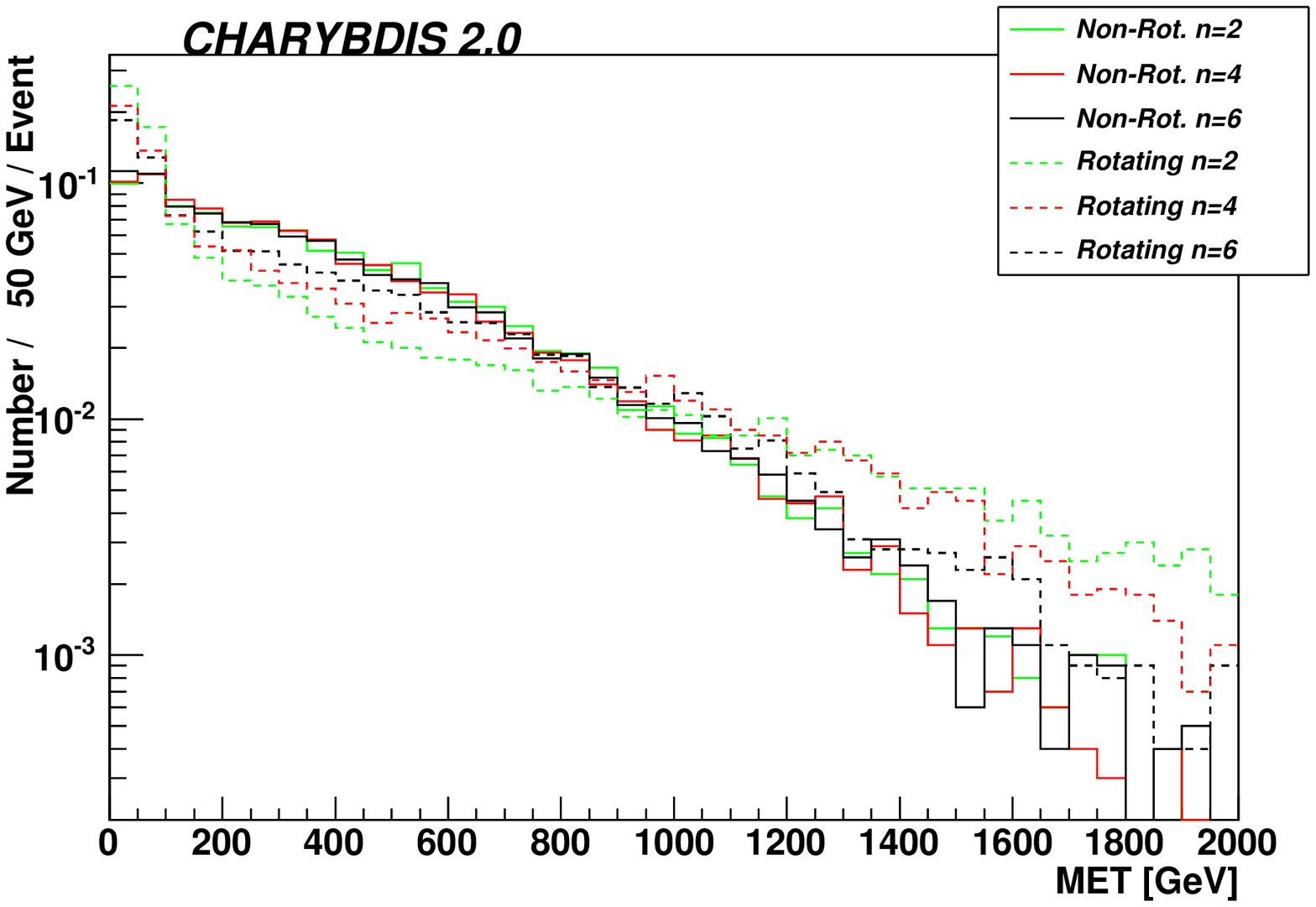} \includegraphics[scale=0.39]{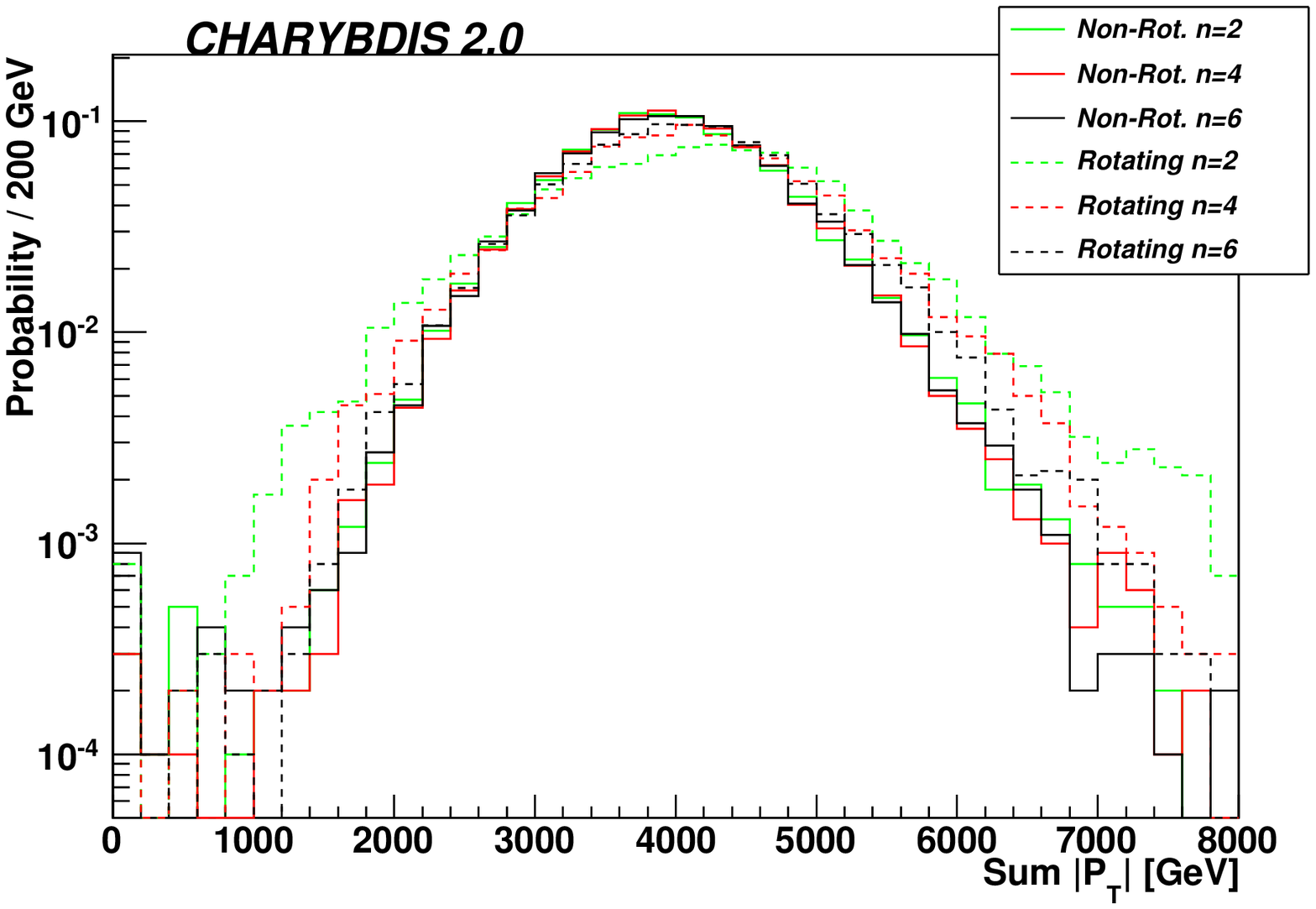}
\caption{Missing transverse energy and scalar $P_T$ sum for rotating and non-rotating black hole samples after \vb{AcerDET} fast detector simulation. Samples used the \vb{NBODYAVERAGE} criterion for the remnant phase and include a simulation of the mass and angular momentum lost during production and balding.}
\label{fig:spin_EventVar}
\end{figure}

Similar trends can be seen in event variables such as missing transverse energy (MET) and $\Sigma |P_T|$. The reduced particle multiplicity increases the probability of minimal or no MET, where no neutrinos are present in the event (neither directly emitted by the black hole, nor in weak decays of other primary emissions). The greater energy of the Hawking emissions increase the very high MET tail: a neutrino emitted by a rotating black hole is likely to have higher energy and momentum. The result is a flatter, longer tail for the rotating case, extending further beyond 1 TeV, as shown in Fig.~\ref{fig:spin_EventVar}. 

When compared to the number of primary emissions from the evaporation, a greater number of detector objects (leptons, photons, hadronic jets) are observed following fast detector simulation.
Neutrinos emitted by the black hole will not be seen experimentally, whereas a single heavy quark or vector boson will result in the detection of multiple particles or jets of hadrons. Equally, the transverse momentum spectrum observed experimentally will be slightly softer in general than that of the primary particles emitted by the black hole, due to secondary emissions and radiation.

\subsection{Particle production probabilities}

\begin{figure}[h]
\includegraphics[scale=0.39]{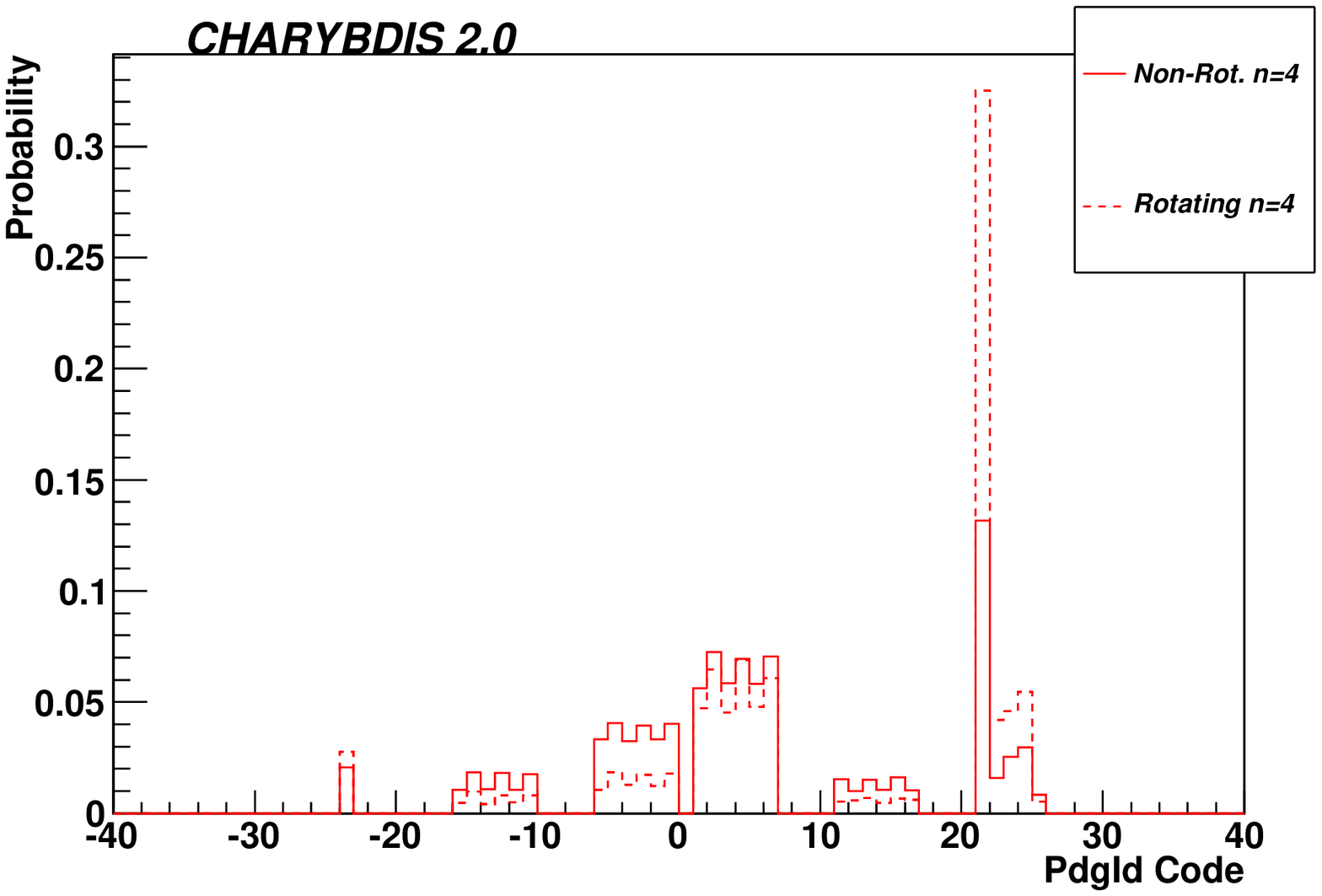} \includegraphics[scale=0.39]{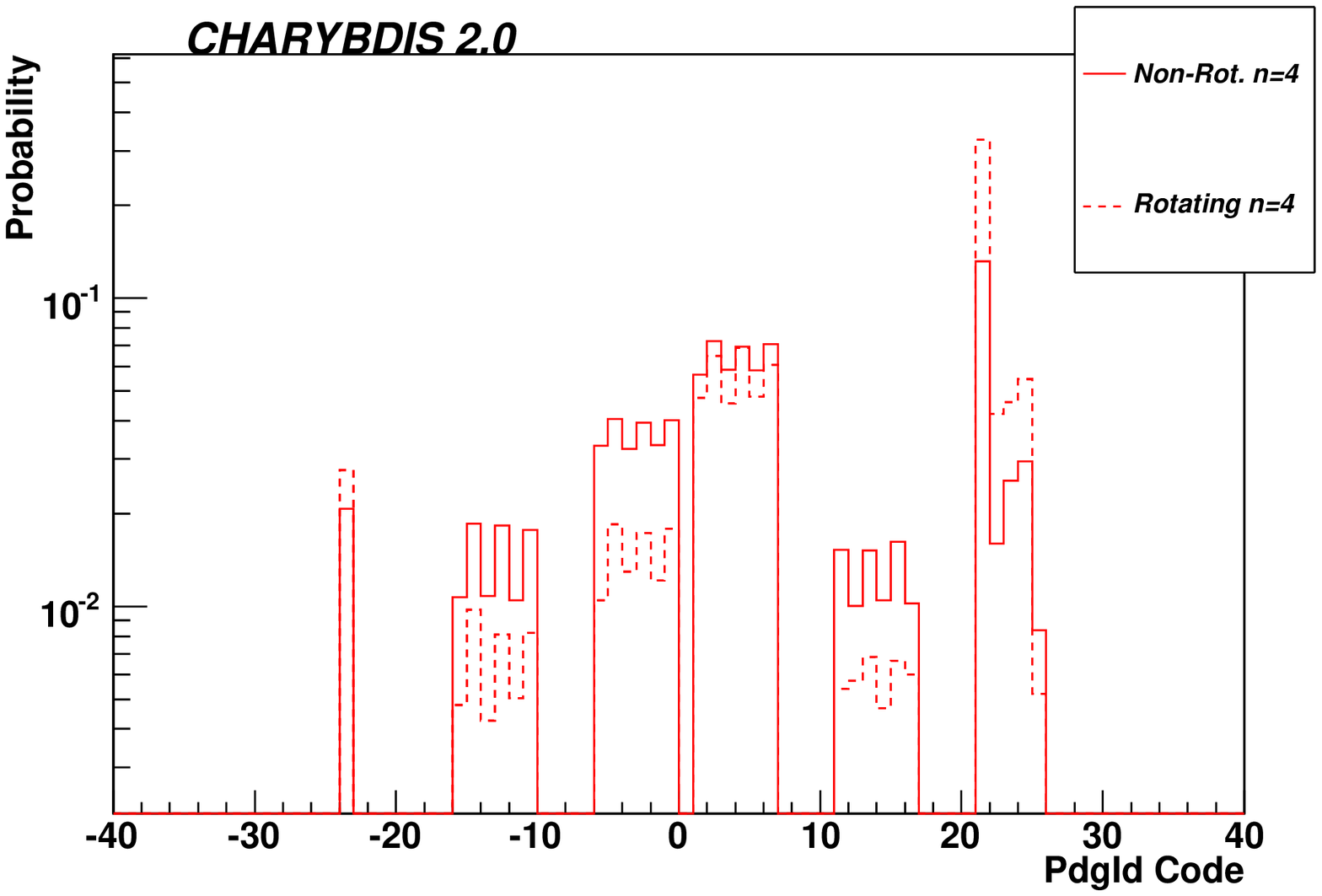} \\
\includegraphics[scale=0.39]{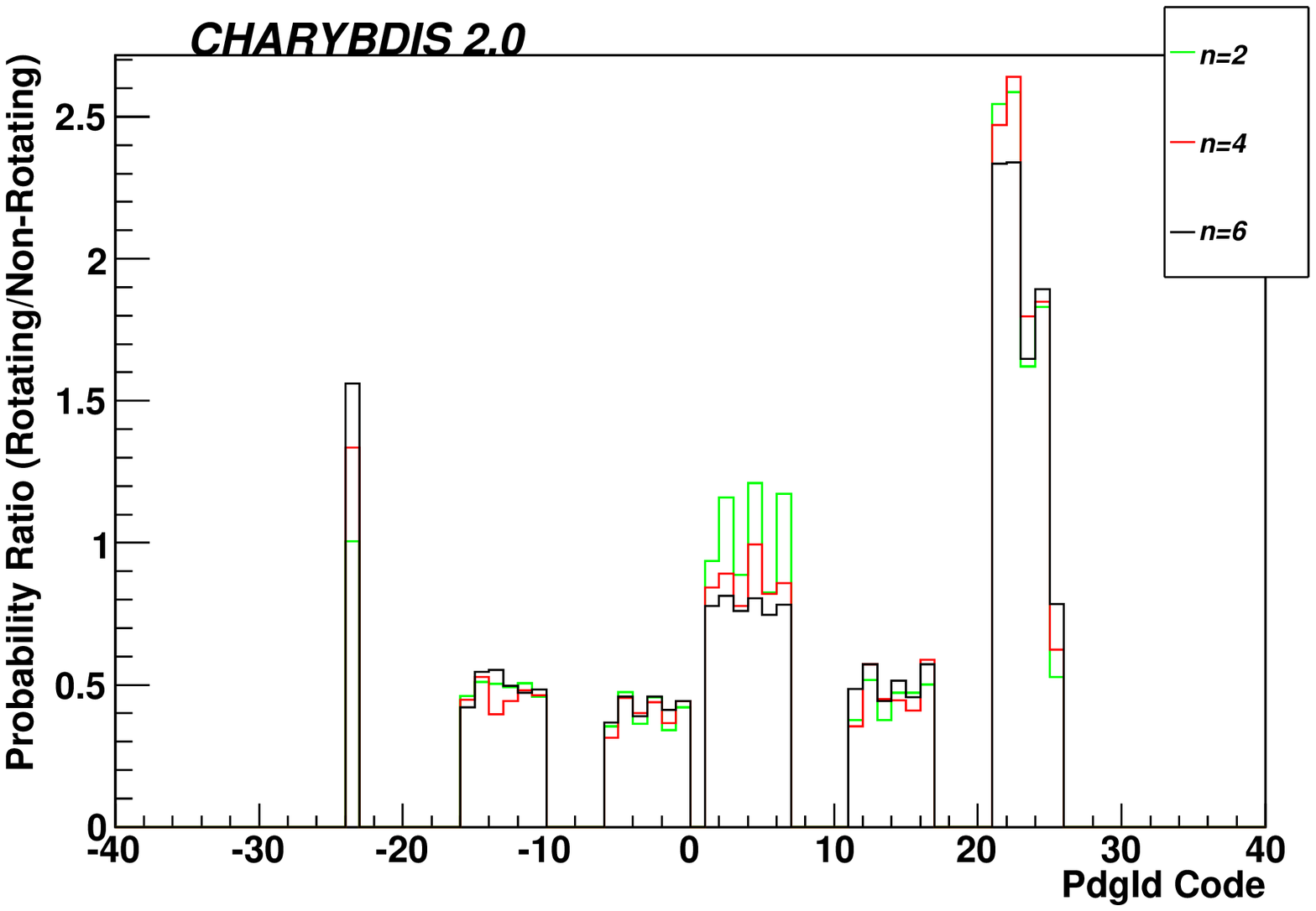} \includegraphics[scale=0.39]{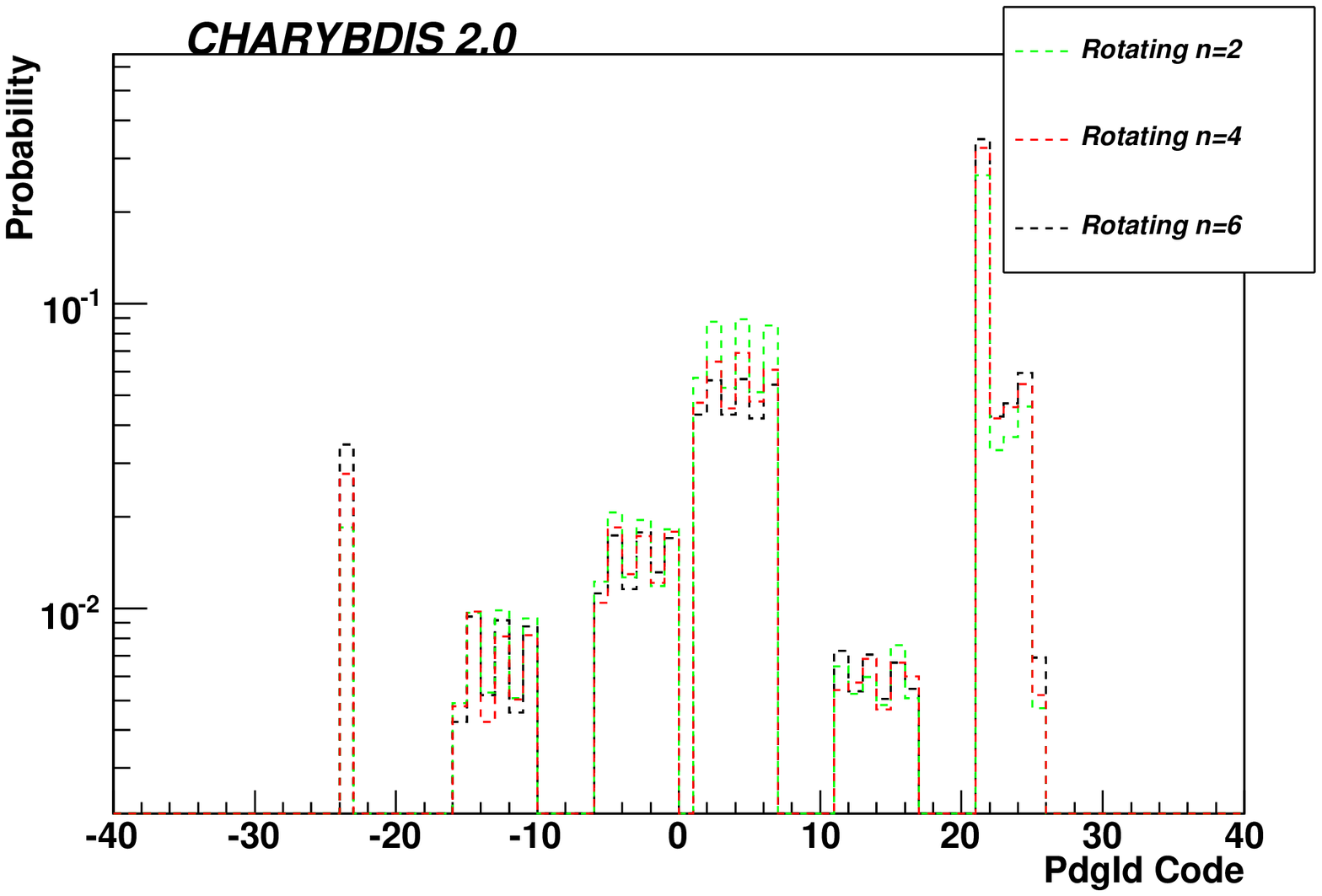}
\caption{Particle emission probabilities for rotating and non-rotating black holes with $n=4$ (top) and for rotating black holes with different $n$ (bottom) on linear and logarithmic scales. The lower left figure shows the fractional enhancement of specific particle emission from rotating black holes, as compared to non-rotating ones. The x-axis shows the PDG identification code for the SM particles, as defined in~\cite{Amsler:2008zzb}.}
\label{fig:spin_PdgId}
\end{figure}

Black hole rotation has a large effect on the particle production probabilities (Fig.~\ref{fig:spin_PdgId}).
The most dramatic of these is the enhanced emission coefficient for vector particles. This is due to the larger fluxes and agrees with the greater differential power fluxes per degree of freedom shown in Fig.~\ref{fig-emission-6d} (see vertical axis).

The greater proportion of vector emissions would provide strong evidence of rotating, rather than Schwarzschild, black holes. However such measurements are difficult to make in practice -- at the LHC it will not be possible to distinguish gluon jets from quark ones. Though highly boosted vector bosons provide experimental challenges, Z bosons can often be studied via their leptonic decay modes. Perhaps the most accessible other means to investigate black hole rotation might be the study of the photon multiplicity or its ratio to other particles, TeV-energy photons being one manifestation of black holes reproduced by neither other new physics scenarios nor SM backgrounds. 
Another experimental difficulty for the detection and the isolation of the black-hole signal is that rotation decreases the probability of producing a lepton -- often useful in reducing jet-like SM backgrounds to black hole events~\cite{Aad:2009wy}.

The emission probabilities for each particle species are largely independent of the number of extra dimensions, which primarily affects the emission energy and multiplicity, so that a reproduction of the distribution of particle species would be powerful evidence of black holes.

The particle-antiparticle imbalance in Fig.~\ref{fig:spin_PdgId} is chiefly caused by the (usually positively charged) input state. According to the mechanism described in Sect.~\ref{sec:spin}, up-type quarks and down-type antiquarks are favoured, so as to meet the constraints of charge balance. 
Similarly, the net positive baryon number of the input state and the need to conserve baryon number for hadronisation leads to a preference for quarks over antiquarks. 
The apparent increase in this with rotation is a reflection of the reduced particle multiplicity: with fewer particles amongst which to share the charge imbalance, the effect is magnified.
This is potentially a source of uncertainty since, unlike charge, black holes do not have to conserve baryon or lepton quantum numbers. At present we are constrained to conserve these by the needs of hadronisation generators.

\subsection{Black hole evolution}
\begin{figure}[t]
\includegraphics[scale=0.39]{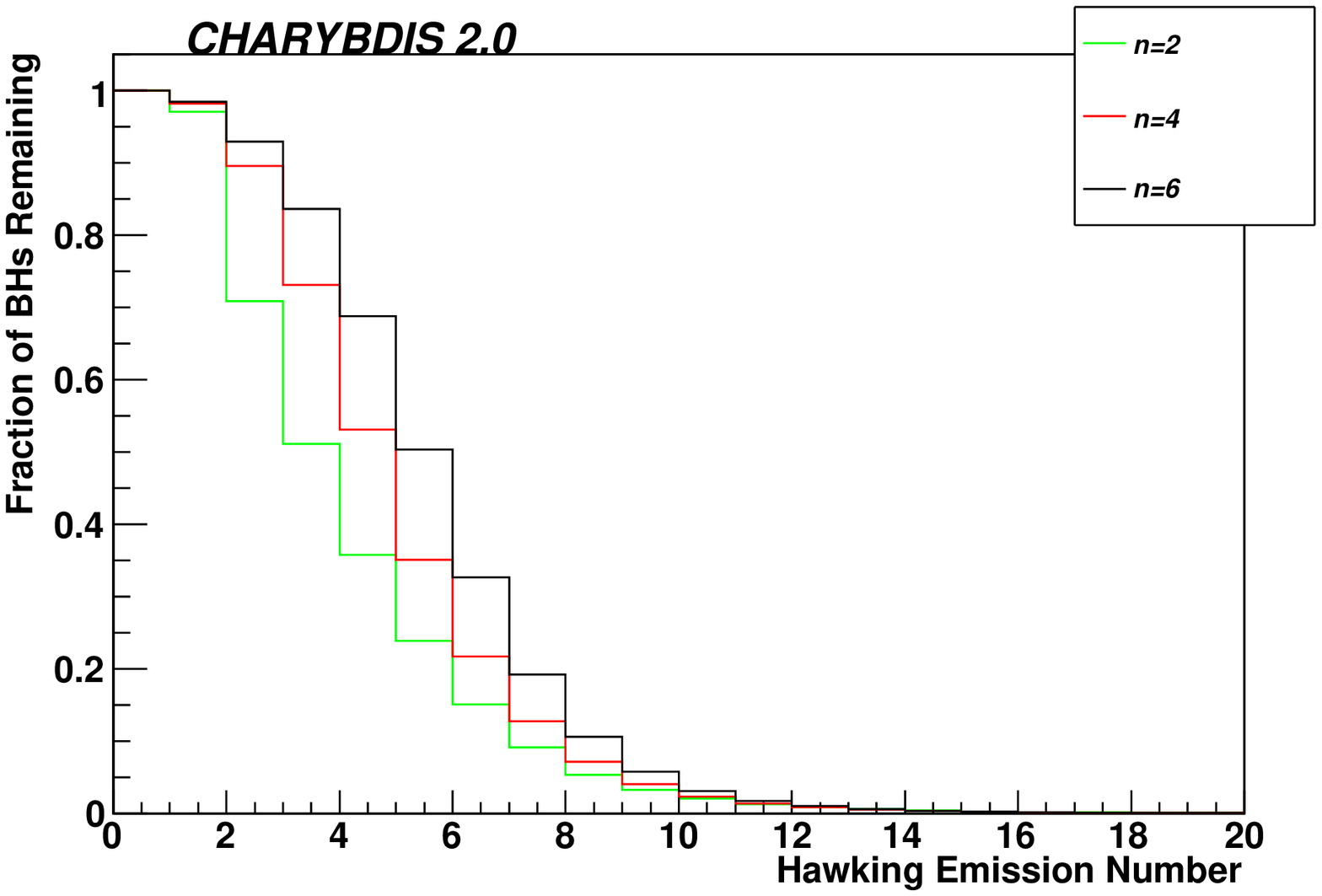} \includegraphics[scale=0.39]{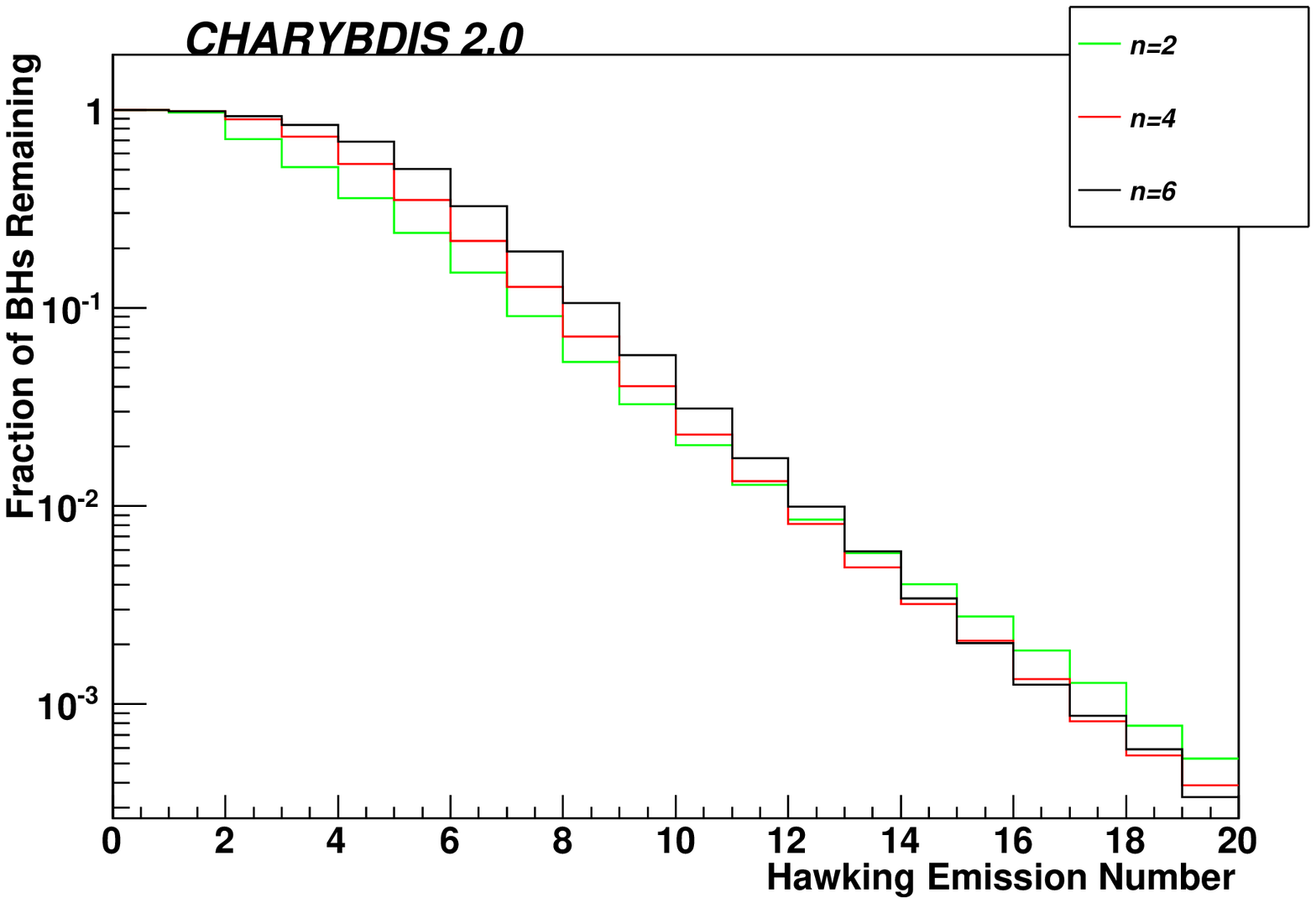} \\
\includegraphics[scale=0.39]{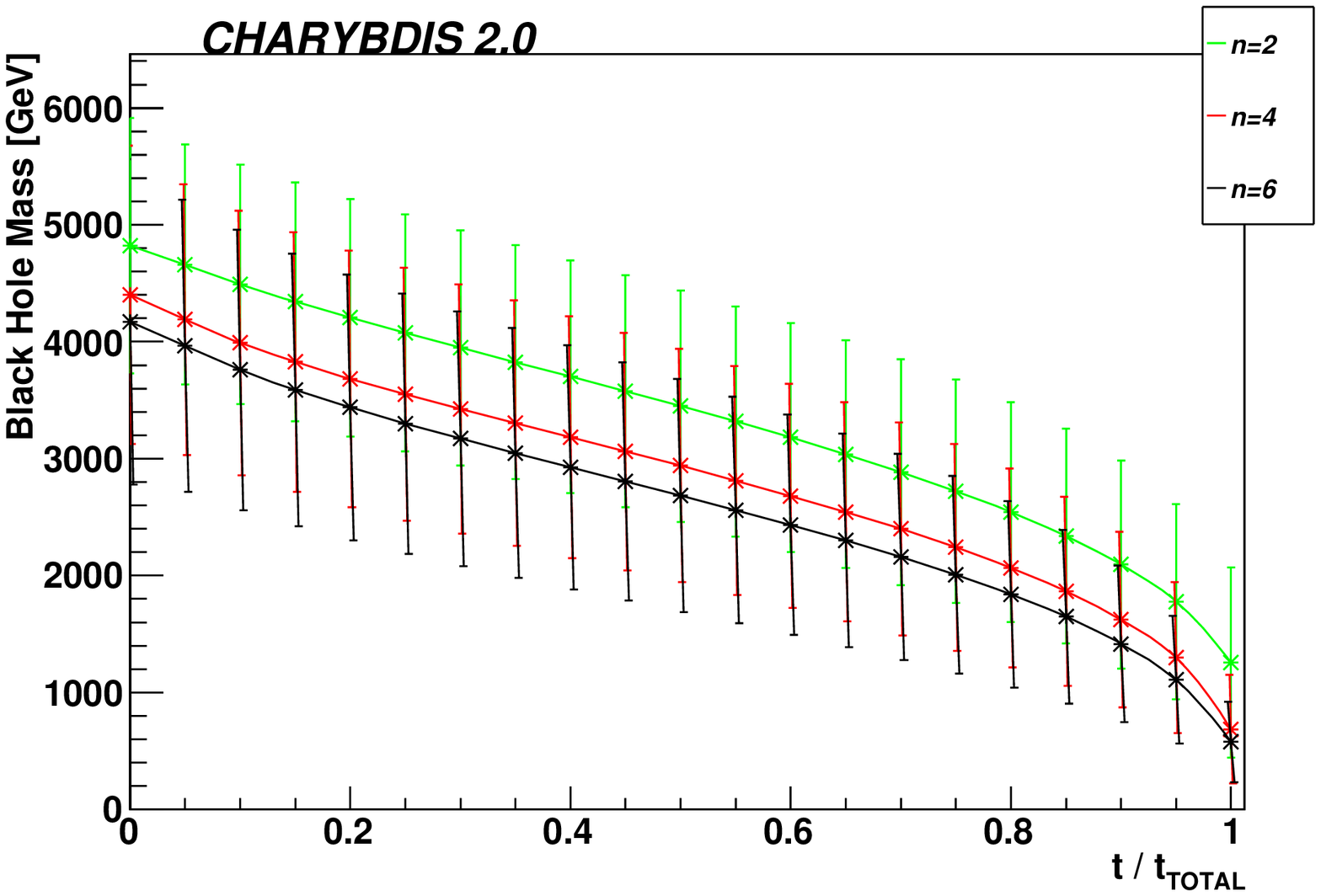} \includegraphics[scale=0.39]{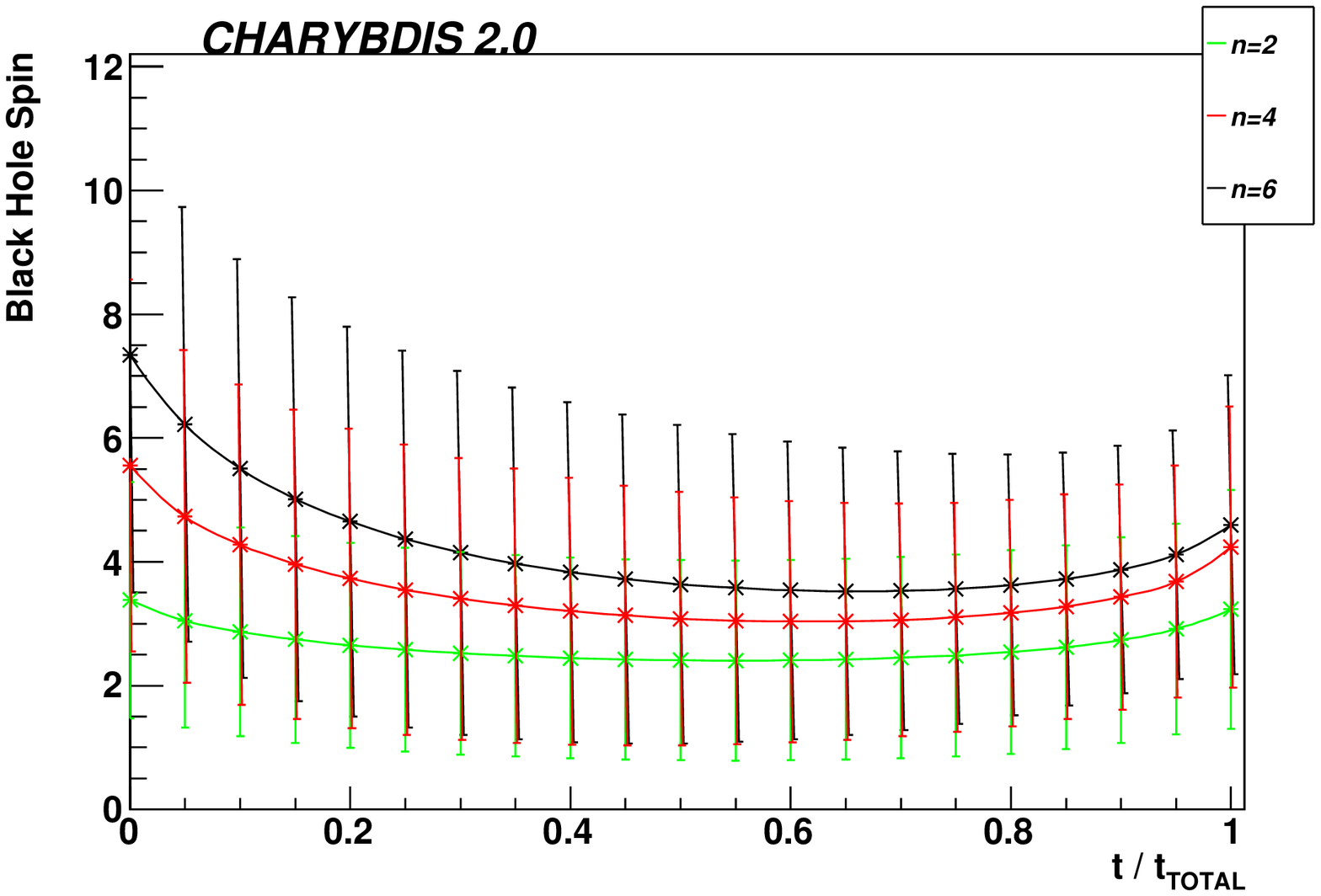} \\
\includegraphics[scale=0.39]{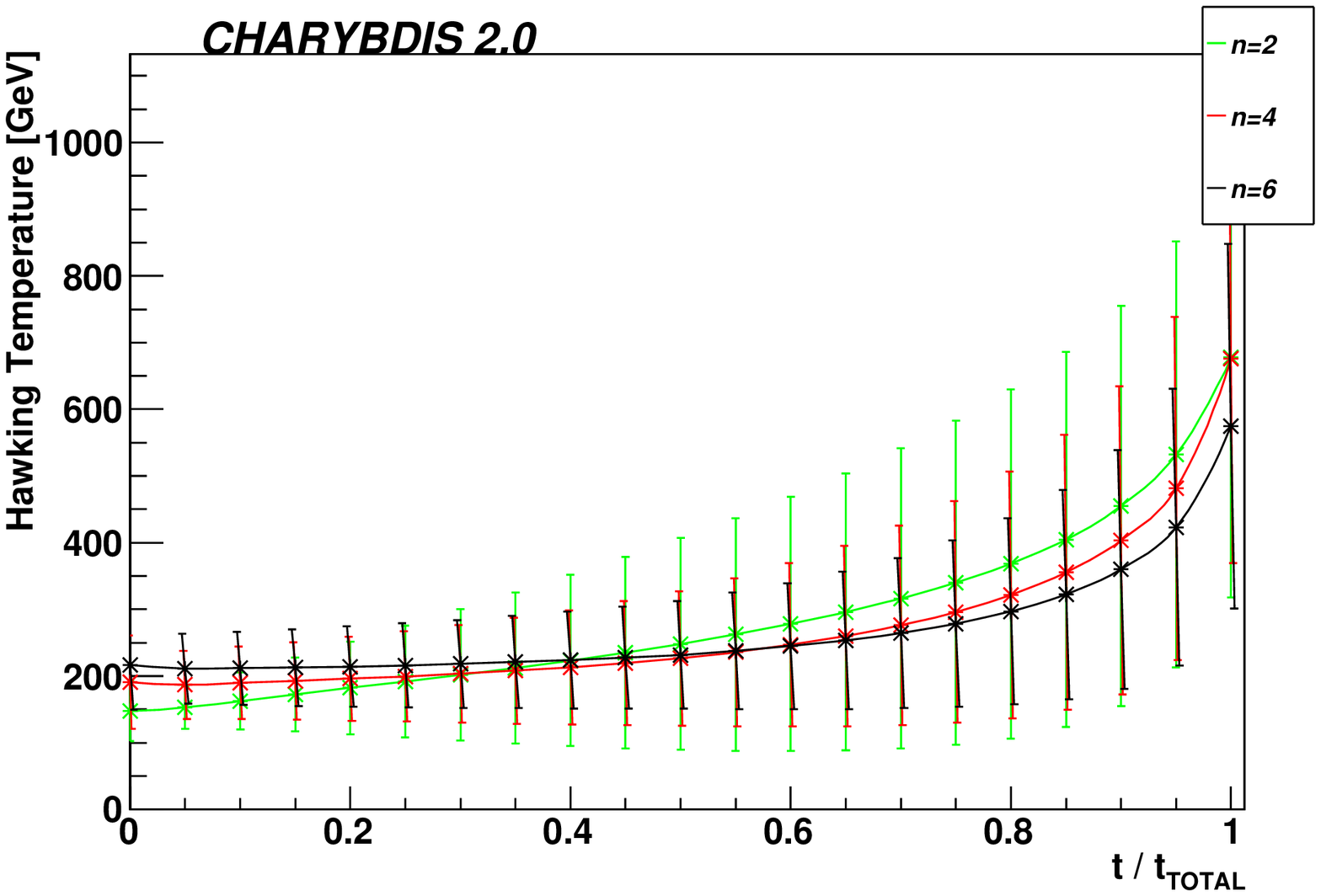} \includegraphics[scale=0.39]{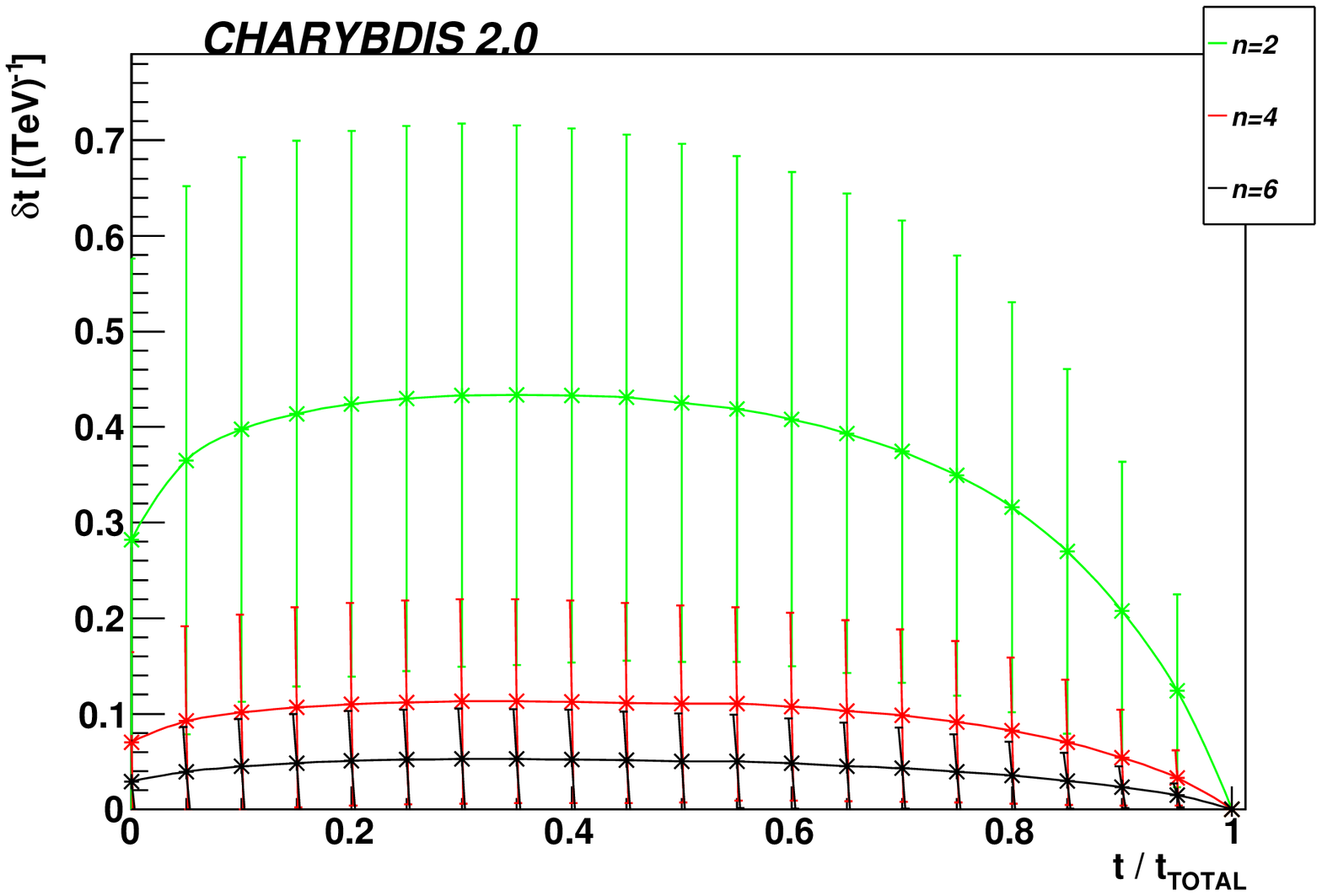}
\caption{The evolution of black hole parameters during the Hawking evaporation phase. The lower two rows show the mean value tendency curves of the relevant parameter as a function of fractional time $t/t_{Total}$ through the evaporation. Error bars indicate the standard deviation of the distribution. The bottom right plot indicates the typical time interval $\delta t$ between emissions; its large standard deviation is due to a long tail for the distributions at each $t/t_{Total}$.}
\label{fig:evo_plots}
\end{figure}
As the Hawking emission proceeds, the black hole evolves, becoming lighter, hotter and losing angular momentum as detailed in Fig.~\ref{fig:evo_plots}. Without the simulation of losses in production/balding, the black hole spins down more quickly than it loses mass; for large black hole angular momentum, energetic emissions at high $m$ are highly favoured. Turning on the simulation suppresses initial states with high black hole angular momentum. 
Consequently this effect is reduced in magnitude, though the majority of the black hole angular momentum is still lost before its mass. 

Initially the high angular momentum of the black hole leads to a high emission flux so the typical time interval between emissions is reduced relative to a non-rotating BH. 

The higher spin term in the Planckian factor causes there to be fewer, more energetic emissions for few extra dimensions. Half the mass is lost in the first 3 emissions for $n=2$, compared with 4 ($n=4$) and 5 for $n=6$. The distribution does have a substantial tail however, with 1\% of black hole events producing more than 11 primary emissions. 

\begin{figure}[t]
\includegraphics[scale=0.375]{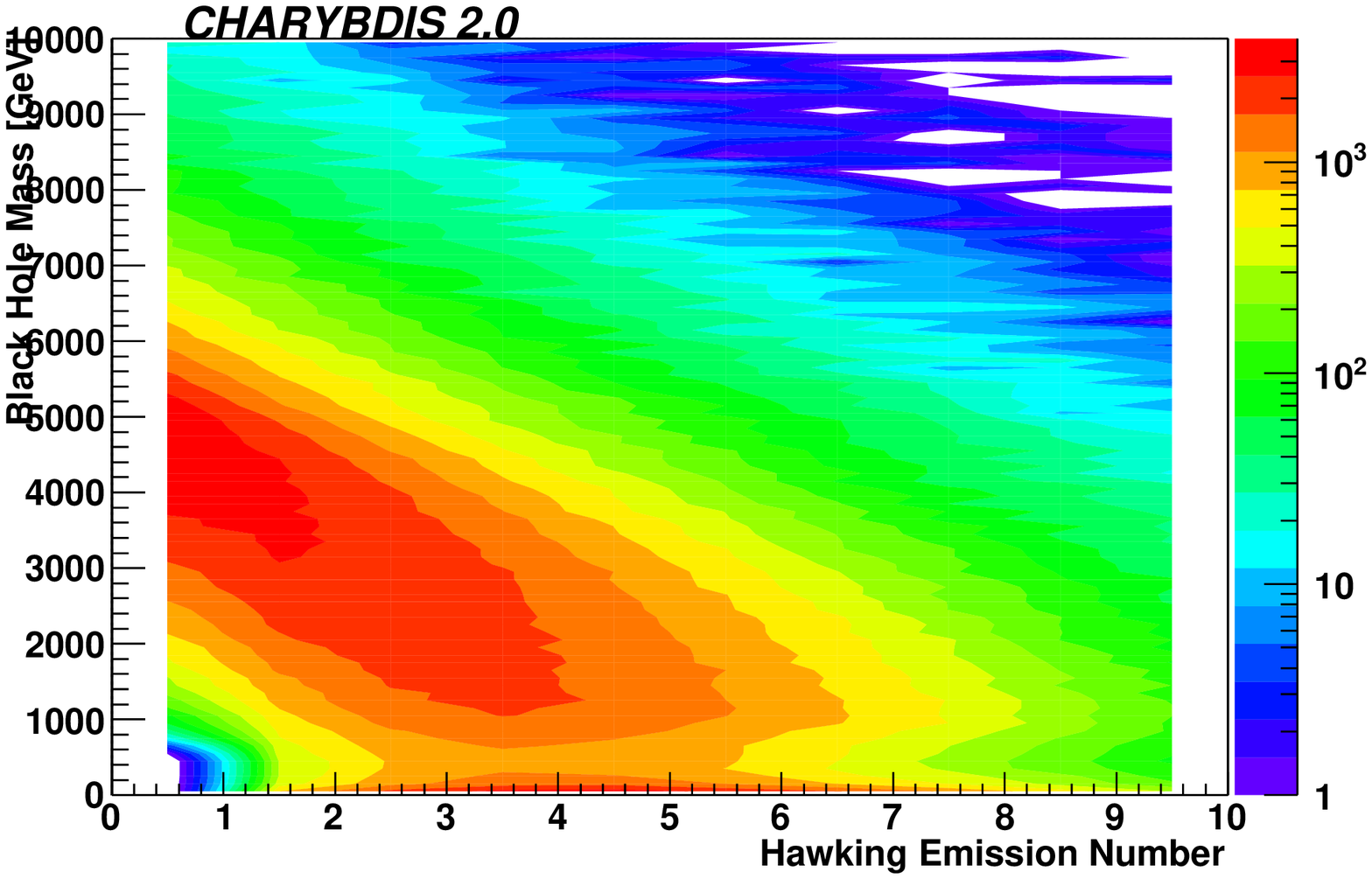} \includegraphics[scale=0.375]{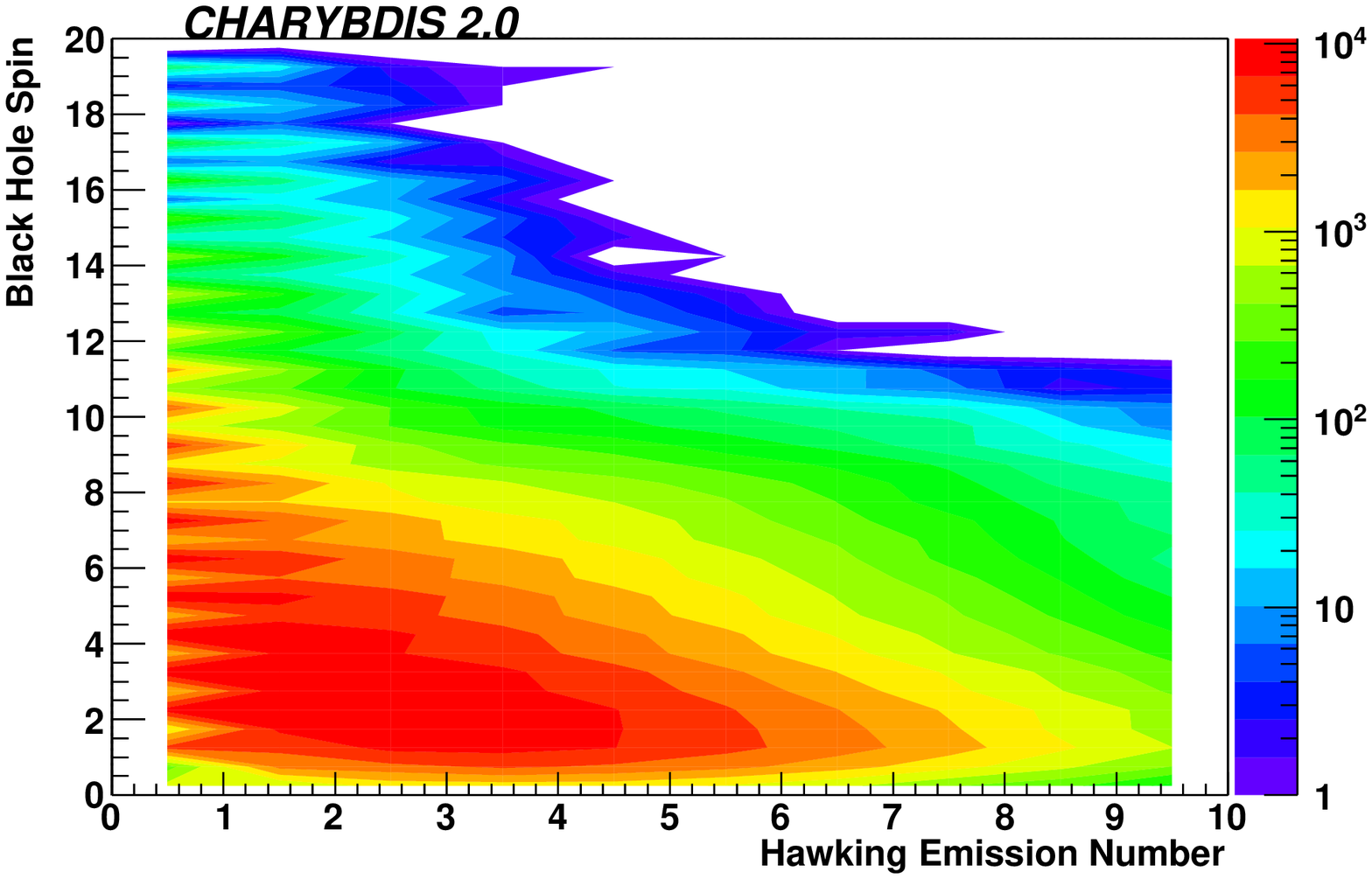} \\
\includegraphics[scale=0.375]{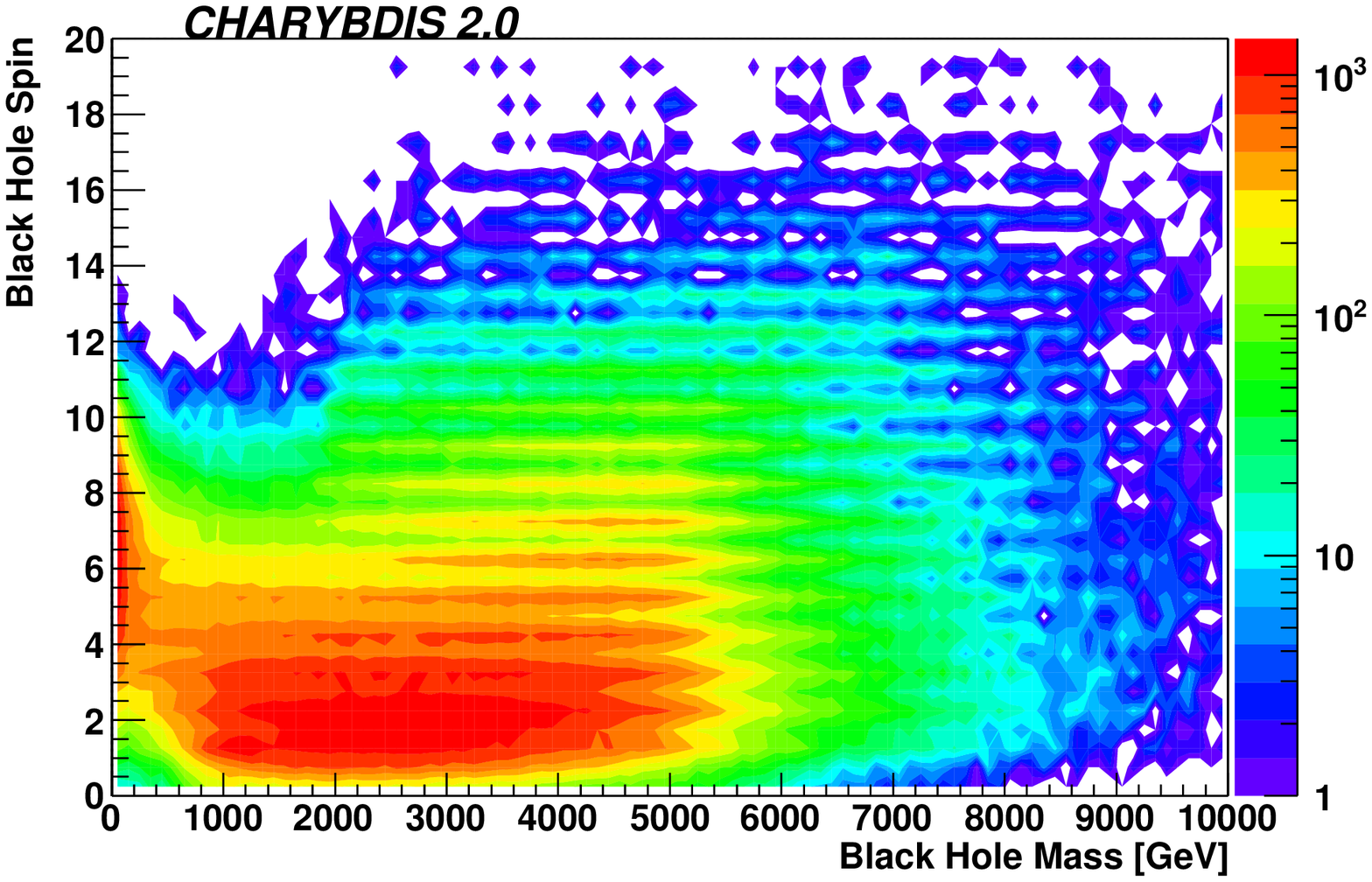} \includegraphics[scale=0.375]{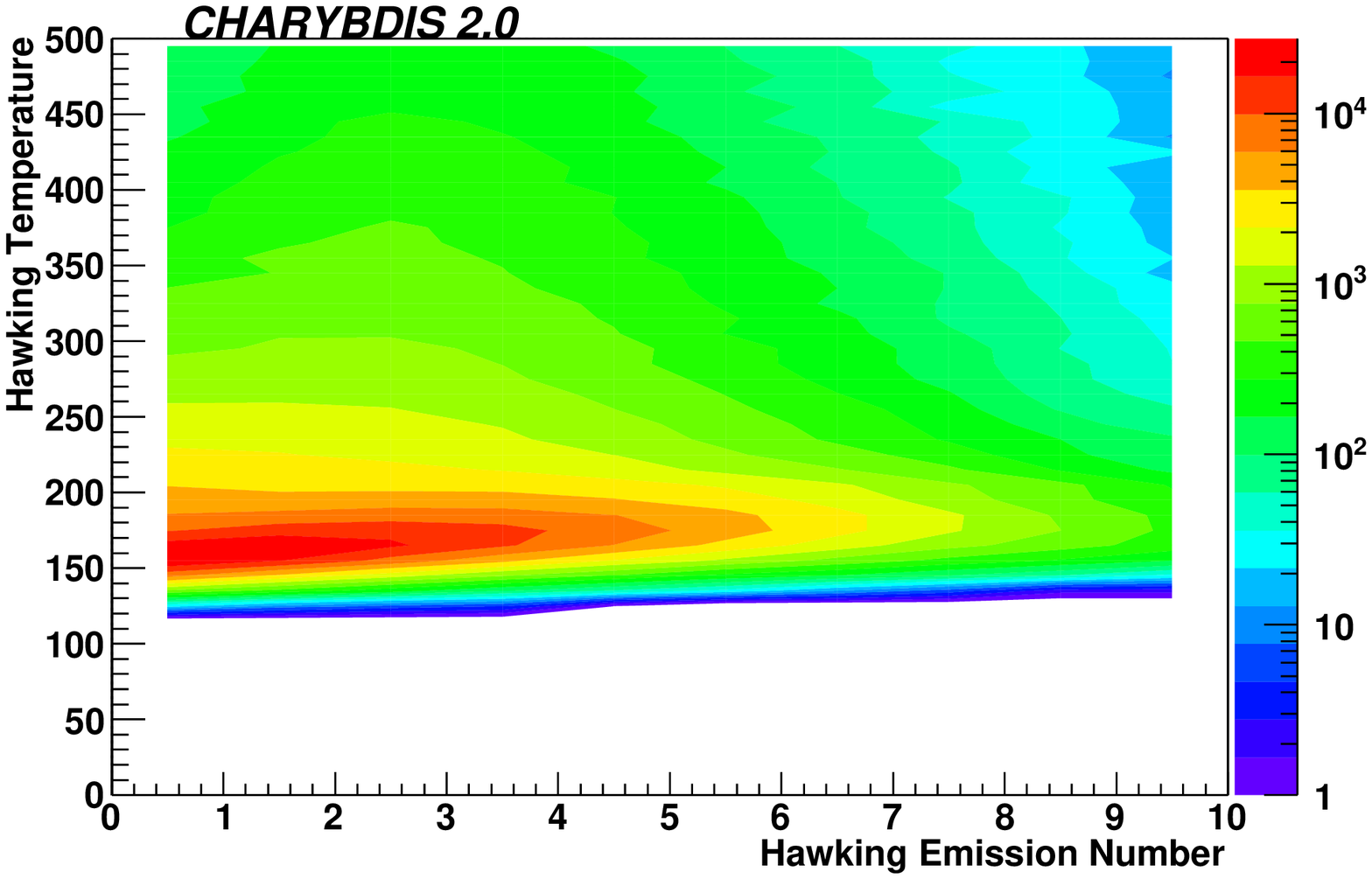}
\caption{2D contour plots showing how the black hole mass, angular momentum evolve with successive Hawking emissions, and their correlations for samples with $n=4$ extra dimensions, using the \vb{NBODYAVERAGE} criterion for the remnant phase and including a simulation of the mass and angular momentum lost during production and balding.} 
\label{fig:n4_evo_contz}
\end{figure}

As is shown in the contour plots of Fig.~\ref{fig:n4_evo_contz}, black holes with high initial angular momentum tend to lose much of it during their first few emissions, whereafter further emissions decrease the black hole mass more smoothly, whilst its angular momentum stays relatively low, but non-zero. Thus the spin-down phase persists throughout the black hole decay -- only a small proportion of black holes settling into a Schwarzschild, non-rotating state. This is in direct agreement with the theoretical plots in Figs.~\ref{maps_backreaction} and~\ref{maps_backreaction_geo}.

As the black hole becomes lighter, its temperature rises, as does its oblateness ($a_*$) and the time interval between emissions drops. These effects are gradual except when the black hole mass becomes very low, at the end of the Hawking radiation phase. At this point we have reached the remnant phase. 

The striated, lined structure in the angular momentum plots is due to the initial state: the probability of a quark-gluon interaction, and correspondingly a half-integer angular momentum state is much lower than the integer state above and below it.

\subsection{Remnant options}

\vb{CHARYBDIS2} includes several models for the remnant phase. Both fixed and variable body decays have parameter switches to enable the systematics to be studied, as detailed in Sect.~\ref{sec:term}. 

\begin{figure}[t]
\includegraphics[scale=0.39]{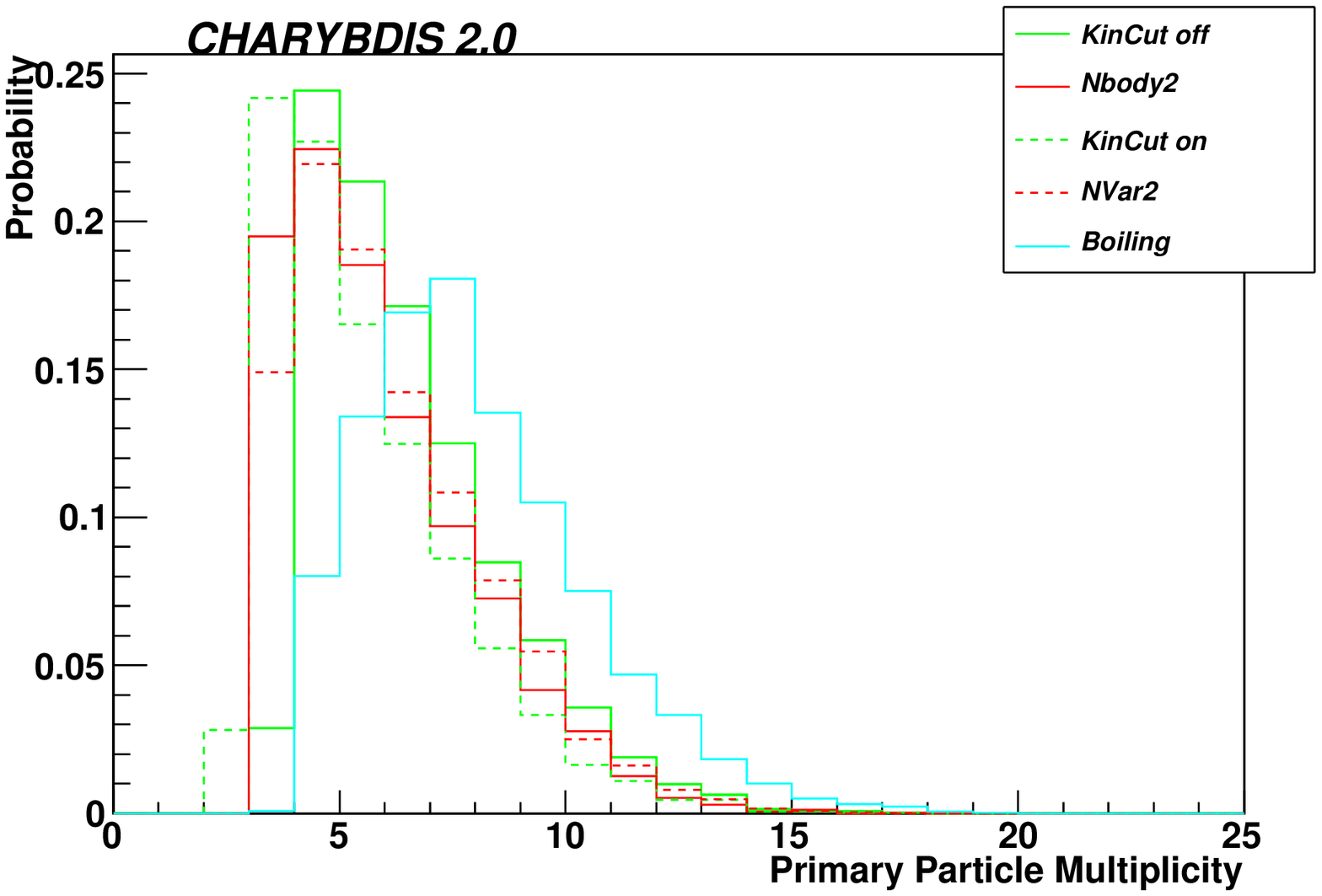} \includegraphics[scale=0.39]{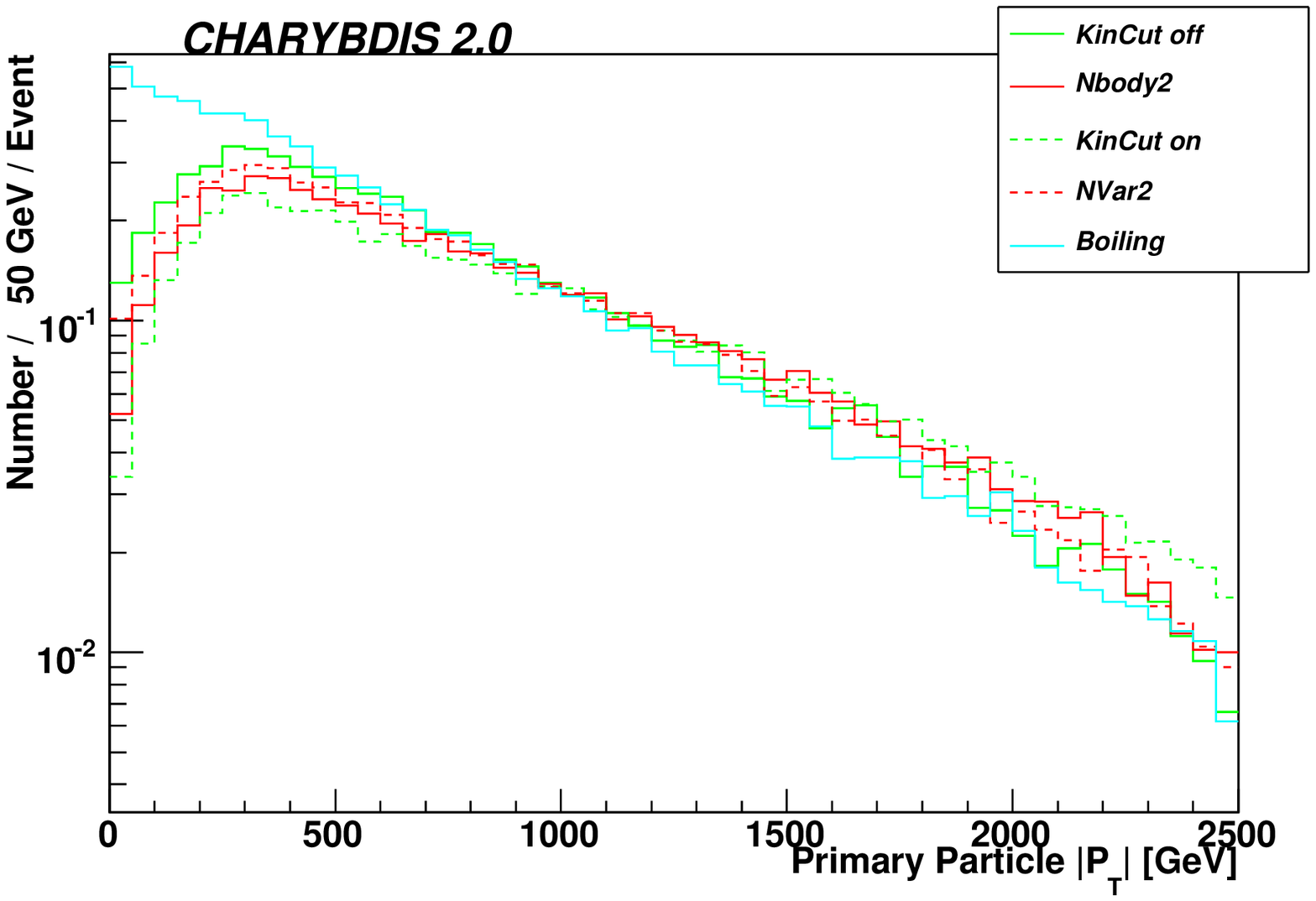} \\
\includegraphics[scale=0.39]{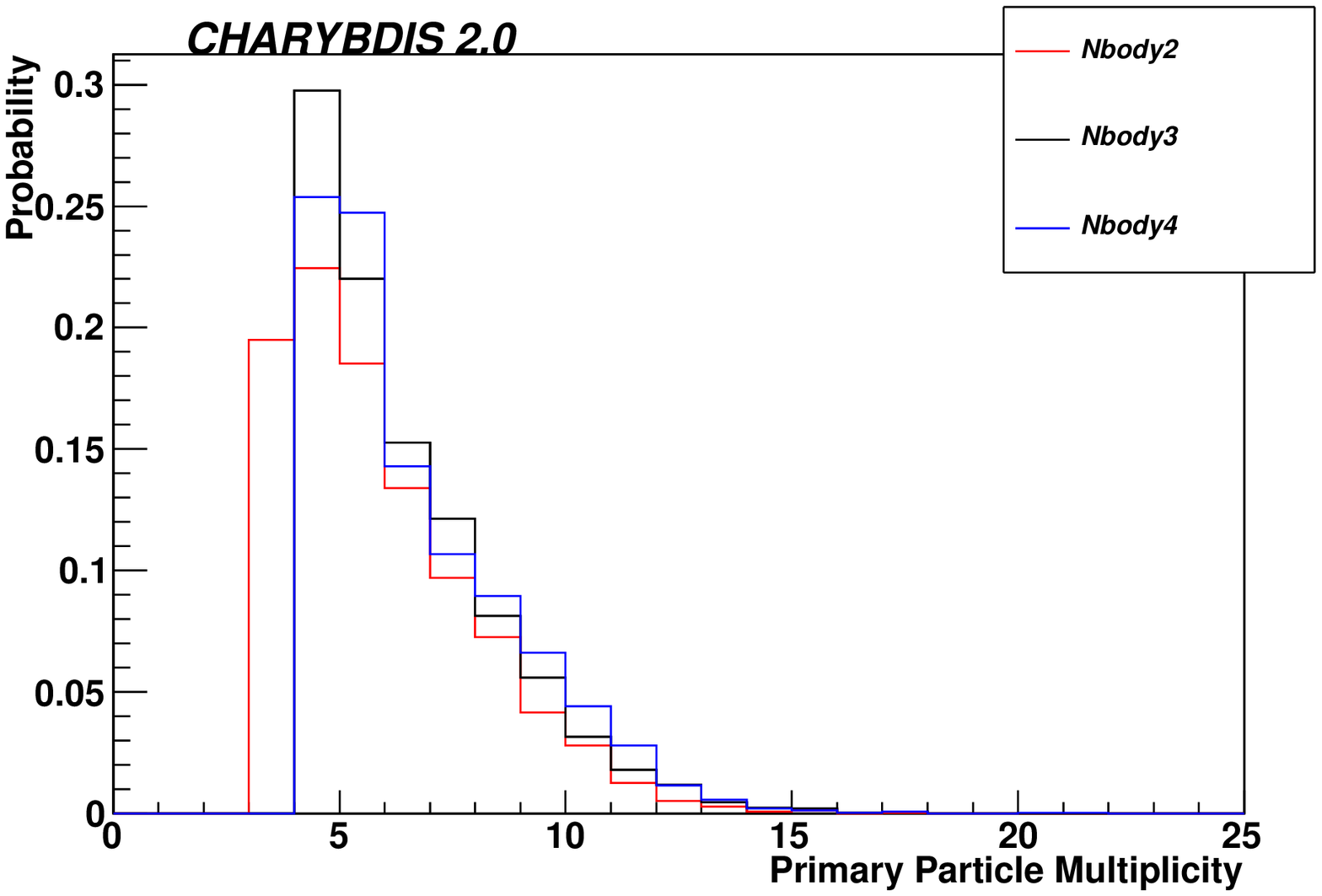} \includegraphics[scale=0.39]{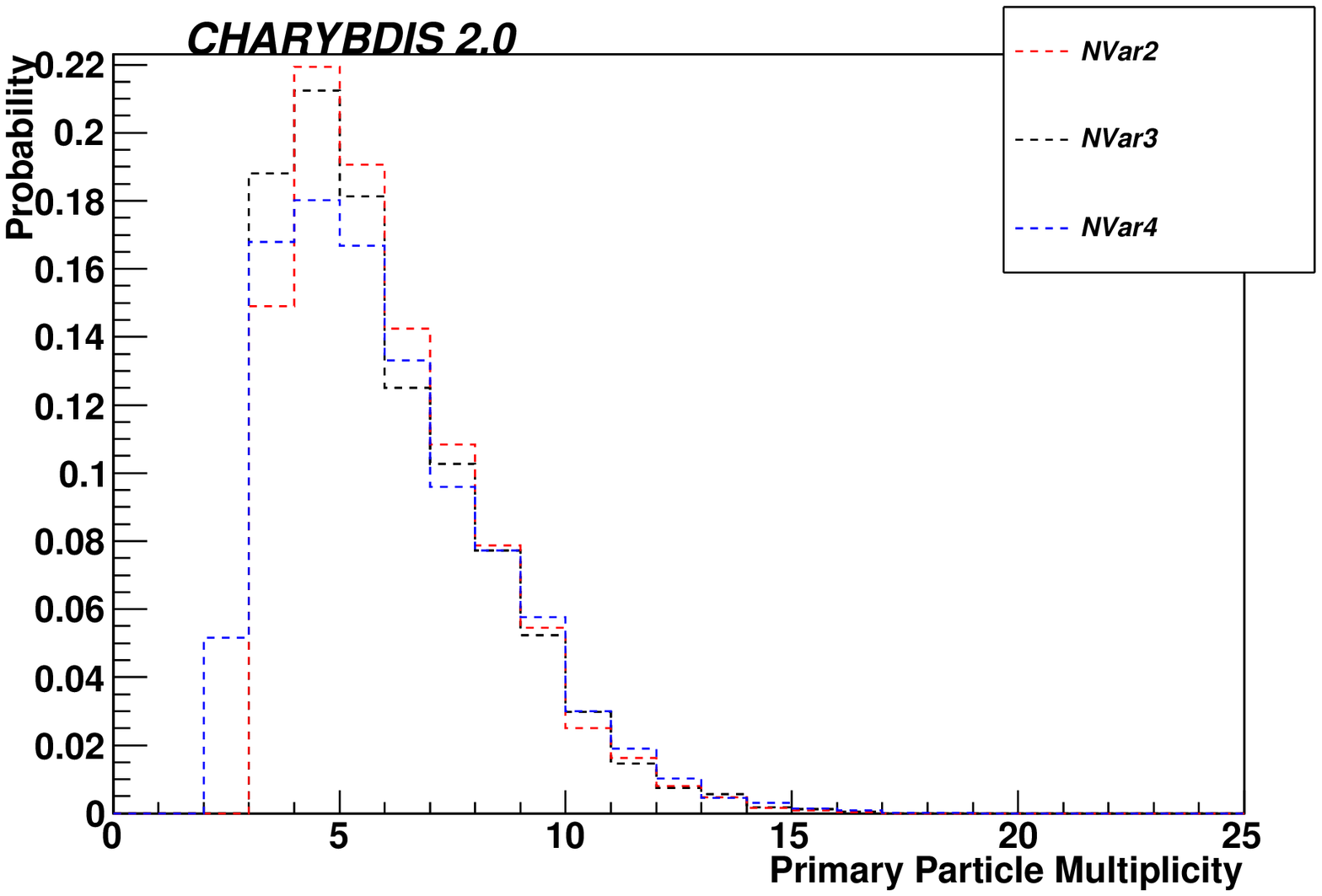}
\caption{Primary particle multiplicity and $|P_T|$ distributions for black hole samples with $n=2$, using a wide range of remnant options, as defined in Table~\ref{tab:remn}. }
\label{fig:remn_mult}
\end{figure}

The fixed multiplicity model, present in \vb{CHARYBDIS} and optional in \vb{CHARYBDIS2}, is linked to the choice of the variable \vb{KINCUT}. If \vb{KINCUT=.FALSE.}, proposed decays that are kinematically disallowed are ignored; if \vb{KINCUT=.TRUE.}, their proposal terminates the evaporation phase. The former choice will give a greater number of less energetic particles, as evidenced by Fig.~\ref{fig:remn_mult} which constrasts a range of remnant models defined in Table~\ref{tab:remn}.

\vb{CHARYBDIS2} uses the \vb{NBODYAVERAGE} remnant criterion as a default, where the fluxes are used to calculate the expected number of further emissions. This provides a physically motivated model. Using this criterion with either a fixed 2-body (``nbody2'') or a variable multiplicity remnant model (``nvar2'') gives a distribution lying between the upper and lower values obtained using the older model switches (upper plots of Fig.~\ref{fig:remn_mult}), indicating good control over the uncertainties mentioned in Sect.~\ref{sec:term}. The string-motivated boiling model gives a slightly higher multiplicity, since successive emissions are produced until the remnant mass drops below the remnant minimum mass, resulting in a greater number of softer particles produced in the remnant phase.

The new \vb{NBODYAVERAGE} model is also more robust with respect to changes in the number of particles produced in the remnant phase. This is because the flux calculation allows the spin-down phase to be terminated whenever the expected number of further emissions is fewer than that selected for the remnant phase. This is illustrated in the lower plots of Fig.~\ref{fig:remn_mult}, where changing the number of particles produced in the remnant phase results in similar multiplicities and spectra; events with 4-body remnant decays do not always have two more particles than their 2-body analogues.

\begin{figure}[t]
\includegraphics[scale=0.39]{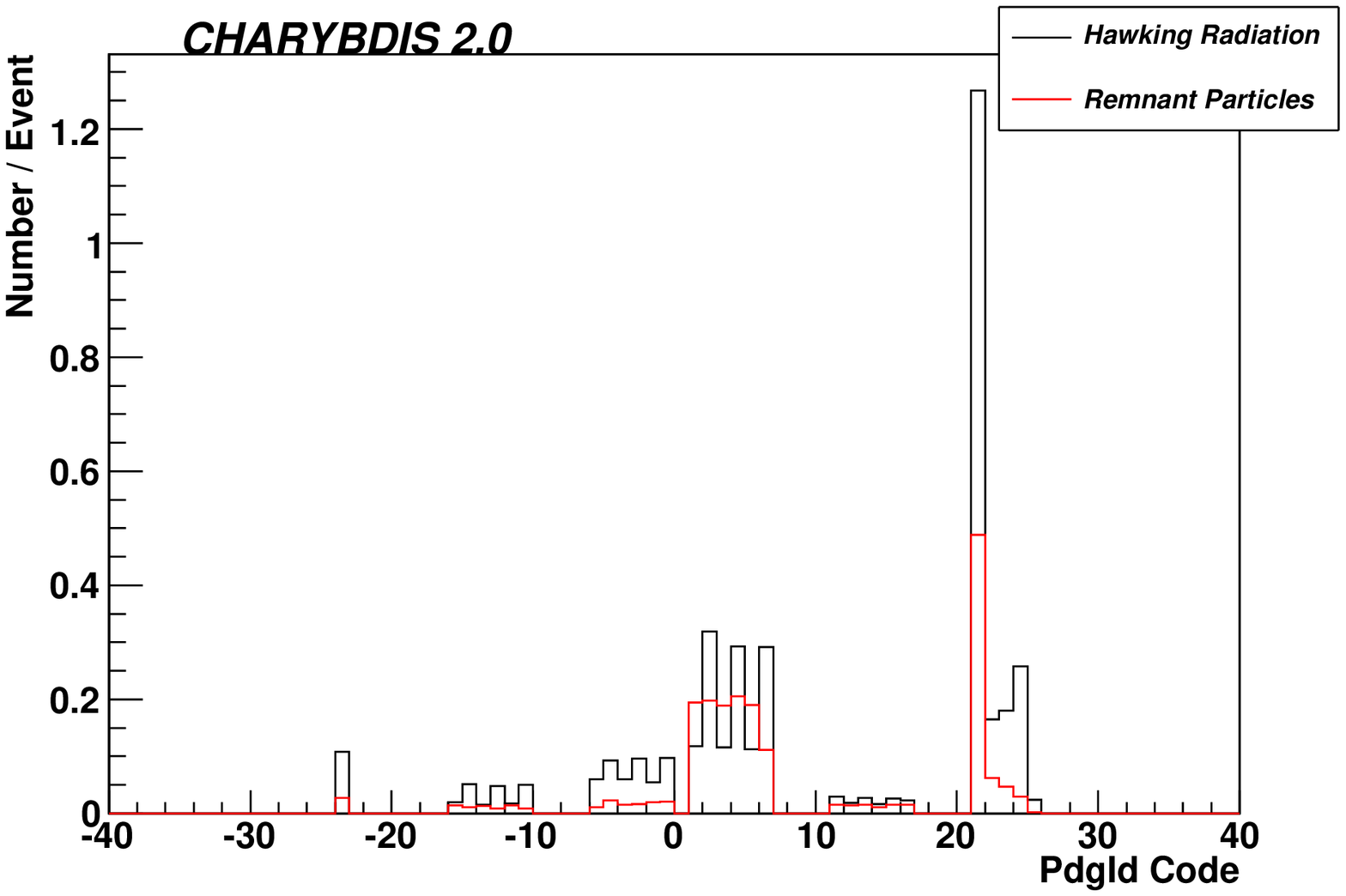} \includegraphics[scale=0.39]{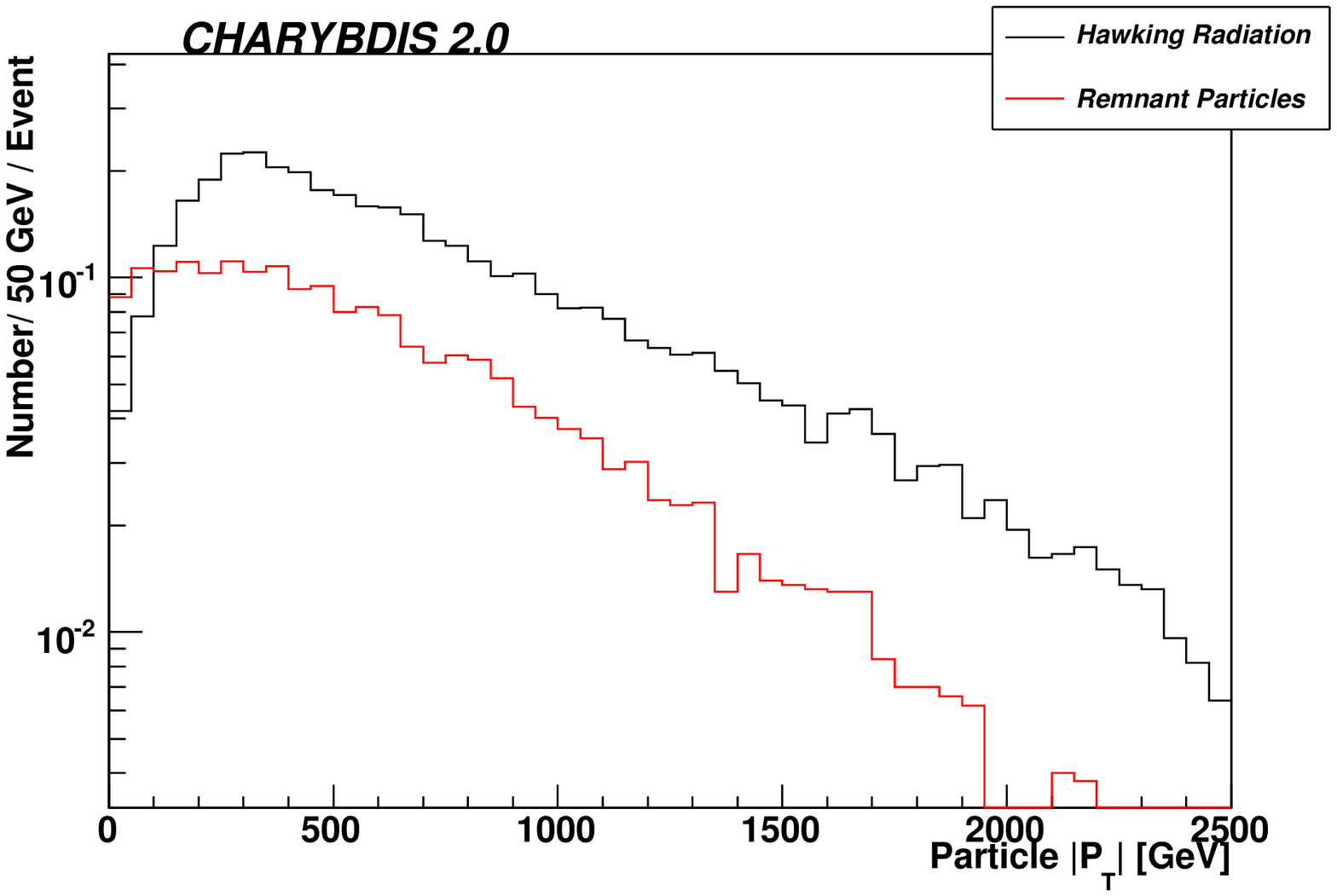} \\
\includegraphics[scale=0.39]{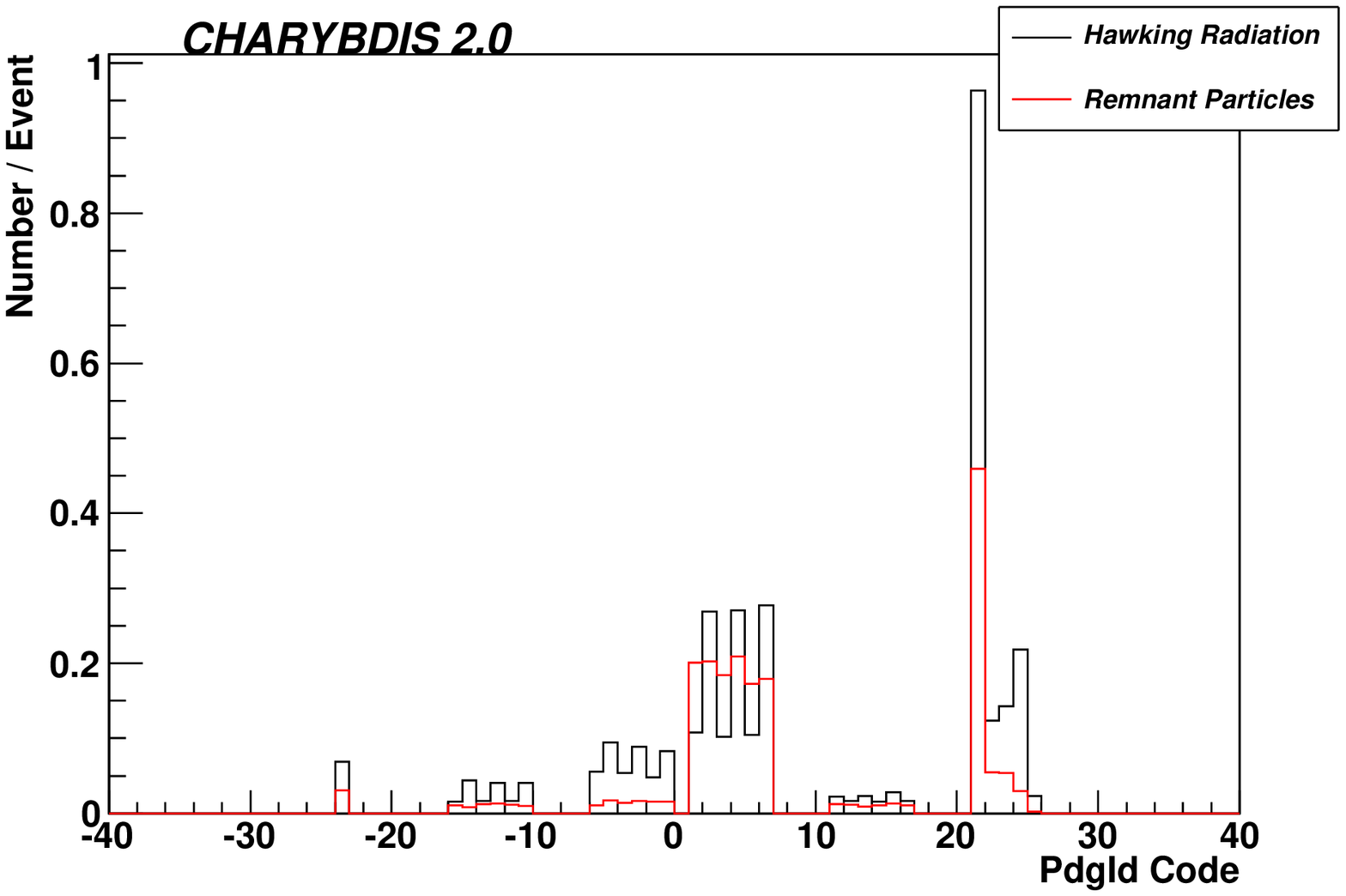} \includegraphics[scale=0.39]{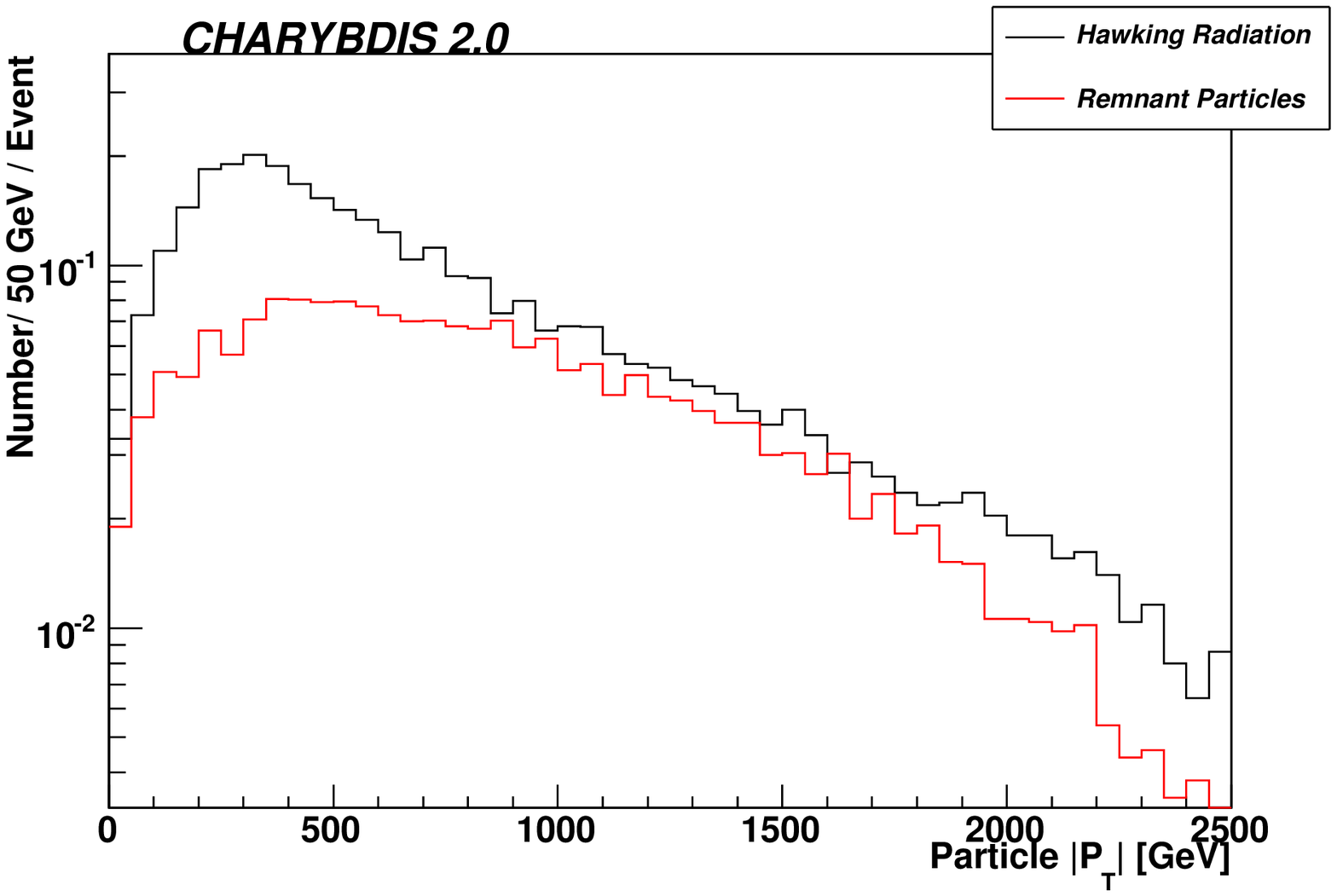}
\caption{Particle type and $P_T$ plots for 2-body remnant decays using the old model ``Kincut off'' (top) and the new model ``nbody2'' (bottom) as defined in Table~\ref{tab:remn}. Distributions are normalised per event.}
\label{fig:remn_HvsR}
\end{figure}

Another advantage of the \vb{NBODYAVERAGE} method is that by using the integrated power and flux, the spin-down phase is terminated at a point that allows a smoother transition to the remnant phase, as shown by their spectra in Fig.~\ref{fig:remn_HvsR}, where the \vb{NBODYAVERAGE} (lower) method gives a more concordant distribution of particle transverse momenta. Performing a remnant decay only when the mass drops below the Planck mass gives a much softer momentum spectrum, in contrast to the high energies favoured by light rotating black holes. The option to start the remnant decay based on the drop of $\left<N\right>$ provides a smoother transition, since the final decay particles will have a harder spectrum, more similar to the Hawking phase. Emissions in the remnant phase are predominantly coloured, with positive baryon number favoured, so as to meet the constraints of baryon number conservation. 

\subsection{Mass reconstruction}
\begin{figure}[t]
\includegraphics[scale=0.39]{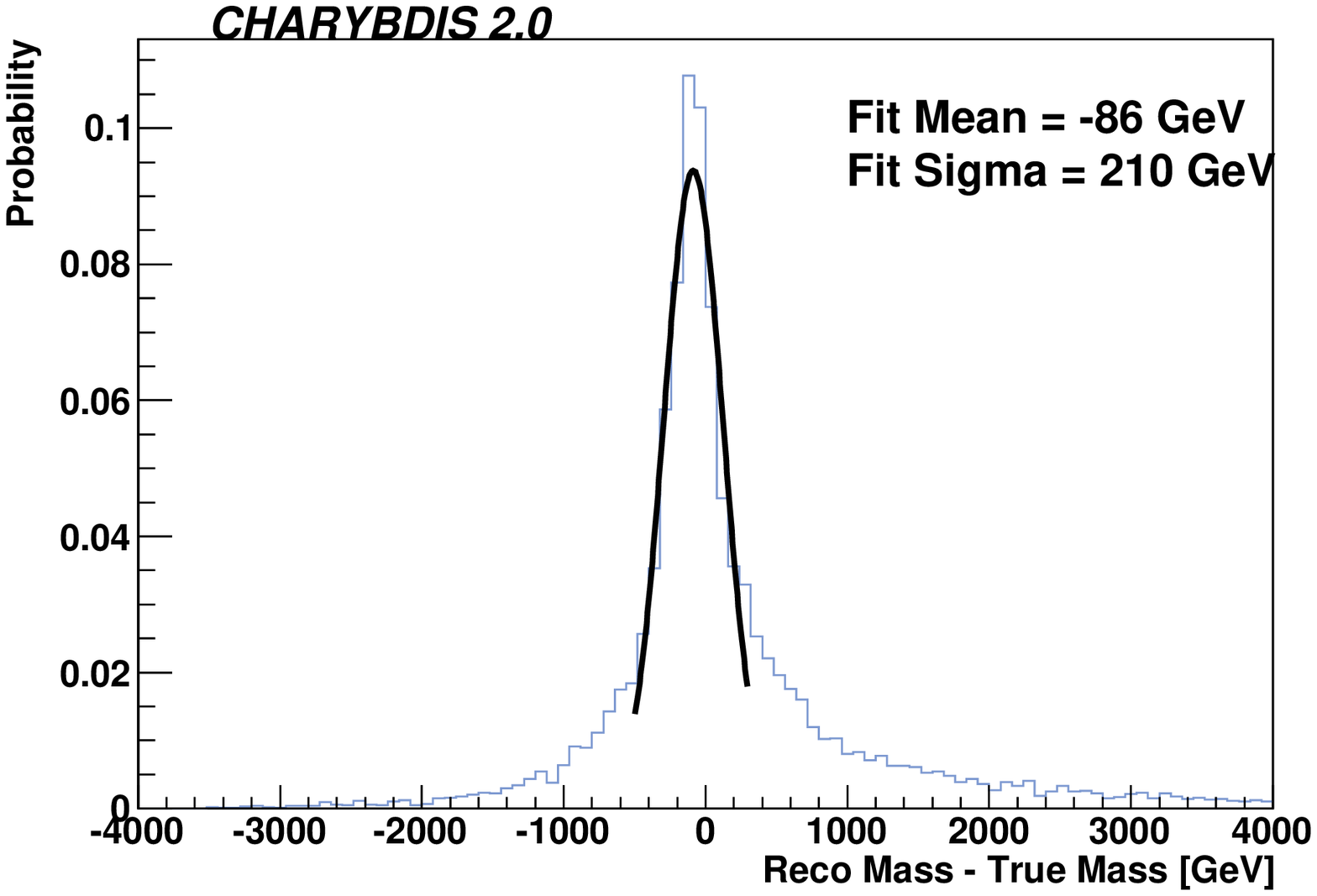} \includegraphics[scale=0.39]{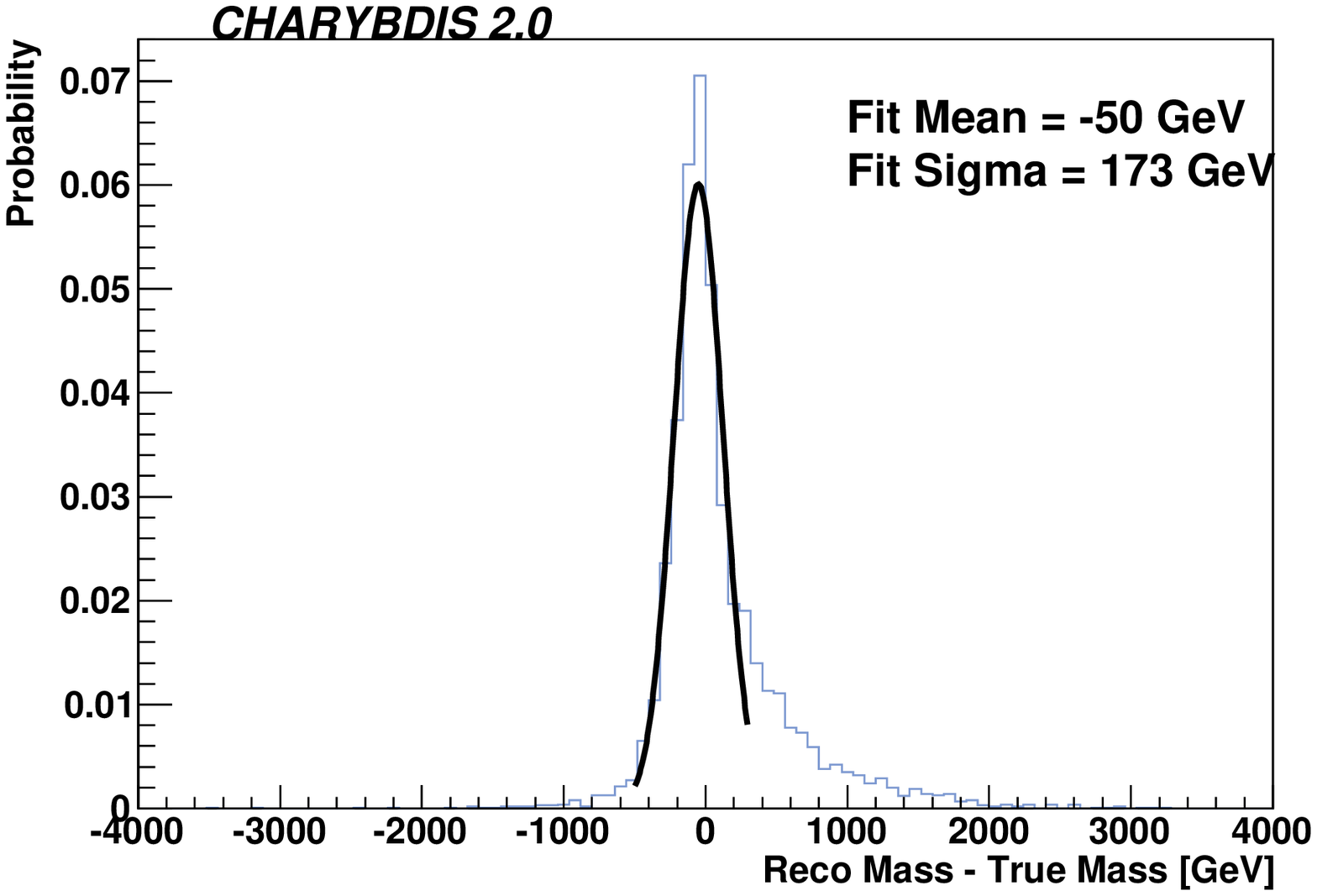}
\caption{Sample black hole mass resolutions after \vb{AcerDET} detector simulation with $n=2$ and no balding simulation for all events (left) and after a cut of MET$<100$~GeV (right). The fits are indicative of the resolution in the peak and do not model the non-Gaussian tails which remain.}
\label{fig:prod_massres}
\end{figure}
\begin{figure}[t]
\includegraphics[scale=0.39]{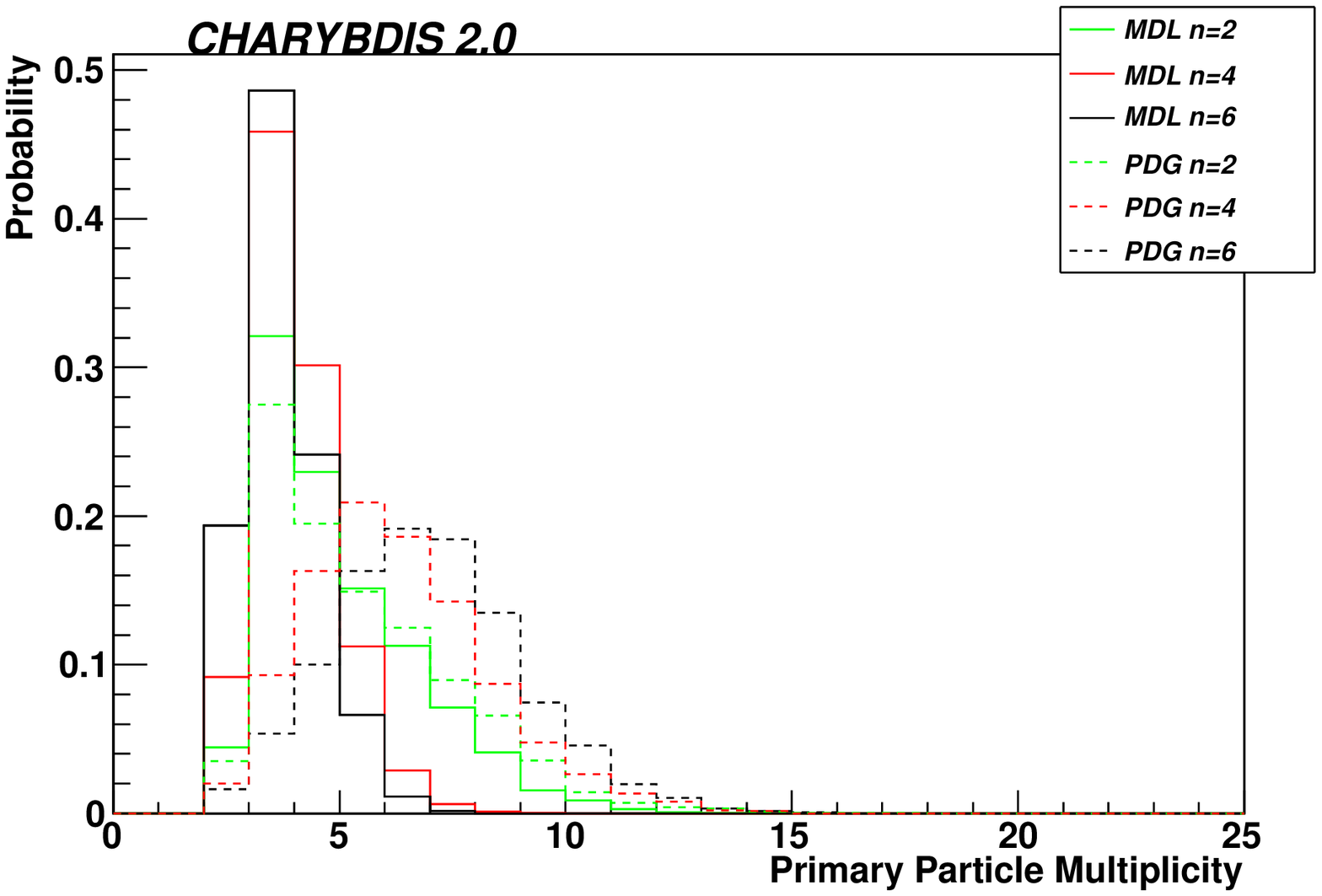} \includegraphics[scale=0.39]{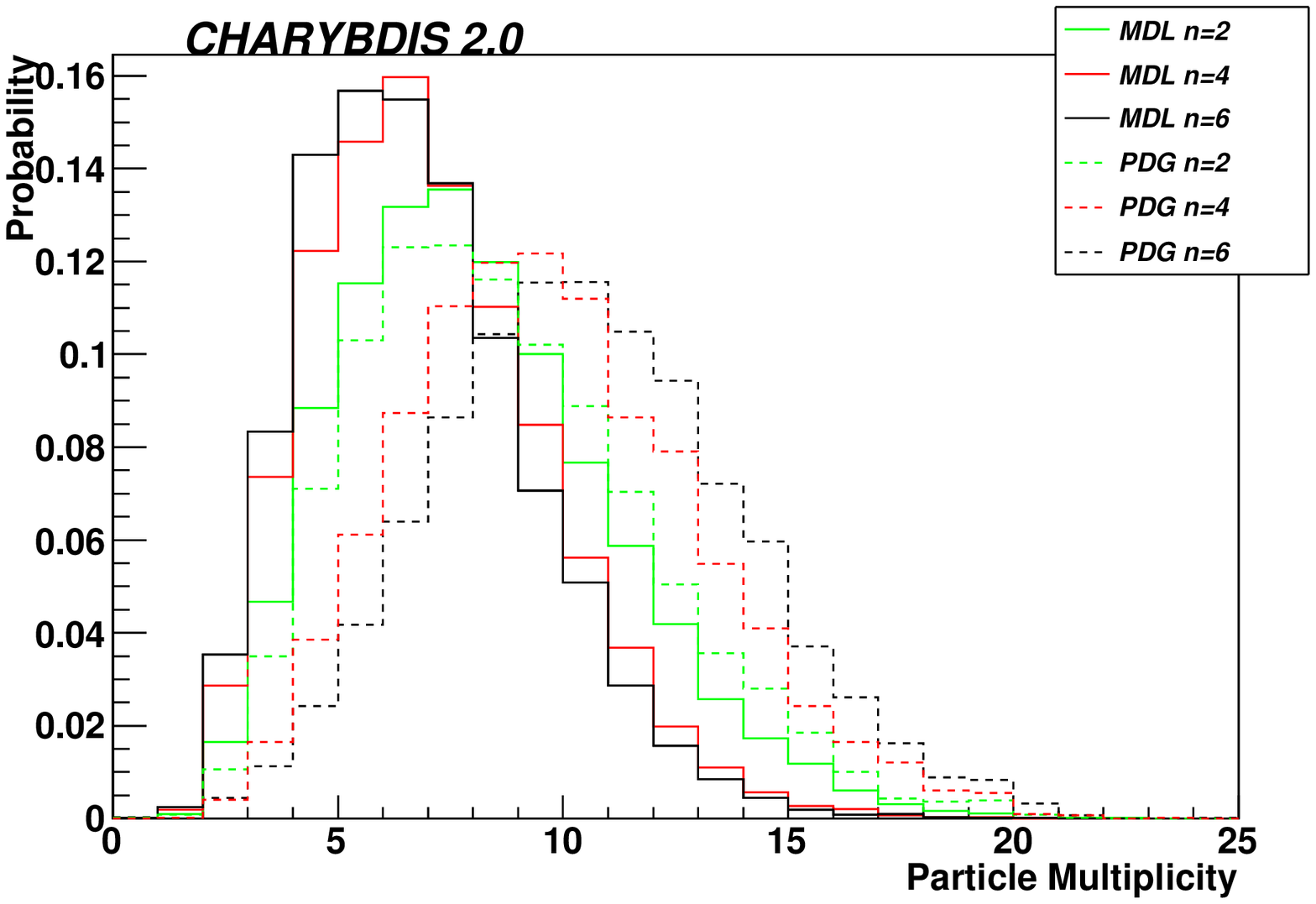} \\
\includegraphics[scale=0.39]{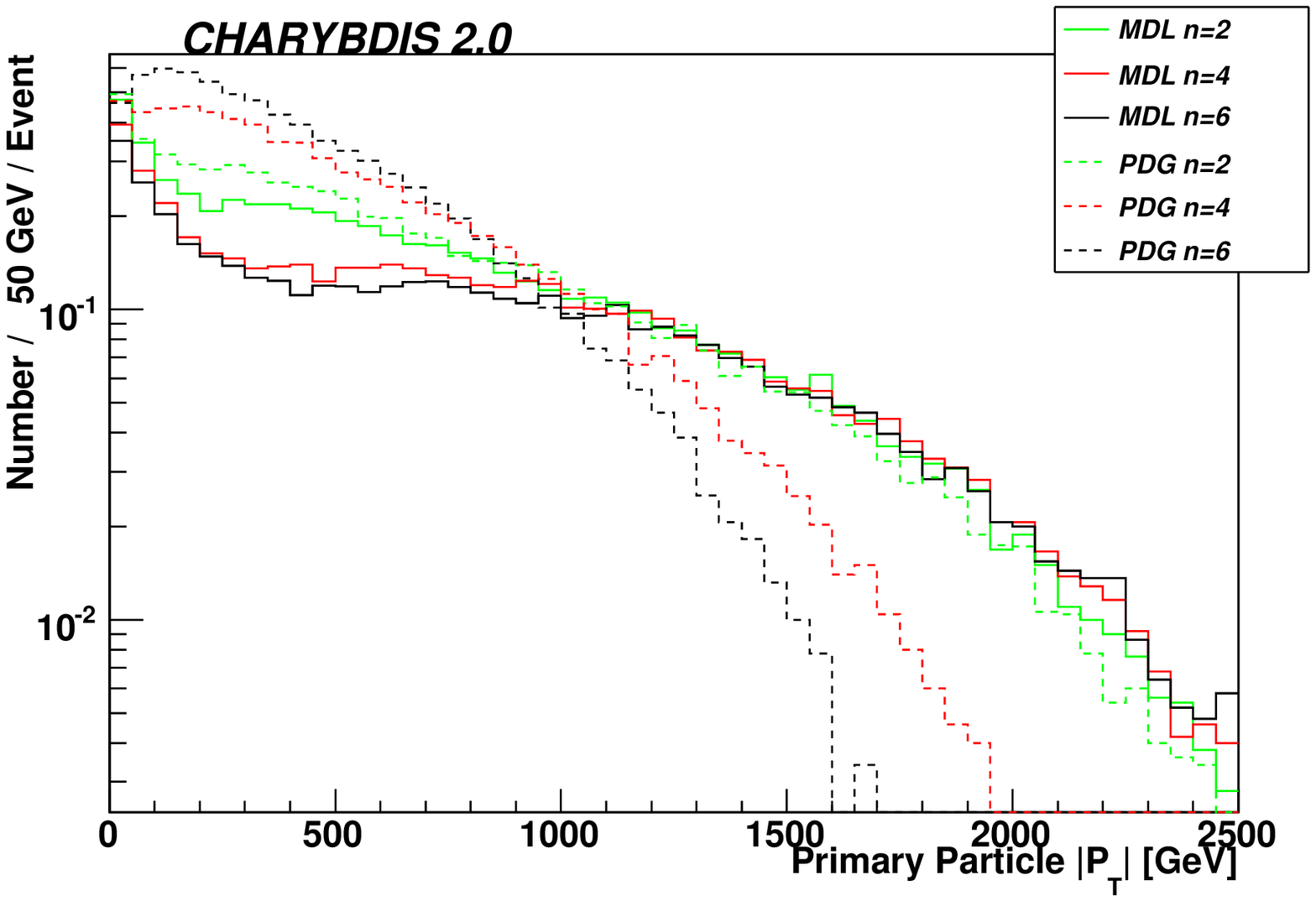} \includegraphics[scale=0.39]{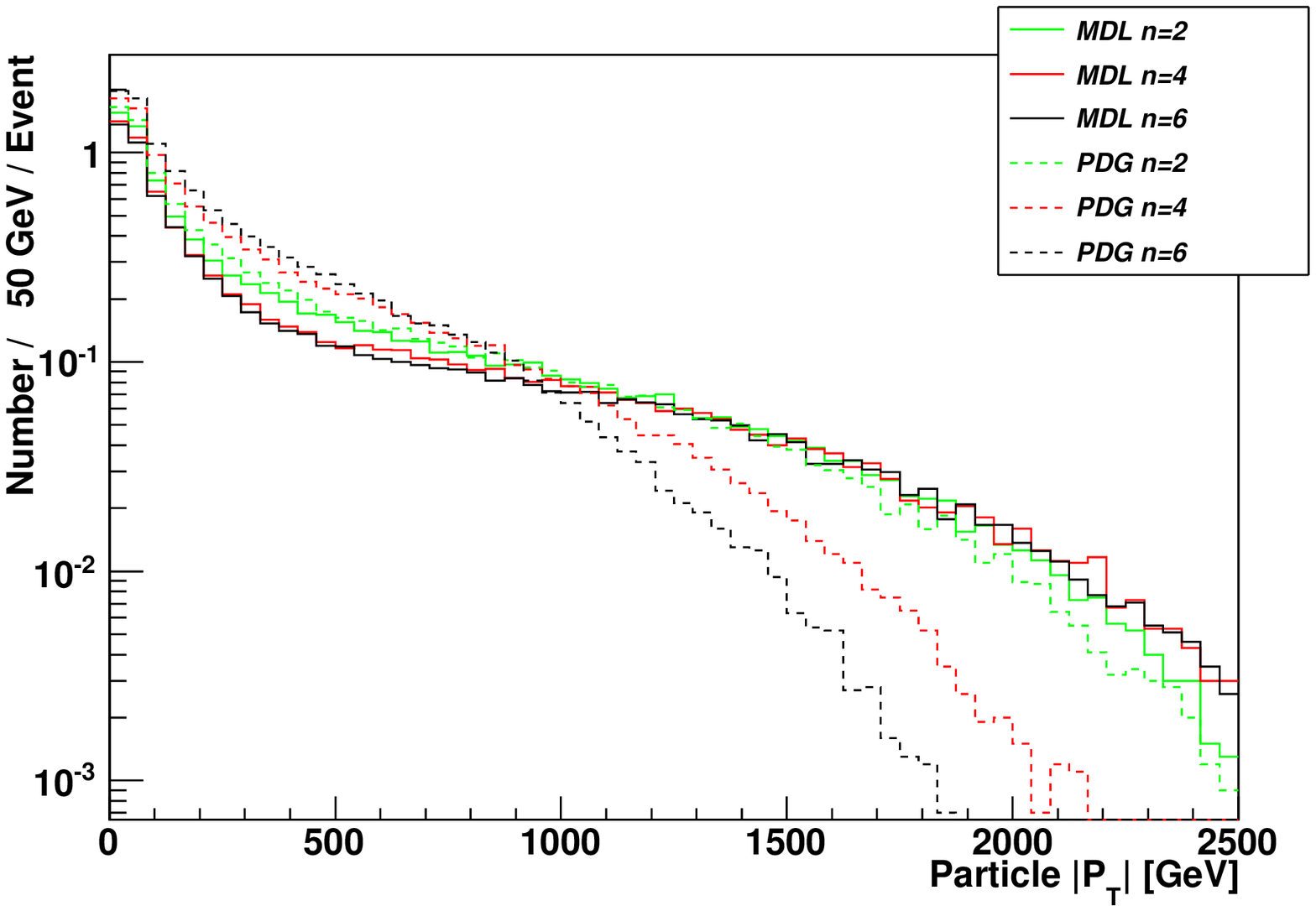}
\caption{Particle Multiplicity distributions and $P_T$ spectra at generator level (left) and after \vb{AcerDET} detector simulation (right) for a 1 TeV Planck mass in PDG and Dimipoulos-Landsberg ``MDL'' conventions.}
\label{fig:planck_basics}
\end{figure}
In principle, it is possible to reconstruct the black hole mass by combining the 4-momenta of all particles observed in the event and missing transverse energy. Mass resolutions of 200-300 GeV can be achieved for some samples as shown in Fig.~\ref{fig:prod_massres}, but there is significant variation with different samples and black hole parameters. Events with large amounts of MET (particularly from multiple sources) tend to be more poorly reconstructed. Invoking a 100~GeV cut on MET results in better reconstruction at the cost of some signal efficiency. Such a cut may not be entirely conservative however, for there may be additional sources of MET neglected in our simulation, such as that from Hawking emission of gravitons.

\subsection{Planck mass conventions}
Although a one-to-one mapping between differing conventions for the Planck scale is  straightforward to make, naively changing the convention without compensating for the Planck mass value leads to large apparent differences in the output distribution, due to their different $n$ dependence. This is particularly the case for high numbers of extra dimensions for which the definitions vary most widely, as shown in Fig.~\ref{fig:planck_basics} where the Dimopoulos-Landsberg and PDG Planck mass definitions are contrasted.

A 1 TeV Planck mass in the Dimopoulos-Landsberg convention equates to a larger value in the PDG convention, especially for large $n$ (the increase ranges from a factor of $1.1$ at $n=1$ to $2.7$ at $n=6$). Consequently they have a higher Hawking temperature and far fewer particles are emitted, but with higher energies and transverse momenta.


\section{Conclusions}\label{sec:conc}
We have presented in detail the physics content of the new black hole
event generator \vb{CHARYBDIS2}, together with some results illustrating
important features of the simulation of the different phases of black
hole production and decay.  The main new features compared to most earlier
generators, including \vb{CHARYBDIS}, are: detailed modelling of the
cross section and the loss of energy and angular momentum during formation
of the black hole (the so-called balding phase), based on the best available
theoretical information; full treatment of angular momentum during the evaporation
phase, including spin of the incoming partons, rotation of the black
hole, and anisotropy and polarisation of all Standard-Model fields
emitted on the brane; and finally a variety of options for the Planck-scale
termination phase, ranging from a stable remnant to a variable-multiplicity
model connecting smoothly with the evaporation phase.

Our main finding is that angular momentum has strong effects on the properties
of the final state particles in black-hole events.  Even after allowing for a substantial
loss of angular momentum in the balding phase, the isotropic evaporation of
a spinless Schwarzschild-Tangherlini black hole is not a good approximation,
nor is the notion of a rapid spin-down phase followed by mainly isotropic
evaporation at foreseeable energies. Although the Hawking temperature
does not depend strongly on the angular momentum of a spinning black hole of
a given mass, there is a strong bias in the emission spectra towards higher-energy
emissions into higher partial waves, which help the black hole to shed its
angular momentum. The resultant spectra are flatter, with substantial tails beyond 1~TeV.
As a consequence of these more energetic emissions, rotating black holes emit with reduced
multiplicity relative to their non-rotating counterparts. However the absolute multiplicity can still be large.

The preferential equatorial emission of scalar, fermionic and high energy vector particles leads to slightly less central distributions at detector level. This effect is reduced by the evolution of the spin axis during evaporation (away from the initial orientation perpendicular to the beam direction).
The emission of polarised higher-spin fields is favoured, compared to the spinless case, leading to increased vector emission and marking a further departure from a purely democratic distribution of particle species. This shows little dependence upon the number of dimensions.

These findings will complicate the interpretation of black-hole events, should
they occur at the LHC or future colliders.  While the basic signature of energetic,
democratic emission of all Standard-Model species and large missing energy
remains valid, the deduction of the fundamental Planck scale and the number
of extra dimensions will be more difficult than was anticipated in earlier
studies~\cite{Harris:2004xt}.  On the other hand, many interesting new and potentially
observable features emerge, such as the different angular distributions
and polarisation of particles of different spins. We intend to
investigate possible analysis strategies in a future publication.

In view of the important effects of angular momentum, we would counsel against
the use of black-hole event generators that neglect these effects.  This leaves
\vb{BlackMax} and \vb{CHARYBDIS2} and the generators of choice for future studies.
Both programs take black-hole angular momentum fully into account, but they have
other features and emphases that are complementary.  In the formation phase,
\vb{BlackMax} uses a geometrical approximation for the cross section and parametrizes
the loss of energy and angular momentum as fixed fractions of their initial-state
values, whereas \vb{CHARYBDIS2} incorporates a more detailed model based on the
Yoshino-Rychkov bounds and comparisons with other approaches.  The treatment of
the evaporation phase in the two programs appears broadly similar, but \vb{BlackMax}
has options for brane tension and split branes, and for extra suppression of emissions
that would spin-up the black hole, while \vb{CHARYBDIS2} includes treatment of the
polarisation of emitted fermions and vector bosons.  The conservation of quantum
numbers is also treated somewhat differently.  At the Planck scale, \vb{BlackMax}
emits a final burst of particles with the minimal multiplicity needed to conserve
quantum numbers, whereas \vb{CHARYBDIS2} has a wider range of options.

A deficiency of both programs is the absence of gravitational radiation in the
evaporation phase.  This is because the greybody factors have not yet been computed
for this case,  due to extra theoretical difficulties in the separation of variables.  Unlike
Standard-Model particles, gravitons will necessarily be emitted into the bulk, giving
rise to a new source of lost energy and the possibility of recoil off the brane. In 
the non-rotating case, it is known~\cite{Cardoso:2005mh,Creek:2006ia} that bulk graviton
emission is small for low numbers of extra dimensions, but increases rapidly in higher
dimensions due to the growing number of polarisation states.  However, the large number
of Standard-Model degrees of freedom ensures that brane emission remains dominant.
Clearly a full treatment of the rotating case is desirable, but there is hope that
the effects will not be too significant, taking into account the uncertainties in
energy loss already allowed for in the formation phase. Furthermore, as already
discussed in Sect.~\ref{sec:spin}, we expect the residual SM charges of the black
hole to prevent recoil off the brane in all but a small fraction of events.

In summary, the simulation of black hole production and decay at hadron colliders has
seen rapid advances in recent years, but there remain substantial challenges in both
theoretical understanding and data analysis, should such events be seen. It is
hoped that \vb{CHARYBDIS2} will serve as a convenient basis for continuing theoretical
refinement and as a useful tool in the design of analysis strategies for this
exotic and complex possibility.


\section*{Acknowledgements}

We thank colleagues in the Cambridge SUSY Working Group for helpful
discussions. SRD and MC gratefully acknowledge financial support from the Irish Research Council for Science, Engineering and Technology (IRCSET). SRD and MC would like to thank Elizabeth Winstanley and Panagiota Kanti for helpful advice and guidance. MC was partially funded by Funda\c c\~ao para a Ci\^encia e Tecnologia (FCT) - Portugal through PTDC/FIS/64175/2006.  MS was supported by the FCT grant SFRH/BD/23052/2005.


\section*{Appendices}
\appendix
\section{Conventions}\label{app:conventions}
Throughout this article, unless stated otherwise, all theoretical expressions are in natural units where the Planck constant and speed of light are respectively $\hbar=c=1$. We keep the dependence on the ($4+n$)-dimensional Planck mass explicit in all expressions and adopt as reference convention, the PDG definition\footnote{See for example the extra dimensions section of the PDG review~\cite{Amsler:2008zzb}} which uses the Einstein-Hilbert action
\begin{equation}\label{EH_convention}
S_{EH}=\dfrac{1}{2}\hat{M}_{D}^{2+n}\int d^{(4+n)}x\sqrt{-g}\mathcal{R}_{(4+n)}=\dfrac{1}{16\pi G_D}\int d^{(4+n)}x\sqrt{-g}\mathcal{R}_{(4+n)} \ \ ,
\end{equation}
to set the reduced Planck mass $\hat{M}_{D}$. The Planck mass $M_D$ is then defined as 
\begin{equation}
M_{D}^{2+n}=(2\pi)^{n}\hat{M}_{D}^{2+n} \ .
\end{equation}
An alternative convention is obtained by defining
\begin{equation}
M_{4+n}^{2+n}=2M_{D}^{2+n} \ .
\end{equation}
This is the Giddings-Thomas convention~\cite{Giddings:2001bu} used internally in \vb{CHARYBDIS2}.

\section{The Yoshino-Rychkov mass/angular momentum bounds}\label{app:TS_method}

This appendix is divided into two sections. The first is a brief review of the method in the Yoshino-Rychkov paper~\cite{Yoshino:2005hi}, focusing on the key equations we used to implement the Yoshino-Rychkov boundary curves in \vb{CHARYBDIS2}. The second section explains how \vb{CHARYBDIS2} calculates the boundary curve $\xi_b(\zeta)$ for a given $D$ and $b$.

\subsection*{Summary of the Yoshino-Rychkov method}

It should be noted that in the Yoshino-Rychkov calculation, the usual assumptions are made -- namely, that the extra dimensions may be regarded as infinite in extent, and that the only effect of the brane's gravitational field is to restrict the black hole produced to lie on the brane (the `probe brane' approximation). 

The first stage of the calculation is the choice of metric for each of the colliding partons. Yoshino and Rychkov use the Aichelburg-Sexl metric~\cite{Aichelburg:1970ef}. This is the metric produced by boosting a Schwarzschild-Tangherlini metric to the speed of light, whilst reducing its associated mass to zero in such a way that its energy remains finite. It is appropriate to an ultrarelativistic point particle with no spin or charge, and so the Yoshino-Rychkov calculation neglects the effects of spin, charge, and finite size of the partons.

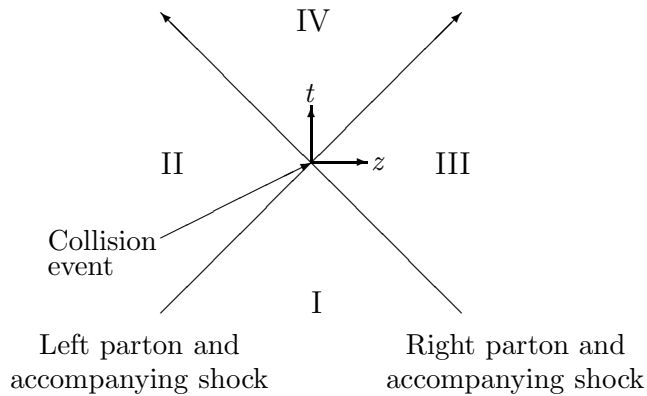
\begin{figure}[h]
\centering
\setlength{\unitlength}{0.5cm}
\begin{picture}(18,11)
\put(4,2){\vector(1,1){8}}
\put(12,2){\vector(-1,1){8}}
\put(8,6){\vector(1,0){1.5}}
\put(8,6){\vector(0,1){1.5}}
\put(8,2){\large{I}}
\put(4,5.7){\large{II}}
\put(11.3,5.7){\large{III}}
\put(7.6,9.5){\large{IV}}
\put(0,0){\shortstack{Left parton and \\ accompanying shock}}
\put(10,0){\shortstack{Right parton and \\ accompanying shock}}
\put(9.6,5.8){\em{z}}
\put(7.8,7.6){\em{t}}
\put(1,3){\shortstack[l]{Collision \\ event}}
\put(4,4){\vector(2,1){4}}
\end{picture}
\caption{\label{fig:spacetime} Spacetime regions in a parton-parton collision. In this diagram, the z axis is defined to lie along the direction of motion of the left parton, and $D-2$ spacelike dimensions are suppressed.}
\end{figure}

In the spacetime outside the future lightcone of the collision event (i.e. regions I, II and III of Fig.~\ref{fig:spacetime}), the metric for the complete system is obtained by combining two Aichelburg-Sexl metrics corresponding to partons travelling in opposite directions (in the centre of mass frame). This gives the correct spacetime outside region IV of Fig.~\ref{fig:spacetime}, because the colliding partons are taken as travelling at the speed of light, so there can be no interaction between their gravity waves before the collision. 

The next stage in the calculation is the selection of a spacetime slice somewhere in the union of regions I, II and III, on which one will look for an apparent horizon (AH). An AH is a surface whose outgoing null geodesic congruence has zero expansion. Assuming the cosmic censorship hypothesis~\cite{Penrose:1969qg}, an event horizon (EH) must be present outside any AH; thus finding an AH is sufficient to show that a black hole forms. Furthermore, the fact that the EH must lie outside the AH, combined with the area theorem~\cite{Hawking:1971qh} which states that the EH area never decreases, can typically be used to set bounds on the mass and angular momentum of the formed black hole.

The slice used by Yoshino and Rychkov is the futuremost slice outside region IV - i.e. the boundary of this region. This slice was chosen because it gives the most restrictive, and therefore best, bounds on the maximum impact parameter for black hole formation $b_{max}$, and thus on the cross section $\sigma = \pi b_{max}^2$. It also gives the best bounds on the mass and angular momentum trapped in the black hole following production. The reasoning behind these statements is detailed more fully below.

To obtain a bound $b_{YRmax}$ on the maximum impact parameter for black hole formation for a given $D$, the impact parameter is increased at the given $D$ until one no longer finds an AH - the critical value of $b$ at which an AH no longer appears is then $b_{YRmax}$. This will be a lower bound - there may be impact parameters greater than $b_{YRmax}$ for which an AH forms, but only `after' the slice considered.

From the above, the best lower bound on $b_{max}$ that can be achieved (given that we are not able to look for AHs in region IV) is obtained by using the futuremost slice outside region IV - i.e. the Yoshino-Rychkov slice. The bounds that Yoshino and Rychkov have obtained for $b_{max}$, and the corresponding results for $\sigma$, are given in Table~II of~\cite{Yoshino:2005hi}.

We now discuss the calculation of the mass and angular momentum bound for a given $b$ and $D$ in the Yoshino-Rychkov method. In a trapped surface method, this is achieved by calculating the $D-2$ dimensional area corresponding to the AH, $A_{AH}$. Since the AH has the true black hole EH outside it, and the black hole EH area can never decrease according to the area theorem, it is normally the case that the $D-2$ dimensional area of the final produced black hole EH, $A_{EH}$, should be greater than $A_{AH}$:
\begin{equation} \label{TSareabound}
A_{AH}\leq A_{EH}\;.
\end{equation}
Now, we expect the horizon area of a black hole to be linked to the mass contained within, so we can convert the above formula into one in terms of mass. We define the AH mass as the mass of a Schwarzschild-Tangherlini black hole with area $A_{AH}$:
\begin{equation} \label{AHmass}
M_{AH}=\dfrac{(D-2)\Omega_{D-2}}{16\pi G_D}\left(\dfrac{A_{AH}}{\Omega_{D-2}}\right)^{(D-3)/(D-2)}\;.
\end{equation}
In \eqref{AHmass}, $G_D$ is the $D$-dimensional gravitational constant as in Eq.~\eqref{EH_convention}, and $\Omega_{D-2}$ is the $(D-2)$-area of a unit sphere, given by
\begin{equation}
\Omega_{p}=\dfrac{2\pi^{\frac{p+1}{2}}}{\Gamma\left[(p+1)/2\right]}\;.
\end{equation}
With the definition of the AH mass given in \eqref{AHmass}, the equivalent to \eqref{TSareabound} must be
\begin{equation} \label{TSmassbound}
M_{AH}\leq M_{irr} \ \ ,
\end{equation}
where $M_{irr}$ is the irreducible mass of the produced Myers-Perry black hole -- this is the mass of a Schwarzschild-Tangherlini black hole having the same horizon area as the Myers-Perry black hole. Since it is defined in terms of the area of the Myers-Perry black hole it is a function of both $M$ and $J$. 

Equation \eqref{TSmassbound} then represents the trapped surface bound on the mass and angular momentum trapped in the black hole during production, with the equation of the boundary itself being characterised by
\begin{equation} \label{TSmassboundary}
M_{AH} = M_{irr}\;.
\end{equation}
To convert \eqref{TSmassboundary} into a boundary line in the $(M,J)$ plane (which can be scaled to a boundary line in the $(\xi, \zeta)$ plane, using notation from Sect.~\ref{sec:prodmeth}) we first need an equation for the irreducible mass of a Myers-Perry black hole with mass $M$ and angular momentum $J$. This may be extracted from the definition of $M_{irr}$:
\begin{equation}\label{Mirr_Area}
A_{Myers-Perry}(M,J)=A_{Schwarzschild}(M_{irr})=\Omega_{D-2} r_S^{D-2}(M_{irr})\;.
\end{equation}
Computing the left hand side of \eqref{Mirr_Area} using the Myers-Perry metric~\cite{Myers:1986qa}, we find the link between $M$, $J$ and $M_{irr}$ of a Myers-Perry black hole:
\begin{equation} \label{Mirrdef}
r_S^{D-2}(M_{irr})=r_S^{D-3}(M)r_H(M,J)\;.
\end{equation}
In the above, $r_H(M,J)$ is the Myers-Perry horizon radius, given by
\begin{equation} \label{Kerrradiusdef}
r_H^2(M,J)+\left[ \dfrac{(D-2)J}{2M}\right]^2 = r_H^{D-3}(M)r_H^{5-D}(M,J)\;,
\end{equation}
whilst $r_S(M)$ is the horizon radius of a Schwarzschild-Tangherlini black hole of mass $M$, given by
\begin{equation} \label{Schwarzradiusdef}
r_S(M)=\left[\dfrac{16\pi G_D M}{(D-2)\Omega_{D-2}}\right]^{1/(D-3)}\;.
\end{equation}
We now combine equations \eqref{TSmassboundary} and \eqref{Mirrdef} to show that a point on the bound with mass $M$ has horizon radius $r_{Hb}(M)$ given by
\begin{equation} \label{Rkonbound}
r_{Hb}(M) = \dfrac{r_S^{D-2}(M_{AH})}{r_S^{D-3}(M)}\;.
\end{equation}
We insert this result into equation \eqref{Kerrradiusdef}, and then make repeated use of equation \eqref{Schwarzradiusdef} to replace all $r_S$ symbols in the result by the explicit expression for this quantity. After some rearrangement, we discover that a point on the bound with mass $M$ has angular momentum $J_b$ where
\begin{equation} \label{expbound1}
J_b(M) = \dfrac{2M_{AH}}{(D-2)}\left[\dfrac{16\pi G_D M_{AH}}{(D-2)\Omega_{D-2}}\right]^{1/(D-3)}\sqrt{(M/M_{AH})^{D-2}-1}\;.
\end{equation}
This is an explicit equation for the trapped surface boundary line in the $(M,J)$ plane. One can rearrange \eqref{expbound1} to make $M$ the subject, to obtain an alternative form for the equation of the boundary line:
\begin{equation} \label{expbound2}
M_b(J) = M_{AH}\left\{ 1+\left[\dfrac{(D-2)J}{2M_{AH}}\right]^2
\left[\dfrac{(D-2)\Omega_{D-2}}{16\pi G_D M_{AH}}\right]^{2/(D-3)} \right\}^{1/(D-2)}\;.
\end{equation}
To produce valid bounds from the approach of finding an $(M,J)$ bound outlined above, it is necessary that an arbitrary surface outside the AH in the slice used has a larger area. Unfortunately, for the particular case of the Yoshino-Rychkov slice, this does not hold. However, Yoshino and Rychkov found a different area, $A_{lb}$, which they demonstrated would be a true lower bound on $A_{EH}$ ($A_{lb}$ is equal to twice the area of the intersection of the AH with the transverse collision plane). The calculation of the $(M,J)$ bounds then proceeds as described above, with $M_{lb}$ and $A_{lb}$ replacing $M_{AH}$ and $A_{AH}$ in equations \eqref{TSareabound} - \eqref{expbound2}. 

The Yoshino-Rychkov slice should give the best bounds in this case as well. This is because the use of the futuremost slice outside region IV should give the largest AH areas obtainable outside this region~\cite{Yoshino:2005hi}. As a result, one expects to get the largest, and therefore strictest, lower bounds on $M_{irr}$ using this slice.

\subsection*{Calculation of the mass/angular momentum boundary}

If we rewrite equation \eqref{expbound2} in terms of the fractions of initial state mass and angular momentum $\xi$ and $\zeta$, replacing all $M_{AH}$ symbols in the equation by $M_{lb}$ (in accordance with the fact that the Yoshino-Rychkov calculation uses the latter quantity), then we obtain
\begin{equation} \label{expboundred}
\xi_b(\zeta) = \xi_{lb}\left\{ 1+\left[\dfrac{(D-2) b}{4\xi_{lb}}\right]^2
\left[\dfrac{(D-2)\Omega_{D-2}}{32\pi G_D \mu \xi_{lb}}\right]^{2/(D-3)} \right\}^{1/(D-2)}\;.
\end{equation}
In \eqref{expboundred}, $\mu$ stands for the energy of each colliding parton in the centre of mass frame. This equation can be used to calculate the Yoshino-Rychkov bound $\xi_b(\zeta)$ for a given $b$ and $D$, provided one is able to obtain $\xi_{lb}$ for the $b$ and $D$ values used.

The problem of implementing the Yoshino-Rychkov bound in \vb{CHARYBDIS2} is then one of ensuring that the program has a means of obtaining $\xi_{lb}$ for all values of $D$ and $b$ ($5\leq D \leq 11$, $0 \leq b \leq b_{YRmax}(D)$). Calculating $\xi_{lb}$ exactly (i.e. using the full analytic method) for a given $b$ and $D$, requires repeating the entire Yoshino-Rychkov calculation each time -- i.e.\ find the AH numerically, compute the area of its intersection with the transverse collision plane $A_{lb}$, and then convert this to an $\xi_{lb}$ value using an equation similar to \eqref{AHmass}. Implementing the full calculation in \vb{CHARYBDIS2}, to be repeated on an event by event basis, would slow down the program considerably. Instead, we use the $\xi_{lb}$~vs.~$b$ data files for $D=5$ to $D=11$ that had already been generated by Yoshino and Rychkov, and which were used to produce the $D=5$ to $D=11$ plots in figure~10 of~\cite{Yoshino:2005hi}. \vb{CHARYBDIS2} computes the value of $\xi_{lb}$ for a given $b$ and $D$ by linear interpolation on the points in these data files. Linear interpolation provides sufficient accuracy due to the close spacing (in $b$) of the points in the data files.

\section{Details of the `constant angular velocity' bias}\label{app:prodprobdist}

This section describes the implementation of the constant angular velocity bias in the simulation of the black hole production phase.  With the bias off (\vb{CVBIAS=.FALSE.}), the values of ($\xi$,$\zeta$)\footnote{$\xi,\zeta$ are the trapped mass and angular momentum fractions passed to the routine which imposes the Yoshino-Rychkov boundary condition.} are simply those generated from the linear ramp distributions described in Sect.~\ref{sec:prodmeth}. When the bias is turned on, the $(\xi,\zeta)$ point to be passed to the Yoshino-Rychkov boundary routine is obtained in a more complex fashion, which is outlined below.

First, a point is generated using the linear ramp distributions as before. The horizon angular velocity and $a_*$ value corresponding to the point, $\Omega(\xi,\zeta)$ and $a_*(\xi,\zeta)$, are calculated using the standard equations, and compared to those of the initial state, $\Omega(1,1)$ and $a_*(1,1)$. In particular, the quantities $|\Omega(\xi,\zeta)-\Omega(1,1)|/\Omega(1,1)$ and $|\log[a_*(\xi,\zeta)/a_*(1,1)]|$ are calculated, and compared to the values of the constants $\Delta$ and $\Lambda$ respectively (whose values will be discussed shortly). Note that $|\Omega(\xi,\zeta)-\Omega(1,1)|/\Omega(1,1)$ and $|\log[a_*(\xi,\zeta)/a_*(1,1)]|$ are essentially both measures of the differences between the values at the point and the initial state values.

The point is then assigned a number $\alpha(\xi,\zeta)$ between $0$ and $1$, whose value largely depends on whether both $|\Omega(\xi,\zeta)-\Omega(1,1)|/\Omega(1,1) \leq \Delta$ {\em and\/} $|\log[a_*(\xi,\zeta)/a_*(1,1)]| \leq \Lambda$ or not (i.e. whether $\Omega(\xi,\zeta)$ and $a_*(\xi,\zeta)$ are sufficiently close to $\Omega(1,1)$ and $a_*(1,1)$ or not). If one or both of the conditions are not satisfied, then the point is assigned a constant $k < 1$ (as defined below). If both conditions are satisfied, the point is assigned the value of a function $\chi(\xi,\zeta)$. The function $\chi(\xi,\zeta)$ has the key properties $k < \chi(\xi,\zeta) \leq 1$, and approaches $1$ as $\Omega(\xi,\zeta)$ gets closer to $\Omega(1,1)$. The details of our choices for $k$ and $\chi(\xi,\zeta)$ will be discussed shortly.

A random number $\beta$ is generated according to a uniform distribution between $0$ and $1$. If $\alpha(\xi,\zeta) > \beta$ then the point is accepted, otherwise it is rejected. If the point is rejected, further $(\xi,\zeta)$ points have to be generated by the ramp distributions, and put through the above procedure, until a point is accepted. The final point is passed to the Yoshino-Rychkov boundary routine.

It is reasonably clear that this procedure for generating a $(\xi,\zeta)$ point (to be passed to the Yoshino-Rychkov boundary routine) is equivalent to a procedure which generates a point from a biased probability distribution of the form asserted in Sect.~\ref{sec:prodmeth}. To be specific, the biased probability distribution resembles the basic ramp probability distribution, but all points whose $\Omega$ and $a_*$ values are sufficiently close to those of the initial state have had their probabilities enhanced. The enhancement is greater the closer $\Omega(\xi,\zeta)$ is to $\Omega(1,1)$.

We now discuss our choices for the function and the parameters used in the above procedure. A suitable choice for the function $\chi(\xi,\zeta)$, which has the properties stated, is based on the Breit-Wigner form (note that we introduce a further 'width' parameter $\Gamma$):

\begin{equation}
\chi(\xi,\zeta) = \dfrac{\Gamma^2/4}{([\Omega(\xi,\zeta)-\Omega(1,1)]/\Omega(1,1))^2+\Gamma^2/4}\;.
\end{equation}

The constant $k$ may then be fixed by imposing continuity on $\alpha(\xi,\zeta)$, such that the biased probability distribution represented by the above procedure does not possess any sudden jumps. Note that we hope the dividing curves between the enhanced region and the unenhanced regions to be  $|\Omega(\xi,\zeta)-\Omega(1,1)|/\Omega(1,1) = \Delta$ on either side of the `connected' curve $\Omega(\xi,\zeta) = \Omega(1,1)$ (which is connected to the point $\xi=1,\zeta=1$). As explained in the main text, the $a_*$ condition is only present to remove probability enhancement around the other `disconnected' $\Omega(\xi,\zeta) = \Omega(1,1)$ curve, and should not interfere with the angular velocity based enhancement around the right curve.

On the basis of these assumptions, continuity of $\alpha(\xi,\zeta)$ is assured by taking

\begin{equation}
k = \chi(\xi,\zeta)|_{|\Omega(\xi,\zeta)-\Omega(1,1)|/\Omega(1,1) = \Delta} = \dfrac{\Gamma^2/4}{\Delta^2+\Gamma^2/4}\;.
\end{equation}
Note that with this choice of $\chi(\xi,\zeta)$ and $k$, $\Gamma$ is a constant which sets the `width' of the probability peak around the connected $\Omega(\xi,\zeta) = \Omega(1,1)$ curve, and  the `height' of this peak. The smaller $\Gamma$ is, the sharper and stronger the probability enhancement around the appropriate curve. The default value of $\Gamma$ is 0.4 for a probability enhancement which is not too strong - however, the value of this variable could potentially be changed.

The values of $\Delta$ and $\Lambda$ were chosen by looking at a large number of individual $(b,D)$ cases used by \vb{CHARYBDIS2}, and trying to find a suitable combination of values that gave enhancement of a suitable region only around the connected curve. This procedure resulted in the choice $\Delta = 0.2$ and $\Lambda = 0.4$ (these values should not be changed, as small changes can cause drastic changes in the regions where the probability is enhanced). 

A final comment is appropriate explaining the slightly peculiar form of the $a_*$ condition for probability enhancement -- $|\log[a_*(\xi,\zeta)/a_*(1,1)]| \leq \Lambda$. The reason for this form is because, when trying to find a condition that discriminated points around the connected curve from those around the disconnected curve, we noticed that those around the disconnected curve, had $a_*$ values that were one order of magnitude (or more) away from the $a_*$ values of the points around the connected curve. To remove the enhancement around the disconnected curve, a condition based on the logarithm of the ratio $a_*(\xi,\zeta)/a_*(1,1)$ is then more appropriate.

\section{Evaluation of spheroidal wave functions}\label{app:spher}
In a paper by Leaver~\cite{Leaver:1985ax} the spheroidal wave functions for arbitrary spin are expanded as a series in  $x=\cos\theta$ around $x=-1$. In this paper there is an extra parameter which can take either sign. We can alternatively expand around $x=+1$. So in general, we can construct three more expansions. Even though all four of them converge uniformly in $x\in \left[-1,1\right]$, the numerical errors behave differently if we consider a fixed region. Therefore it is useful to look at all the options and use different expansions for different regions. The only extra complication will be to match them in a common region.

The possible expansions are
\begin{equation}
S_k^{s,p}(c,x)=e^{px}\left(1+x\right)^{\alpha}\left(1-x\right)^{\beta}\sum_{n=0}^{+\infty}b_{n}\left(1+sx\right)^{n}
\end{equation}
where $\alpha,\beta$ are chosen as to reproduce the correct behaviour around the regular singular points $x=\pm1$,
\begin{eqnarray}
&\alpha=\left|\dfrac{m-h}{2}\right|\ ,\ \ &\beta=\left|\dfrac{m+h}{2}\right|\;, 
\end{eqnarray}
and $s=\pm1$ for expansions around $x=-1$ and $x=1$ respectively. Substituting into \eqref{angular_equation} we obtain the recurrence relation (for $n>0$),
\begin{equation}\label{recurrence}
b_{n}=\left(\frac{1}{2}+\dfrac{\epsilon_1}{n+\sigma}-\dfrac{\epsilon_2}{n(n+\sigma)}\right)b_{n-1}+\left(1+\dfrac{\gamma}{n}\right)\dfrac{sp}{n+\sigma}b_{n-2}
\end{equation}
and the simplifying condition $p^2=c^2\Rightarrow p=\pm c$. We can set $b_{-1}=0$ and choose the normalisation $b_0=1$. The parameters above are given by the following expressions:
\begin{eqnarray}
\sigma&=&\alpha(s+1)+\beta(1-s) \nonumber\\
\epsilon_1&=&\dfrac{\alpha(1-s)+\beta(1+s)-1-4sp}{2} \nonumber\\
\epsilon_2&=&\dfrac{\mathcal{A}_k+h(h+1)+c^2-(\alpha+\beta)(\alpha+\beta+1)}{2}+\alpha+\beta \nonumber \\
&&+p\left[\alpha(s+1)+\beta(s-1)-s+sh\,\textrm{sign}(p)\right]\nonumber \\
\gamma&=&\alpha+\beta-1+h\,\textrm{sign}(p)\;.
\end{eqnarray}
For almost all values of $\mathcal{A}_k$, the ratio $b_{n+1}/b_{n}$ in \eqref{recurrence} goes to $1/2$ as $n\rightarrow \infty$. So as the order increases, the remainder will behave as (when $N\rightarrow\infty$)
\begin{equation}\label{error}
\sum_{n=N}^{+\infty}b_{n}\left(1+sx\right)^{n}\rightarrow b_{N}\left(1+sx\right)^{N}\frac{2}{1-sx} \ \ ,
\end{equation}
which diverges at $x=s$ and goes exponentially faster to zero as we move closer to $x=-s$. So if we do not tune to particular values of $\mathcal{A}_k$ this is how the numerical errors will behave. However, from the analytical point of view we still want to have an expansion which converges uniformly, which is not the case in general as suggested by \eqref{error}. Therefore we need to know how the ratio $b_{n+1}/b_n$ behaves for large $n$. Following~\cite{Baber:1935} we determine this by assuming that for large $n$
\begin{equation}
\dfrac{b_{n+1}}{b_n}\sim\dfrac{b_n}{b_{n-1}}\sim k(n)+\ldots
\end{equation} 
where $\ldots$ denotes subleading contributions. Inserting this in \eqref{recurrence} and solving for $k$ we get two possible behaviours
\begin{equation}
k(n)\sim \left\{\begin{array}{l}
\frac{1}{2}+\frac{2sp}{n} \vspace{2mm}\\
-\frac{2sp}{n}
\end{array}\right.\rightarrow \left\{\begin{array}{l}
\frac{1}{2}\vspace{2mm}\\
0
\end{array}\right. .
\end{equation}
It is straightforward to check that uniform convergence for $x\in[-1,1]$ is only achieved in the second case. A solution with these convergence properties is called a minimal sequence solution~\cite{Leaver:1985ax}. Furthermore, there is a theorem (see Theorem 1.1 of~\cite{Gautschi:1967} and the reasoning in~\cite{Baber:1935}) which ensures that the sequence obtained from the eigenvalue problem in \eqref{recurrence}, with the initial conditions given above, is a minimal sequence. Therefore, after determining the eigenvalue we can compute the solution (up to a normalisation) by fixing $b_0=1$. 

However, from the numerical point of view, a small error in the eigenvalues implies that $b_{n+1}/b_n$ will fail to go to zero when numerically evaluated through \eqref{recurrence}. So the remainder will actually behave as \eqref{error}, since in practice we are approximating the eigenfunction by a nearby function, for which the sequence of expansion coefficients behaves as a dominant solution of \eqref{recurrence} when $n$ is large\footnote{See for example~\cite{Gautschi:1967} for a discussion of numerical issues when generating minimal sequences which are solutions of three-term recursion relations.}.
Thus the numerical radius of convergence will be a bit smaller and the expansion will fail at $x=s$.

We avoid the above convergence problem by using two expansions, one around each singular point $x=\pm1$, and then matching the normalisation in a region where both converge appropriately. A simple procedure for matching follows from the observation that
\begin{eqnarray}
S_+&=&AS_- \nonumber\\
\Rightarrow \log{|A|}&=&\log{|S_+|}-\log{|S_-|} \ \ ,
\end{eqnarray}
where $A$ is a constant and $S_\pm$ denotes expansion around $x=\pm 1$ respectively. We can find $|A|$ by averaging the quantity on the right-hand side over an overlap region. Furthermore we can estimate the relative matching error through the quadratic deviation,
\begin{eqnarray}
\dfrac{\Delta|A|}{|A|}&=&\Delta\log{|A|}\nonumber\\
&\sim&\sqrt{\left<\log^2|A|\right>-\left<\log|A|\right>^2}
\end{eqnarray}
where $\langle\ldots\rangle$ denotes average over the points used.
In practice we use points in the overlap region $x\in[-0.25,0.25]$.

For the expansions above to be useful we need to estimate an order of truncation, $n_{\textrm{trunc}}$. A simple condition is that the ratio of consecutive terms $b_{n+1}/b_{n}$ is approximately constant above the order of truncation. This is equivalent (in the region $sx<0$ where there is strong convergence) to an exponentially suppressed upper bound on the remainder when $n_{\textrm{trunc}}$ is large (see \eqref{error}). From this and \eqref{recurrence}
\begin{equation}
n_\textrm{trunc}\sim\max\{[j+|h|+4c],[\sqrt{|\mathcal{A}_k|+|h|(|h|+1)+(c+|m|+2|h|+3)^2}]\} \ .
\end{equation}
An initial estimate of the truncation error is obtained from \eqref{error}.
Another criterion for truncation is that the first neglected higher order term is small compared to the truncated sum. 

In practice, after the spheroidal function is calculated with a certain truncation, then ten more terms in the series are included and the two estimates are compared. Simultaneously, the first neglected term is compared with the estimate for the sum. If these errors are still large then ten more terms are calculated. This procedure is repeated until the desired accuracy is obtained.

To compute the eigenvalues $\mathcal{A}_k$ efficiently, the
recurrence relation can be put in a symmetric tridiagonal form by performing a change of basis,
\begin{equation}
a_{n}=x_{n}b_{n}
\end{equation}
such that the coefficient of the $b_{n-2}$ term for a certain order $n$ is the same as the coefficient of the $b_{n-1}$ term at order $n-1$. This is possible if
\begin{equation}
x_{n}=\sqrt{\dfrac{n(n+\sigma)}{-sp(n+1+\gamma)}}x_{n-1}=\sqrt{\dfrac{n(n+\sigma)}{c(n+1+\gamma)}}x_{n-1} \ \ ,
\end{equation}
were we have taken (without loss of generality) $x_0=1$ and the convenient choice $p=-sc$. Furthermore we checked that $x_n\neq0$ $\forall n\geq1$, so the transformation is well defined. Thus the recurrence relation takes the form
\begin{multline}
\sqrt{cn(n+\sigma)(n+1+\gamma)}a_{n}-n\left(\dfrac{n+\sigma}{2}+\epsilon_1\right)a_{n-1} \\
+\sqrt{c(n-1)(n-1+\sigma)(n+\gamma)}a_{n-2}=-\epsilon_2a_{n-1}.
\end{multline}
This is a tridiagonal symmetric eigenvalue problem which can be solved numerically very efficiently, by starting with a truncation order (as estimated above) and checking for precision by repeating the calculation with some more corrections.

\section{Decomposition of spheroidal waves into plane waves}\label{app:sphroidal_plane}

It is well known (see for example~\cite{Peskin:1995ev}) that in the massless limit, the Dirac fields decomposes into two independent fields with helicities $h=\pm 1$. If we denote each of those 2-component spinors by $\Psi_{h}$, then their equations of motion in flat space become
\begin{equation}
\left(\partial_t+h\mathbf{\sigma}\cdot\mathbf{\partial}\right)\Psi_h=0 \ ,
\end{equation}
where $\mathbf{\sigma}=(\sigma^1,\sigma^2,\sigma^3)$ are the Pauli matrices. Then whichever spatial coordinates $\mathbf{x}$ we decide to use, the field operator will have the expansion (in terms of positive and negative energy solutions)
\begin{equation}
\hat{\Psi}_h=\sum_\lambda\dfrac{1}{\sqrt{2\omega_\lambda}}\left[\hat{a}_{h,\lambda}\psi_{h,\lambda}(\mathbf{x})e^{-i\omega_\lambda t}+\hat{b}^\dagger_{h,\lambda}\psi_{-h,\lambda}(\mathbf{x})e^{i\omega_\lambda t}\right] \ ,
\end{equation}
where $\hat{a}_{h,\lambda},\hat{b}_{h,\lambda}^{\dagger}$ are respectively, the usual fermionic annihilation and creation operators for particles and anti-particles; $\lambda$ is an unspecified complete set of quantum numbers; $\psi_{h,\lambda}$ are spinorial normal modes $\left(\omega+ih\mathbf{\sigma}\cdot\mathbf{\partial}\right)\psi_{h,\Lambda}=0$, normalised according to
\begin{equation}\label{norm_spinors}
\int \mathrm{d}^3\mathbf{x}\, \psi_{h,\lambda}^{\dagger}\psi_{h',\lambda'}=\delta_{h,h'}\delta(\lambda,\lambda') \ ;
\end{equation}
$\delta$ is a Dirac delta function such that
\begin{equation}
\sum_{\lambda'}f(\lambda')\delta(\lambda,\lambda')=f(\lambda) \ ;
\end{equation}
and the sum sign can represent integration for continuous quantum numbers.
Similarly to Klein-Gordon theory, it is possible to define a scalar product between 2-spinors $\psi,\chi$ using a bilateral derivative:
\begin{equation}
\left(\psi,\chi\right)= i\int d^3x\left(\psi^{\dagger}\partial_t\chi-\partial_t\psi^{\dagger}\chi\right) \ \ .
\end{equation}
Then we find an expression for the operators
\begin{eqnarray}
\hat{a}_{h,\lambda}&=&\left(\psi_{h,\lambda}e^{-i\omega_\lambda t},\hat{\Psi}_h\right) \\
\hat{b}_{h,\lambda}^\dagger&=&\left(\psi_{-h,\lambda}e^{i\omega_\lambda t},\hat{\Psi}_h\right) \ .
\end{eqnarray} 
$\Psi_h$ can also be expanded in a basis of spinors with a different set of quantum numbers $\gamma$, which may be associated with a different choice of coordinates. The previous expressions will then give a Bogoliubov transformation between operators in one basis and the other: 
\begin{eqnarray}
\hat{a}_{h,\lambda}&=&\sum_\gamma\dfrac{1}{\sqrt{2\omega_\gamma}}\left[\left(\psi_{h,\lambda}e^{-i\omega_\lambda t},\psi_{h,\gamma}e^{-i\omega_\gamma t}\right)\hat{a}_{h,\gamma}+\left(\psi_{h,\lambda}e^{-i\omega_\lambda t},\psi_{-h,\gamma}e^{i\omega_\gamma t}\right)\hat{b}^\dagger_{h,\gamma}\right] \\
\hat{b}_{h,\lambda}^\dagger&=&\sum_\gamma\dfrac{1}{\sqrt{2\omega_\gamma}}\left[\left(\psi_{-h,\lambda}e^{i\omega_\lambda t},\psi_{h,\gamma}e^{-i\omega_\gamma t}\right)\hat{a}_{h,\gamma}+\left(\psi_{-h,\lambda}e^{i\omega_\lambda t},\psi_{-h,\gamma}e^{i\omega_\gamma t}\right)\hat{b}^\dagger_{h,\gamma}\right] \ .
\end{eqnarray} 
The second term of the first expansion, and the first term in the second one will be zero. This is because they are responsible for particle creation, which does not occur in this change of basis, given that we are in the same Lorentz frame in Minkowski space-time. This can be seen explicitly for our case of a transformation between plane wave states and spheroidal states. First note that in the Kinnersley basis the plane wave spinors take the form~\cite{Dolan:2006thesis} 
\begin{equation}
\chi_{\mathbf{p}}^{+}=\left(\begin{array}{c}
e^{i\frac{\tilde{\phi}}{2}}\cos\frac{\theta_{\mathbf{p}}}{2}\cos\frac{\theta}{2}-e^{-i\frac{\tilde{\phi}}{2}}\sin\frac{\theta_{\mathbf{p}}}{2}\sin\frac{\theta}{2} \ \ \\
e^{i\frac{\tilde{\phi}}{2}}\cos\frac{\theta_{\mathbf{p}}}{2}\sin\frac{\theta}{2}+e^{-i\frac{\tilde{\phi}}{2}}\sin\frac{\theta_{\mathbf{p}}}{2}\cos\frac{\theta}{2} \ \ \end{array}\right) \ \ ,\end{equation}
 \begin{equation}
\chi_{\mathbf{p}}^{-}=\left(\begin{array}{c}
e^{i\frac{\tilde{\phi}}{2}}\cos\frac{\theta_{\mathbf{p}}}{2}\sin\frac{\theta}{2}+e^{-i\frac{\tilde{\phi}}{2}}\sin\frac{\theta_{\mathbf{p}}}{2}\cos\frac{\theta}{2}\\
e^{i\frac{\tilde{\phi}}{2}}\cos\frac{\theta_{\mathbf{p}}}{2}\cos\frac{\theta}{2}-e^{-i\frac{\tilde{\phi}}{2}}\sin\frac{\theta_{\mathbf{p}}}{2}\sin\frac{\theta}{2}\end{array}\right) \ \ ,\end{equation}
where $\Omega_{\mathbf{p}}=\left(\theta_{\mathbf{p}},\phi_{\mathbf{p}}\right)$ defines the orientation of the momentum vector $\mathbf{p}$, $\tilde{\phi}=\phi-\phi_{\mathbf{p}}$ and we have omitted the $e^{i\mathbf{p}\cdot \mathbf{x}}$ dependence. As long as we are looking at a fixed plane wave, $\phi$ can be shifted by choosing a new origin, which amounts to setting $\phi_{\mathbf{p}}=0$. The upper/lower component of $\chi^{\pm}_{\mathbf{p}}$ has the interesting property of being invariant under the exchange $\theta_{\mathbf{p}}\leftrightarrow \theta$. Using the asymptotic form \eqref{spheroidal_coordinates} for the Cartesian coordinates in terms of spheroidal coordinates we obtain
\begin{equation}
\mathbf{p}\cdot\mathbf{x}=\cos\theta_{\mathbf{p}}\cos\theta\sqrt{(ar)^2+c^2}+ar\sin\theta_{\mathbf{p}}\sin\theta\cos\phi \ ,
\end{equation} which is again symmetric under the exchange of $\theta$'s. On the other hand, the spheroidal spinors in the Kinnersley basis have the form~\cite{Casals:2006xp}
\begin{equation}
\chi_{\lambda}^{\pm}=e^{im\phi}\left(\begin{array}{c}
_{-}R_{j,m}^{\pm}(r)S_{-\frac{1}{2},j,m}(c,\cos\theta)\\
_{+}R_{j,m}^{\pm}(r)S_{\frac{1}{2},j,m}(c,\cos\theta)\end{array}\right) \ .
\end{equation} 
We are seeking for a relation between plane waves and spheroidal waves. This can be achieved by writing down the general decomposition
\begin{equation}
\chi_{\mathbf{p}}^{\pm}e^{i\mathbf{p}.\mathbf{x}}=\sum_{j,m} c_\lambda(\mathbf{p})\chi_{\lambda}^{\pm}(\mathbf{x}) \ ,
\end{equation}
where now $\lambda = \{\omega,j,m\}$. But we know that the upper/lower component of the left hand side spinor with helicity $\pm$ is invariant under exchange of $\theta$'s, so the $c_\lambda$ prefactor must be proportional to the upper/lower spheroidal function with argument $\theta_{\mathbf{p}}$. Furthermore we put back the $\phi_{\mathbf{p}}$ dependence by shifting $\phi$ to obtain
\begin{equation}
\chi_{\mathbf{p}}^{h}e^{i\mathbf{p}.\mathbf{x}}=\sum_{j,m}\tilde{c}_\Lambda\cdot{}S_{-h,j,m}^\star(c,\Omega_{\mathbf{p}})\chi_{\Lambda}^{h}(\mathbf{x}) \ ,
\end{equation} 
where the $\phi_{\mathbf{p}}$ dependence in $S_{-h,j,m}^\star(c,\Omega_{\mathbf{p}})$ is implicit. An integral relation is obtained when multiplying by an appropriate spinor, integrating over $\mathbf{x}$ and using the normalisation condition \eqref{norm_spinors}
\begin{equation}
\int d^3\mathbf{x} \chi_\Lambda^{h'}(\mathbf{x})\chi_{\mathbf{p}}^{h}e^{i\mathbf{p}.\mathbf{x}}=\tilde{c}_\Lambda\cdot{}S^\star_{-h,j,m}(c,\Omega_{\mathbf{p}})\delta_{h,h'}\delta(\omega-\omega_{\mathbf{p}}) \ .
\end{equation}
Finally, we can use this in the Bogoliubov transformations to obtain
\begin{eqnarray}
\hat{a}_{h,\lambda}^\dagger &\propto& \int d\Omega_{\mathbf{p}}\cdot{}S_{-h,j,m}(c,\Omega_{\mathbf{p}})\hat{a}_{h,\mathbf{p}}^\dagger \\
\hat{b}_{h,\lambda}^\dagger &\propto& \int d\Omega_{\mathbf{p}}\cdot {}S^\star_{h,j,m}(c,\Omega_{\mathbf{p}})\hat{b}_{h,\mathbf{p}}^\dagger \ , 
\end{eqnarray} where the prefactor is independent of the angular variables. These expressions show how a state of a particle/anti-particle with helicity $h$ decomposes into plane wave states with the same helicity and momentum orientations $\Omega_{\mathbf{p}}$. The probability of a certain orientation is given by the square modulus of the spheroidal function with spin weight $\mp h$.

Since massless scalar, spinor and vector perturbations all follow the same master equation, these conclusions apply similarly to the remaining cases.


\end{document}